\documentclass[preprint2]{emulateapj}

\newcommand{\bzcat}{Roma-BZCAT}

\newcommand{\chn}{{\it Chandra}}

\newcommand{\fer}{{\it Fermi}}

\newcommand{\swf}{{\it Swift}}

\newcommand{\xmm}{{\it XMM-Newton}}
\newcommand{\wse}{{\it WISE}}

\usepackage{graphicx}
\usepackage{longtable}
\usepackage{mdwlist}
\usepackage{natbib}
\usepackage{longtable}
\usepackage{rotating}
\usepackage{morefloats}
\usepackage{threeparttable}
\usepackage{wasysym}
\usepackage{txfonts}
\usepackage{gensymb}
\usepackage{multirow}
\usepackage{upgreek}
 \usepackage{booktabs}
\usepackage[verbose]{placeins}
\usepackage{blindtext}

\slugcomment{version \today: nuria}
\shorttitle{Optical spectroscopic observations of blazars}
\shortauthors{\'Alvarez Crespo et al. 2014}
 
\begin{document}
\title{Optical spectroscopic observations of $\gamma$-ray blazar candidates V. TNG, KPNO and OAN observations of blazar candidates of uncertain type in the Northern Hemisphere}

\author{
N. \'Alvarez Crespo\altaffilmark{1,2},
N. Masetti\altaffilmark{3},
F. Ricci\altaffilmark{4},
M. Landoni\altaffilmark{5},
V. Pati\~no-\'Alvarez\altaffilmark{6},
F. Massaro\altaffilmark{1,2},
R. D'Abrusco\altaffilmark{7},
A. Paggi\altaffilmark{7},
V. Chavushyan\altaffilmark{6}, 
E. Jim\'enez-Bail\'on\altaffilmark{8},
J. Torrealba\altaffilmark{6},
L. Latronico\altaffilmark{2}, 
F. La Franca\altaffilmark{4},
Howard A. Smith\altaffilmark{7}
\&
G. Tosti\altaffilmark{9}
}

\altaffiltext{1}{Dipartimento di Fisica, Universit\`a degli Studi di Torino, via Pietro Giuria 1, I-10125 Torino, Italy}
\altaffiltext{2}{Istituto Nazionale di Fisica Nucleare, Sezione di Torino, I-10125 Torino, Italy}
\altaffiltext{3}{INAF - Istituto di Astrofisica Spaziale e Fisica Cosmica di Bologna, via Gobetti 101, 40129, Bologna, Italy}
\altaffiltext{4}{Dipartimento di Matematica e Fisica, Universit\`a Roma Tre, via della Vasca Navale 84, I-00146, Roma, Italy}
\altaffiltext{5}{INAF-Osservatorio Astronomico di Brera, Via Emilio Bianchi 46, I-23807 Merate, Italy}
\altaffiltext{6}{Instituto Nacional de Astrof\'{i}sica, \'Optica y Electr\'onica, Apartado Postal 51-216, 72000 Puebla, M\'exico}
\altaffiltext{7}{Harvard - Smithsonian Astrophysical Observatory, 60 Garden Street, Cambridge, MA 02138, USA}
\altaffiltext{8}{Instituto de Astronom\'{\i}a, Universidad Nacional Aut\'onoma de M\'exico, Apdo. Postal 877, Ensenada, 22800 Baja California, M\'exico}
\altaffiltext{9}{Dipartimento di Fisica, Universit\`a degli Studi di Perugia, 06123 Perugia, Italy}

\begin{abstract}
The extragalactic $\gamma$-ray sky is dominated by emission from blazars, a peculiar class of active galactic nuclei (AGNs). Many of the $\gamma$-ray sources 
 included in \fer\ -Large Area Telescope Third Source
catalog (3FGL) are classified as a blazar candidate of uncertain type (BCU) because there is no optical spectra available in the literature to confirm their nature.
In 2013 we started a spectroscopic campaign to look for the optical counterparts of the BCUs and of the Unidentified  $\gamma$-ray Sources. 
The main goal of our investigation is to confirm the blazar nature of these sources having peculiar properties
 as compact radio emission and/or selected on the basis of their infrared (IR) colors. Whenever possible we also determine their redshifts. Here we present the results of the
observations carried out in the Northern hemisphere in 2013 and 2014 at Telescopio Nazionale Galilleo (TNG), Kitt Peak National Observatory (KPNO) and Observatorio
Astron\'omico Nacional (OAN) in San Pedro M\'artir. 
In this paper we describe the optical spectra of 25 sources. We confirmed that all the 15 BCUs observed in our campaign and included in our sample
are blazars and we estimated the redshift for 3 of them. In addition, 
we present the spectra for 3 sources classified as BL Lacs in the literature but with no optical spectra available to date. We found that one of them is a quasar (QSO) at a redshift $z$ = 0.208 and the other 2
are BL Lacs. Moreover, we also present 7 new spectra for known blazars listed in the \bzcat\ having an uncertain redshift or being classified as BL Lac candidates. We found that one of them, 
5BZB J0724+2621 is a `changing look' blazar. According to the spectrum available in the literature it was classified as a BL Lac but in our observation we clearly detected
 a broad emission line that lead to classify this source as a QSO
at $z$=1.17.
\end{abstract}


\section{Introduction}
\label{sec:intro}
Blazars are radio-loud active galactic nuclei (AGNs) characterised by non-thermal emission over the entire electromagnetic spectrum, from the radio band to $\gamma$ rays \citep[see e.g. ][]{giommi13}.
They show rapid variability from hours timescales in the optical band up to minutes in the $\gamma$ rays (see e.g. Aharonian 2000; Homan et al. 2002), 
high linear polarization from the radio to the optical frequencies \citep[see e.g.][]{marscher10,agudo14},  compact radio emission (see e.g. Taylor et al. 2007; Lister et al. 2009), flat radio spectra and superluminal motions 
(see e.g. Vermeulen \& Cohen 1994; Kellermann et al. 2013 and references therein).
 The spectral energy distribution is characterised by a double bump, the first component 
 peaking in infrared (IR)/optical wavelengths and the second one in X-rays (for more details see e.g. Giommi \& Padovani 1994, Inoue \& Takahara 1996).
 They are strong $\gamma$-ray emitters reaching luminosities up to $10^{49}\ erg\ s^{-1}$ as reported in both the \fer -LAT First Source Catalog \citep{abdo10} and the \fer -LAT Second Source Catalog \citep{nolan12}.

 According to  Stickel et al. (1991) blazars are divided into two main subclasses: BL Lac objects which present featureless optical spectra or with 
 emission/absorption lines of rest frame equivalent width $EW  < 5\ \AA$, and flat spectrum radio quasars 
 that show quasar-like optical spectra with broad emission lines ($EW > 5\, \AA$).
In the following we label the former class as BZBs and the latter as BZQ  according to the nomenclature adopted in \bzcat\ `Multifrequency Catalog of BLAZARS'  \citep{bzcat4}.
In this catalogue there are listed BL Lac candidates as sources claimed to be BL Lacs in the literature but no optical spectra was found to confirm their classification. 
There are also sources classified as blazars of uncertain type (BZUs), adopted for sources with peculiar characteristics, similar to those previously mentioned but also showing blazar 
activity like occasional presence/absence of broad spectral lines, transition objects between
 a radio galaxy and a BL Lac or galaxies hosting a low luminosity radio nucleus.
 In the latest version of the \bzcat\ \citep{5bzcat} there is one more class,  BL Lacs whose optical spectra exhibit a typical elliptical galaxy spectrum with a low Ca H\&K break contrast, indicated as BZGs.

With a density of the order of 0.1 sources/degree$^2$,
blazars constitute the most numerous population of extragalactic $\gamma$-ray sources, about $38\%$.
However, almost $20\%$ of the sources above 100 MeV in the \fer -LAT Third Source Catalog  \citep{abdo14} are blazar candidates of uncertain type (BCUs). 
They present flat radio spectra and/or X-ray counterpart and have a multifrequency
behaviour similar to blazars but there is no optical spectra to precisely allow this classification \citep{ackermann12}. 
In addition, Unidentified $\gamma$-ray Sources (UGSs) represent $\sim$ $33\%$ of the \fer\ -LAT Third Source Catalog  
and a large fraction of these sources can be associated to blazars  \citep{paper4}. 
Knowing how much of the emission in $\gamma$-rays comes from blazars is important
to set constraints on dark matter scenarios (Mirabal et al. 2012; Berlin \& Hooper 2014),
to discover new classes of $\gamma$-ray emitters, to resolve the $\gamma$-ray sky and to determine the origin of the extragalactic $\gamma$-ray background \citep{ajello15}.
For this purpose, several methods to recognise blazars as the low-energy counterparts of UGSs have been developed. 
For example, in the 3-dimensional IR color space generated by WISE photometry $\gamma$-ray emitting blazars lie in a region distinct from those where most of the other extragalactic sources dominated by thermal emission \citep[][]{ugs2,wibrals} are located. 
In addition, 
radio follow up observations of the Fermi UGSs (e.g., Kovalev 2009; Hovatta et al. 2012, 2014; Petrov et al. 2013; Schinzel et al. 2014)
correlation of the peculiar IR colors with existing radio databases \citep{ugs3} and X-ray follow-up
observations looking for X-ray counterparts (Paggi et al. 2013; Takeuchi et al. 2013) have been performed.
Statistical studies based on $\gamma$-ray source
properties have also allowed us to recognise the nature of
the potential counterparts for UGSs (e.g., Ackermann
et al. 2012; Mirabal et al. 2012; Hassan et al. 2013; Doert \& Errando 2014).
However none of these methods can be conclusive without optical spectroscopic confirmation for both BCUs and UGSs. 

Since 2013 we have been carrying out a spectroscopic campaign to observe the blazar-like sources of uncertain classification as well as those selected according to the methods previously listed.
In this fifth paper of the series, we present the results of optical spectroscopic observations of BCUs carried out in the Northern hemisphere at Kitt Peak National Observatory (KPNO) in Tucson (USA), 
Telescopio Nazionale Galileo (TNG) in La Palma (Spain) and Observatorio Astron\'omico Nacional (OAN) in San Pedro M\'artir (Mexico) between October 2013 and July 2014.
Exploratory program obtained with TNG, OAN and Multiple Mirror Telescope (MMT) were described in Paggi et al. (2014); 
in addition, results for observations carried out in 2013 with SOAR and KPNO were presented in Landoni et al. (2015), Massaro et al. (2015b,2015c) and Ricci et al. (2015).
In this paper we focus mainly on BCUs, although we also had the opportunity to observe \bzcat\ sources.
  
 The paper is organised as follows: Section~\ref{sec:sample} contains the sample selection. We present our dataset and discuss the data reduction procedures in Section ~\ref{sec:obs}. Then in Section ~\ref{sec:results} 
we report the details on the cross-matches with multifrequency databases  and catalogs of the observed targets
and present the results
  of our analysis for different types of sources. Finally Section ~\ref{sec:conclusions}
  is devoted to our summary and conclusions.
 We use cgs units unless otherwise stated. Spectral indices, $\alpha$, are defined by flux density $S_\nu \propto \nu^{-\alpha}$. Flat spectra are defined as $\alpha < 0.5$.

\section{Sample Selection}
\label{sec:sample}

Our final goal is to perform spectroscopic observations of a large sample of $\gamma$-ray 
blazar candidates selected on the basis of the IR colors \citep{massaro11}
and extracted from the WISE Blazar-like Radio-Loud Source (WIBRaLS) Catalog 
\citep{wibrals}. Our  observing strategy, successfully employed during the last years 
\citep[see e.g.][]{landoni14,ricci14}, consists in requesting small subsamples of our main list 
to different telescopes to minimize the impact on their schedule. 
A more comprehensive presentation of our observing strategy, a detailed summary of the results of the spectroscopic observations and their interpretation 
will be presented in a forthcoming paper  \citep{dabrusco16}.

In Figure 1 we show the IR colors of the selected $\gamma$-ray blazar candidates in comparison with those of the known gamma-ray 
blazars associated in the 2FGL that were used to build the training sample (i.e., locus) to perform the all-sky search (see D'Abrusco et al. 2014 for more details).
 
The main criterion followed for the selection of the sources that we observed from the pool of potential targets has been mainly driven by the source visibility 
during the nights obtained at each telescope.
We chose our targets considering the optimal conditions of visibility and airmass 
lower than 1.6. In addition interesting sources, such as blazar with uncertain redshifts
and/or without an optical spectrum available in the literature have been also observed 
during gaps in our observing schedule.
During our observing nights we also decided to re-observe sources when optimal conditions became available.
For this reason, we decided to point at them because given to the optical variability of 
blazars there could be a chance to
observe the source during a low state and thus detect emission and/or absorption
 features that enable a redshift measurement (Vermeulen et al. 1995; Falomo \& Pian 2014).

We updated the $\gamma$-ray classification of the selected sources on the basis of the 
recent release of the 3FGL. Thus all our targets are now
indicated as Blazar candidates of uncertain type (BCU) similar to the old class 
labelled as active galaxies of uncertain type (AGUs) in both the 1FGL and the 2FGL. 

Our sample contains 25 sources grouped as follows:

\begin{itemize}

\item Fifteen sources classified as blazar candidates of uncertain type (BCU) 
in the 3FGL which present IR colors similar to blazars,
flat radio spectrum and/or an X-ray counterpart that appear to have a 
multifrequency behaviour similar to blazars but there are no optical spectra 
to precisely allow these classifications. In particular, 12  
are sources that belong to the WIBRaLS Catalog. The remaining 3 are one 
BZB (WISE J173605.25+203301.1), one BZQ (WISE J043307.54+322840.7)
and one AGU (WISE J065340.46+281848.5), respectively, with no optical spectrum available 
in the literature, they are now grouped with the previous 12 since all of them are 
classified as BCU in the 3FGL.

\item Three sources selected from the 3FGL claimed as BL Lacs in 
the literature but without optical spectra available to confirm it.

\item Two sources that are BL Lac candidates, not necessary $\gamma$-ray emitters, 
classified according to the criteria of the ROMA-BZCAT \citep{massaro11}. 

\item The remaining 5 sources are classified in the Roma-BZCAT as BZBs, but their 
redshifts are still uncertain. 

\end{itemize}

In Table 1 we also report the 1FGL and the 2FGL names together with their old 
classifications and their assigned counterpart. We noticed that the latest associations 
for 3FGL J0653.6+2817 and 3FGL J0433.1+3228, differ from the previous one 
listed in the 3FGL.
However for this two cases the source pointed during our campaign is the 3FGL one as 
reported in Table 2. On the other hand, for 3FGL J1013.5+3440 the source was unidentified 
in the 2FGL catalog and we pointed the potential counterpart WISE J101349.6+344550.8 
\citep[listed in][]{5bzcat} instead of the one assigned in the 3FGL.

In Table  2 we report our results and multifrequency
notes for each source to verify additional information
that can support the blazar-like behaviour.

\begin{figure*}{!}
\begin{center}
\includegraphics[height=7.5cm,width=8.2cm,angle=0]{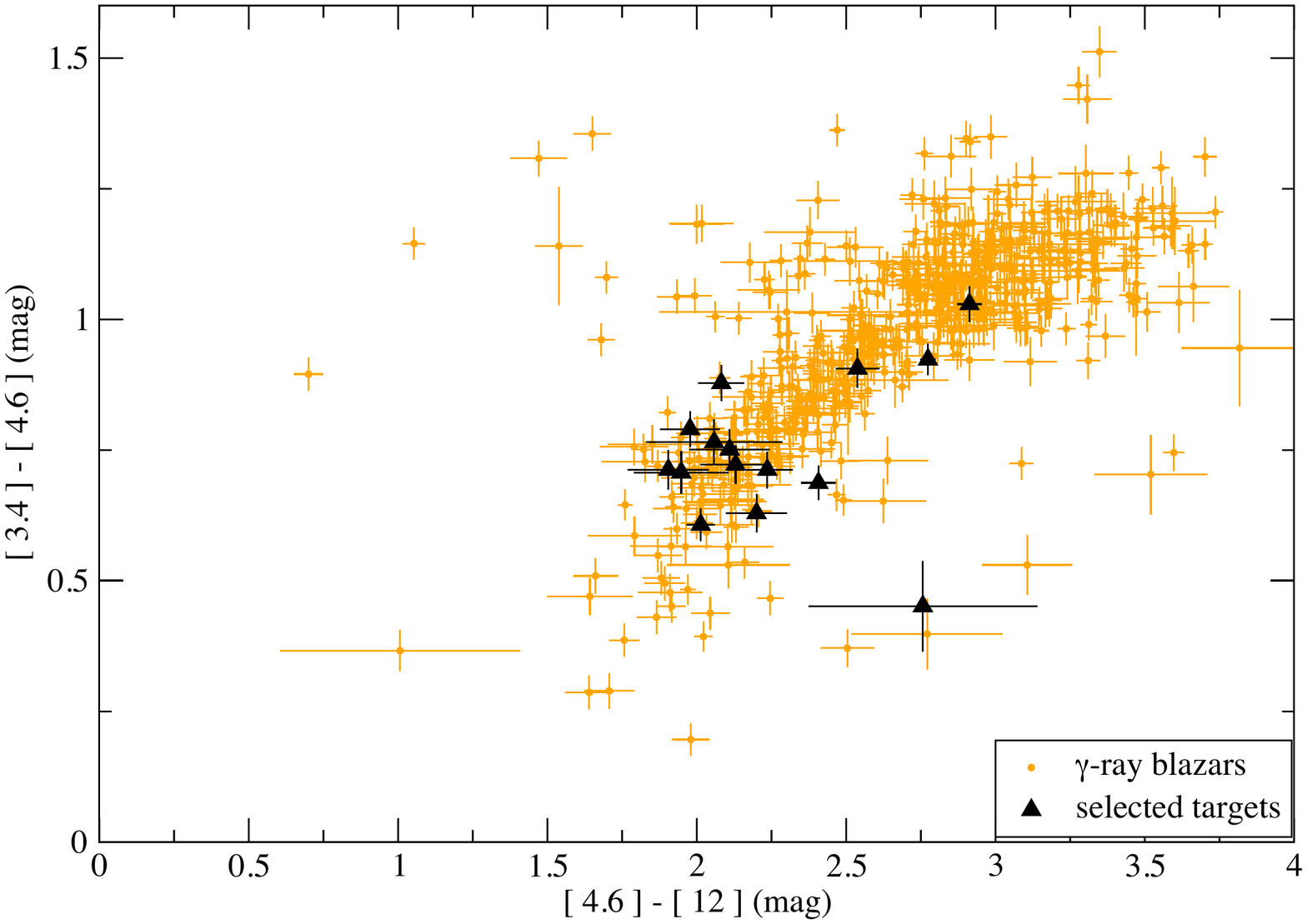} 
\includegraphics[height=7.5cm,width=8.2cm,angle=0]{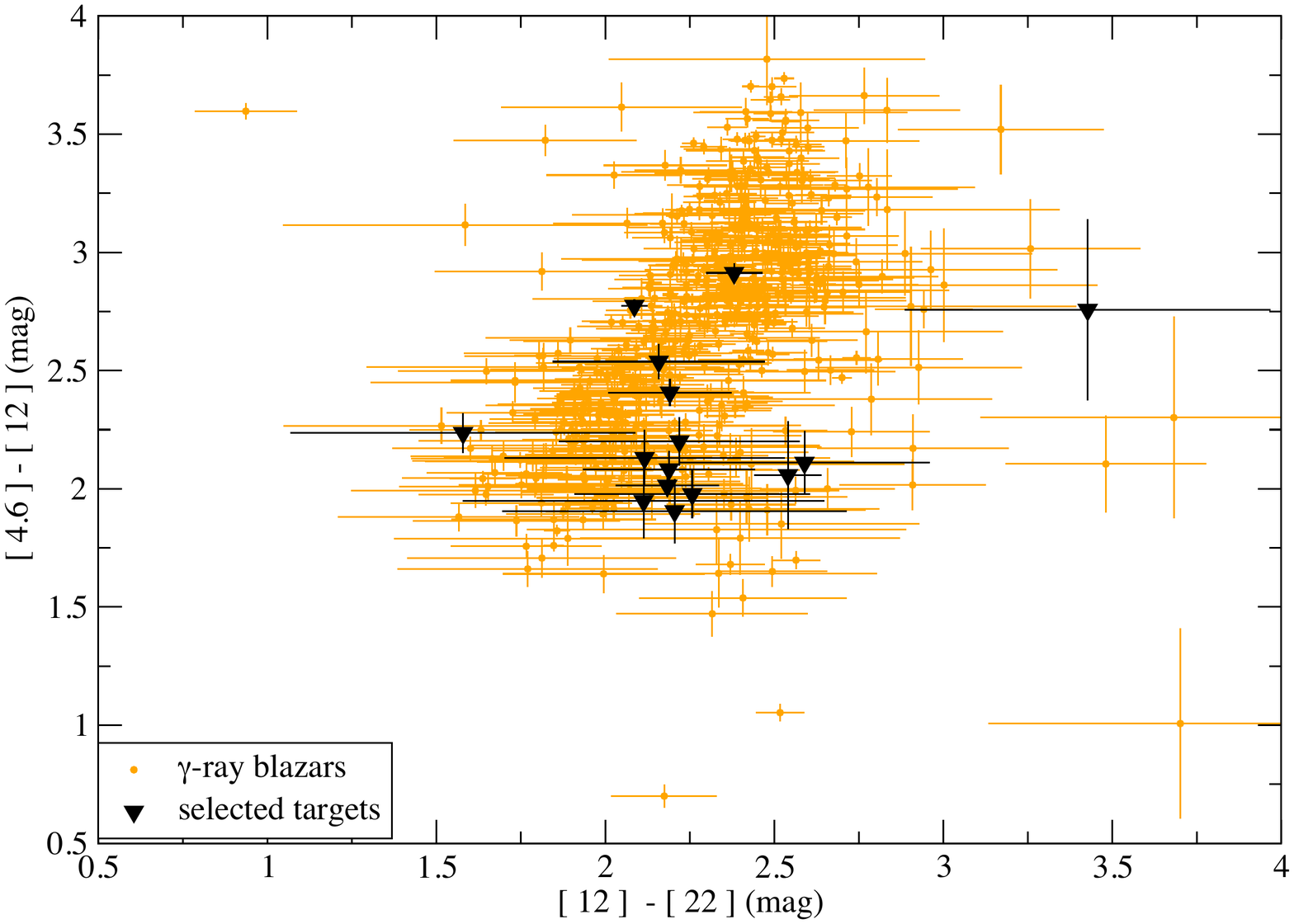} 
\end{center}

\caption{In orange the gamma-ray blazars from the WIBRaLS catalog that define the so called {\it locus} {(i.e., the 3-dimensional region of the IR colors parameter space occupied by the associated gamma-ray blazars in the 2FGL with a WISE counterpart, see D'Abrusco et al. 2014 for additional details).}
\emph{Left:} Projection of the selected targets in the \wse\ gamma-ray strip in the [3.4] - [4.6] versus [4.6] - [12] color-color plane.
\emph{Right:} Projection of the selected targets in the [4.6] - [12] versus [12] - [22] color-color plane.}
\label{fig:wise}
\end{figure*}

\section{Observations and Data Reduction}
\label{sec:obs}

\subsection{Kitt Peak National observatory}
The spectra of twelve objects were obtained in remote observing mode at KPNO Mayall 4-m class telescope using the R-C spectrograph the nights 5th February and 4th June 2014.
We adopted a slit width of 1\arcsec.2 and a low resolution gratings (KPC10A and BL181 depending on the availability at the telescope) 
yielding a dispersion of 3 $\textrm{\AA}$ pixel$^{-1}$ in both cases.
The average seeing during both runs was about 1\arcsec\ and conditions were clear. 
Wavelength calibration was accomplished using the spectra of an Helium-Neon-Argon lamp which guarantees a smooth coverage over the entire range. Due to poor long term stability during each night
we needed to take into account flexures of the instruments and drift, so we took an arc frame before every target to guarantee a good wavelength solution for the scientific spectra. The accuracy reached 
is $\sim$3 $\textrm{\AA}$ rms.

\subsection{Telescopio Nazionale Galileo}
Eight spectra were obtained
using the 3.58-m Telescopio Nazionale Galileo (TNG) located at La Palma, Canary Islands (Spain). Its imaging spectrograph DOLoReS carried a 2048 x 2048 pixel
E2V 4240 CCD; spectra were acquired with the LR-B grism and a 1\arcsec.5 slit width which secured a nominal spectra coverage in the 3500-8200 \AA\ 
range and a dispersion of 2.5 $\textrm{\AA}$ pixel$^{-1}$ The TNG data were acquired between October 2013 and July 2014. 
We adopted the same data reduction procedure for TNG as for KPNO observations. 
 Wavelength calibration was achieved with the spectra of an Helium-Neon  lamp acquired between 2 exposures of a same object.

\subsection{Observatorio Astron\'omico Nacional San Pedro M\'artir}
Five objects were observed with the 2.1 m telescope of the Observatorio Astron\'omico Nacional (OAN) in San Pedro M\'artir (Mexico) on September and October 2014.
The telescope carries a Boller \& Chivens spectrograph and a 1024x1024 pixel E2V-4240 CCD. The slit width was 2\arcsec.5.
The spectrograph was tuned in the $\sim$ 4000 - 8000 \AA\ range with a resolution of 10 $\textrm{\AA}$ pixel$^{-1}$. Wavelength calibration was done using the spectra of a cooper-Helium-Neon-Argon lamp.

The data reduction has been performed according to our standard procedures. Further  details are given in  Masetti et al. (2013), Ricci et al. (2015) and Massaro et al. (2015c).

The set of spectroscopic data acquired was optimally extracted  and reduced following standard procedures with IRAF \citep[]{horne86,tody86}.
For each acquisition we performed bias subtraction, flat field correction and cosmic rays rejection. To remove cosmic rays we achieved 2 or 3 individual exposures for each target and 
averaged them according to their signal to noise ratios.
We then exploited the availability of the two individual exposures in the case of dubious detected 
spectral features to better reject spurious ones.

We dereddened the spectra for the galactic absorption  assuming $E_{B - V}$ values taken by Schlegel et al. (1998) relation.
Although our program did not require precise photometric calibration, we observed a spectrophotometric standard star to
perform relative flux calibration on each spectrum. The overall spectral shape is correct, but the absolute calibration may suffer from
sky condition issues, such seeing and transparency. To detect faint spectral features, especially because our targets might be BL Lac objects, aimed at estimating redshifts, 
we also present normalised spectra.

\begin{sidewaystable}[htbf]
\begin{center}
\caption{bcu = AGN of uncertain type; ugs = unidentified $\gamma$-ray source; bzb = bll = BL Lac; bzq = blazar of QSO type.}
\bigskip
\resizebox{\textwidth}{!}{
\begin{tabular}{lllllllllll}
\noalign{\smallskip}
\hline
\noalign{\smallskip}
1FGL  &  1FGL & 1FGL &                2FGL  &   2FGL &  2FGL & 3FGL & 3FGL  &  3FGL \\
name  &   class   & association  &   name &     class & association & name   & class &association \\ 
\hline
\hline
Fermi Active Galaxies of Uncertain type\\
\hline
              &     &                        &               &     &                       & J0015.7+5552 & bcu & GB6 J0015+5551 \\      
              &     &                        &               &     &                       & J0148.3+5200 & bcu & GB6 J0148+5202   \\    
              &     &                        &               &     &                       & J0145.6+8600 & bcu & NVSS J014929+860114\\  
              &     &                        &               &     &                       & J0219.0+2440 & bcu & 87GB 021610.9+243205 \\
 J0433.5+3230 & bzq & CRATES J0433+3237      &  J0433.7+3233 & bzq & MG2 J043338+3236      & J0433.1+3228 & bcu & NVSS J043307+322840  \\
              &     &                        &  J0653.7+2818 & bcu & NVSS J065345+282010   & J0653.6+2817 & bcu & GB6 J0653+2816       \\
 J0659.9+1303 & ugs &                        &               &     &                       & J0700.2+1304 & bcu & GB6 J0700+1304       \\
              &     &                        &               &     &                       & J0728.0+4828 & bcu & GB6 J0727+4827       \\
 J1322.1+0838 & ugs &                        &               &     &                       & J1322.3+0839 & bcu & NVSS J132210+084231  \\
              &     &                        &               &     &                       & J1434.6+6640 & bcu & 1RXS J143442.0+664031\\
 J1511.8-0513 & ugs &                        &  J1511.8-0513 & ugs &                       & J1511.8-0513 & bcu & NVSS J151148-051345  \\
 J1647.4+4948 & agn & CGRaBS J1647+4950      &  J1647.5+4950 & agn & SBS 1646+499          & J1647.4+4950 & bcu & SBS 1646+499         \\
 J1735.7+2031 & ugs &                        &  J1735.9+2033 & bzb & NVSS J173605+203301   & J1736.0+2033 & bcu & NVSS J173605+203301  \\
              &     &                        &  J1913.4+4440 & bcu & 1RXS J191401.9+443849 & J1913.9+4441 & bcu & 1RXS J191401.9+443849\\
              &     &                        &               &     &                       & J2156.0+1818 & bcu & RX J2156.0+1818      \\
  \hline
  Fermi BL Lacs with no optical spectra\\
  \hline
              &     &                       &  J0103.5+5336  & bcu & 1RXS J010325.9+533721 & J0103.4+5336* & bll & 1RXS J010325.9+533721\\
              &     &                       &  J1013.6+3434  & ugs &                       & J1013.5+3440$^+$  & fsrq& OL 318               \\
              &     &                       &  J2021.5+0632  & ugs &                       & J2021.9+0630* & bll & 87GB 201926.8+061922 \\
\hline
BZB candidates in the Roma-BZCAT \\
\hline
 J0814.5-1011 & ugs &                       &  J0814.0-1006  & bzb & NVSS J081411-101208   & J0814.1-1012 & bll & NVSS J081411-101208  \\
  \hline
  BZBs listed in the the Roma-BZCAT with uncertain z\\
  \hline
 J0333.7+299  & ugs &                       & J0333.7+2918   & bcu & TXS 0330+291          & J0333.6+2916 & bll & TXS 0330+291     \\    
              &     &                       &  J0941.9-0755  & bcu & PMN J0942-0800        & J0942.1-0756* & bll & PMN J0942-0800    \\   
 J2038.1+6552 & ugs &                       &  J2036.6+6551  & bzb & 87GB 203539.4+654245  & J2036.4+6551 & bll & 87GB 203539.4+654245 \\
 J2323.5+4211 & bzb & 1ES 2321+419          &  J2323.8+4212  & bzb & 1ES 2321+419          & J2323.9+4211 & bll & 1ES 2321+419         \\

\hline
\noalign{\smallskip}
\end{tabular}
}
\end{center}
Column description. (1): 1FGL name, (2): 1FGL class, (3): 1FGL association, (4): 2FGL name, (5): 2FGL class, (6): 2FGL association, (7): 3FGL name, (8): 3FGL class, (9): 3FGL association.
($^*$) These objects are classified as BL Lac in the 3FGL because we provided this information while the present paper was in preparation.

($^+$) For this source we did not point the associated 3FGL counterpart but a different radio source that lies within the Fermi positional uncertainty region at 95\% level of confidence.
\end{sidewaystable}

\begin{sidewaystable}[htbf]
\begin{center}
\caption{Description of the selected sample. Our sources are divided in 4 subsamples:
1) Sources classified as active galaxies of uncertain type according to the 
3LAC; 2) \fer\ sources classified as BL Lacs in the literature without optical spectra available; 
3) BZB candidates in the \bzcat.
4)BL Lac candidates, both detected and not detected by \fer\
for which no optical spectroscopic information were found in the literature or BZBs with uncertain/unknown redshift estimate.}
\bigskip
\resizebox{\textwidth}{!}{
\begin{tabular}{lllllllllll}
\noalign{\smallskip}
\hline
\noalign{\smallskip}
3FGL & Alternative & R.A.   & Dec.    & Telescope & Obs. Date   & Exp. & SNR  & multifrequency notes$^*$ & z & class\\
name     & name & (J2000) & (J2000) & & (yyyy-mm-dd) & (sec) & & & & \\ 
\hline
\hline
Fermi Active Galaxies of Uncertain type\\
\hline
 J0015.7+5552& WISE J001540.13+555144.7 & 00:15:40.1 & +55:51:44 & KPNO & 2014-02-05 & 2x1200 & 10 & N, 87, GB, rf, w, M, X, x & ? & BL Lac\\
 J0148.3+5200& WISE J014820.33+520204.9 & 01:48:20.2 & +52:02:06 & OAN & 2014-10-01 & 3x1800 & 26 & L, N, 87, GB, rf, w & ? & BL Lac\\
 J0145.6+8600& WISE J014935.28+860115.4 & 01:49:29.8 & +86:01:14 & TNG & 2013-10-12 & 2x1200 & 55 & W, N, w & 0.15 & BL Lac/galaxy\\
 J0219.0+2440& WISE J021900.40+244520.6 & 02:19:00.4 & +24:45:20 & OAN & 2014-10-02 & 3x1800 & 43 & N, w & ? & BL Lac\\
 J0433.1+3228& WISE J043307.54+322840.7 & 04:33:07.7 &+32:28:40 & KPNO & 2014-02-05 & 2x600 & 5 & N, 87, GB, w, rf & ? & BL Lac\\
 J0653.6+2817& WISE J065340.46+281848.5 & 06:53:40.2 & +28:18:49 & TNG & 2014-03-26 & 2x1200 & 77 & N, w & $>0.45$ & BL Lac/galaxy\\
 J0700.2+1304& WISE J070014.31+130424.4 & 07:00:14.3 & +13:04:24 & KPNO & 2014-02-05 & 2x1350 & 12 & N, w & ? & BL Lac\\
 J0728.0+4828& WISE J072759.84+482720.3 & 07:27:59.9 & +48:27:20 & TNG & 2014-05-07 & 2x1200 & 39 & L, N, 87, GB, w & ? & BL Lac\\
 J1322.3+0839& WISE J132210.17+084232.9 & 13:22:10.2 & +08:42:33 & KPNO & 2014-02-05 & 2x1350 & 13 & N, w, s, g & ? & BL Lac\\
 J1434.6+6640& WISE J143441.46+664026.5 & 14:34:41.8 & +66:40:27 & TNG & 2014-03-26 & 2x1200 & 38 & N, w, X & ? & BL Lac\\
 J1511.8-0513& WISE J151148.56-051346.9 & 15:11:48.6 & -05:13:45 & TNG & 2014-07-02 & 2x1200 & 49 & N, w, SED in Takeuchi+13 & ? & BL Lac\\
 J1647.4+4950& WISE J164734.91+495000.5 & 16:47:34.9 & +49:50:00 & KPNO & 2014-06-05 & 2x300 & 22 & N, 87, GB, c, rf, w, M, s, g, X,  ($z$ = 0.047 Falco+98) & 0.049 & QSO\\
 J1736.0+2033& WISE J173605.25+203301.1 & 17:36:05.3 & +20:33:01 & TNG & 2014-05-31 & 2x1200 & 41 & N, w, U, g, X & ? & BL Lac\\
 J1913.9+4441& WISE J191401.88+443832.2 & 19:14:01.9 & +44:38:33 & OAN & 2014-09-30 & 3x1800 & 23 & N, w & ? & BL Lac\\
 J2156.0+1818& WISE J215601.64+181837.1 & 21:56:01.6 & +18:18:39 & OAN & 2014-10-02 & 3x1800 & 12 & N, U, g, X & ? & BL Lac\\
  \hline
  Fermi BL Lacs with no optical spectra\\
  \hline
 J0103.4+5336& WISE J010325.89+533713.4  & 01:03:25.9 & +53:37:13 & KPNO & 2014-02-05 & 2x600 & 10 & N, 87, w, rf, X & ? & BL Lac\\
 J1013.5+3440& WISE J101336.51+344003.6 & 10:13:36.0 & +34:40:06 & TNG & 2014-03-26 & 2x1200 & 18 & N, w, X & 0.208 & QSO\\
 J2021.9+0630& WISE J202155.45+062913.6  & 20:21:55.5 & +06:29:14 & KPNO & 2014-02-05 & 2x1200 & 35 & Pm, N, 87, w & ? & BL Lac\\
\hline
BZB candidates in the Roma-BZCAT \\
\hline
             & 5BZBJ0724+2621           & 07:24:42.7 & +26:21:31 & KPNO & 2014-02-05 & 2x1800 & 20 & N, w, M,  (White+2000) & 1.17 & QSO\\
 J0814.1-1012& 5BZBJ0814-1012           & 08:14:11.7 & -10:12:10 & KPNO & 2014-02-05 & 2x1200 & 35 & N, A, w, 6 & ? & BL Lac\\
  \hline
  BZBs listed in the the Roma-BZCAT with uncertain z\\
  \hline
 J0333.6+2916& 5BZB J0333+2916          & 03:33:49.0 & +29:16:31 & TNG & 2013-10-12 & 2x1200 & 96 & T, N, 87, GB, rf, U, X, ($z >$ 0.14 Shaw+13) & ? & BL Lac\\
 J0942.1-0756& 5BZB J0942-0759          & 09:42:00.6 & -07:55:17 & KPNO & 2014-02-05 & 2x1350 & 9 & N, w,  ($z >$ 0.46 Shaw+13) & ? & BL Lac\\
             & 5BZG J1414+3430          & 14:14:09.3 & +34:30:57 & KPNO & 2014-02-05 & 2x600 & 5 & N, F, w, M, s, X,  ($z$ = 0.275 Bauer+00) & ? & BL Lac\\
 J2036.4+6551& 5BZB J2036+6553          & 20:36:19.9 & +65:53:14 & KPNO & 2014-02-05 & 2x3000 & 10 & N, 87, w,  ($z >$ 0.3 Shaw+13) & ? & BL Lac\\
 J2323.9+4211& 5BZB J2323+4210          & 23:23:52.1 & +42:10:58 & OAN & 2014-10-25 & 3x1800 & 32 & N, w, M, X,  ($z$ = 0.059 Padovani+95,  Shaw+13,  Massaro+15c) & ? & BL Lac\\

\hline
\noalign{\smallskip}
\end{tabular}
}
\end{center}
Column description. (1): 3FGL name, (2): Alternative name, (3): Right Ascension (Equinox J2000), (4): Declination (Equinox J2000), (5): Telescope: 
Kitt Peak National observatory (KPNO); Telescopio Nazionale Galileo (TNG); Observatorio Astron\'omico Nacional San Pedro M\'artir (OAN),
 (6): Observation Date, (7): Exposure time, (8): Signal to Noise Ratio, (9): multifrequency notes (see Table 3),
(10): redshift, (11): source classification
($^*$) Symbols used for the multifrequency notes are all reported in Table 3
together with the references of the catalogs/surveys.
\end{sidewaystable}

\begin{table}[htbf]
\caption{List of catalogs in which we searched for additional multifrequency information.}
\resizebox{\textwidth}{!}{
\begin{tabular}{llll}
\noalign{\smallskip}
\hline
\noalign{\smallskip}
Survey/Catalog name & Acronym & Reference & Symbol \\
\hline
\hline
VLA Low-Frequency Sky Survey Discrete Source Catalog & VLSS & Cohen et al. (2007) & V \\
Westerbork Northern Sky Survey & WENSS & Rengelink et al. (1997) & W \\
Parkes-MIT-NRAO Surveys & PMN & Wright et al. (1994) & Pm \\
Texas Survey of Radio Sources & TEXAS & Douglas et al. (1996) & T \\
Low-frequency Radio Catalog of Flat-spectrum Sources & LORCAT & Massaro et al. (2014b) & L \\
\hline
NRAO VLA Sky Survey & NVSS & Condon et al. (1998) & N \\
VLA Faint Images of The Radio Sky at Twenty-Centimeter & FIRST & Becker et al. (1995), White et al. (1997) & F \\
87 Green Bank catalog of radio sources  & 87 GB & Gregory et al. (1991) & 87 \\ 
Green Bank 6-cm Radio Source Catalog & GB6 & Gregory et al. (1996) & GB \\
\hline
 WISE  all-sky survey in the Allwise Source catalog & WISE & Wright et al. (2010) & w \\
 Two Micron All Sky Survey & 2MASS & Skrutskie et al. (2006) & M \\
 \hline
 Sloan Digital Sky Survey Data Release 9 & SDSSS DR9 & Ahn et al. (2012) & s \\
 Six-degree-Field Galaxy Redshift Survey & 6dFGS & Jones et al. (2004) & 6 \\
 \hline
 ROSAT Bright Source Catalog & RBSC & Voges et al. (1999) & X \\
 ROSAT Faint Source Catalos & RFSC & Voges et al. (2000) & X \\
\xmm\ Slew Survey & XMMSL & Saxton (2008), Warwick et al. (2012) & x \\
 Deep Swift X-Ray Telescope Point Source Catalog & 1XSPS & Evans et al. (2014) & x \\
 \chn\ Source Catalog & CSC & Evans et al. (2010) & x \\
\hline
\noalign{\smallskip}
\end{tabular}
}
Column description. (1): Survey/Catalog name, (2): Acronym, (3): Reference, (4): Symbol used in multifrequency notes in Table 2.
\end{table}

\section{Results}
\label{sec:results}
Here we report all the results of our optical analysis for the selected sources divided in four groups.
We have searched for additional multifrequency information that can support the blazar-like nature in radio, IR, optical and
X-ray surveys and in both NASA Extragalactic Database (NED) and SIMBAD Astronomical Database. All the selected sources with their corresponding multifrequency notes
 are listed in Table 2. There is also a note in cases where the spectral energy distribution (SED) of the source is presented in Takeuchi et al. (2013) 
 and if the radio counterpart has a flat radio spectrum (marked as `rf'). The surveys and catalogs used to search for the counterparts of our targets are listed in Table 3.
Note that we use the same symbol for the X-ray catalogs of \xmm\,\chn\ and \swf\ because these X-ray observatories perform only pointed  
  observations. There is the possibility that the pointed observation related to the field of the \fer\ source was not requested as follow up but for another reason. Therefore, the discovery
  of the X-ray counterpart for an associated and/or identified source could be serendipitous.
  
\subsection{Gamma-ray Active Galaxies of Uncertain type}
Details of the 15 BCUs observed in our sample are listed below. Multifrequency notes relative to each source are reported in Table 2.  
The spectra of the whole sample are shown in the Figures 3 to 25 together with their finding charts.
 Our spectroscopic observations allow us to confirm that all these sources are BL Lac objects and we have
been able to determine the redshift in one case in which the BL Lac shows absorption features from the host elliptical galaxy. For another source we could fix a lower redshift limit 
due to the presence of an intervening system seen along the line of sight. 

The spectra of WISE J014935.28+860115.4 associated with 3FGL J0145.6+8600 is dominated by the emission of the host elliptical galaxy rather than by non-thermal continuum arising from the jet, the features distinguished are: 
G band, doublet Ca H+K  ($EW_{obs}=0.4-0.7\AA$) and Mg I ($EW_{obs} = 0.6 \AA$). 
These features enable us to estimate a redshift of $z$ = 0.15 (see Figure ~\ref{fig:J0145}). 
In the case of the optical counterpart WISE J065344.26+281547.5 associated with 3FGL J0653.6+2817 the presence of an intervening doublet system
of Mg II ($EW_{obs} = 1.3 - 1.0 \AA$) allows us to set a lower limit on its redshift of $z >$0.45 (see Figure ~\ref{fig:J0653}). 
The spectra of WISE J164734.91+495000.5, counterpart of 3FGL J1647.4+4950 shows a broad emission line that we tentatively identify with H$\alpha$ ($EW_{obs} = 27.7$).
On the basis of our optical spectra and the radio data we classify the source as a FSRQ at a redshift $z$=0.049 (see Figure ~\ref{fig:J1647}). This redshift is consistent with the previous value published
by Falco et al. (1998) where the spectrum is described.
In the rest of the BCUs only  featureless continuum spectra
are shown so we can conclude they are BL Lacs but we did not fix any redshift (Figures ~\ref{fig:J0728} to  ~\ref{fig:J2156}).

\subsection{\fer\ BL Lacs with no optical spectra} 
During our northern campaign we have observed 3 sources claimed in the literature as BL Lacs  but with no optical spectra available at the time of the observation.
For the optical counterparts of the sources 3FGL J0103.4+5336 and 3FGL J2021.9+0630 we were able to confirm the BL Lac nature but not to estimate the redshift because in neither our
 spectra are there features present 
(see Figures ~\ref{fig:J0103} and ~\ref{fig:J2021} respectively). On the basis of our optical spectra we classify the counterpart of 3FGL J1013.5+3440 as a QSO at a redshift $z$ = 0.208, this was possible due to the identification of the lines
[O II] ($EW_{obs} = 17.2 \AA$), [Ne III] ($EW_{obs} = 4.1 \AA$), 
$H\beta$ ($EW_{obs} = 4.5 \AA$), the doublet [O III] ($EW_{obs} = 14.3 -  15.8 \AA$) 
and $H\alpha$ ($EW_{obs} = 48.5 \AA$) (see Figure ~\ref{fig:J1013}). This is consistent with a BZQ classification but the lack of radio data did not permit us
to verify the flatness of the radio spectrum as expected for BZQs.

\subsection{BL Lac candidates in the \bzcat}
Here we discuss the BZB candidates in the \bzcat\ in our observed sample. 
Table 2 reports the \bzcat\ name, that on the \fer\ counterpart when associated with a $\gamma$-ray source in the 3FGL.
The spectra of this sample are shown in Figures ~\ref{fig:b0724} and ~\ref{fig:b0814} together with their finding charts. We classify 5BZB J0724+2621 
as a QSO, not as a BZQ again because of the lack of radio data.  It shows a broad emission line ($EW_{obs} =  26.8 \AA$) 
that we tentatively identify as Mg II, yielding a redshift estimate of $z$ = 1.17 (see Figure ~\ref{fig:b0724}). 
The other source 5BZB J0814-1012 (a.k.a 3FGL J0814.1-1012) shows a featureless continuum which allow us to confirm its classification as a BL Lac (see Figure ~\ref{fig:b0814}).

\subsection{\bzcat\ sources with uncertain nature or unknown redshifts}
In this subsample of 5 sources
we confirmed the BL Lac nature for all of them but, unfortunately, it was not possible to establish their redshifts.
In particular we re-observed 5BZB J0333+2916 (a.k.a 3FGL J0333.6+2916) and confirm its BL Lac nature but not the lower limit of the redshift set before as  $z >$ 0.14 in Shaw et al. (2013a),
because our observations show only a featureless continuum (see Figure ~\ref{fig:J0333}).
The same situation occurs for 5BZB J0942-0759 (a.k.a 3FGL J0942.1-0756), the spectra is dominated by a featureless continuum and it is not possible to determine 
the redshift ($z >$ 0.46 in Shaw et al. 2013a, see Figure ~\ref{fig:J0942}).
 In the case of the source 5BZG J1414+3430  we have not been able
 to confirm the previous redshift value of $z$ = 0.275 in White et al. (2000) (see Figure ~\ref{fig:b1414}).
For the source 5BZB J2036+6553 (a.k.a 3FGL J2036.4+6551) we do not see any features on top of the non-thermal continuum and the confirmation of $z >$ 0.30 given in Shaw et al. (2013a)
was not possible (see Figure ~\ref{fig:J2036}).
Finally for the last source 5BZB J2323+4210 (a.k.a 3FGL J2323.9+4211) the spectra is featureless so it was not possible to confirm the value $z$=0.059 given by Padovani et al. (1995).
A featureless spectra of this source was presented also by Shaw et al. (2013a) and by Massaro et al. (2015c)
(see Figure~\ref{fig:b2323}).

\section{Summary and conclusions}
\label{sec:conclusions}

We present the results of our 2013 and 2014 optical spectroscopic campaign in the Northern hemisphere with the Telescopio Nazionale Galileo (TNG), Kitt Peak National
Observatory (KPNO) and Observatorio Astron\'omico Nacional (OAN) in San Pedro M\'artir. The main goal of our program is to use optical spectroscopy to confirm the nature of sources
selected among the BCUs for having low radio frequency spectra (i.e. below $\sim$1 GHz) or peculiar IR colours.

Confirmation of blazar nature among these objects will improve and refine future associations for the \fer\ catalog. Also, once our campaign is completed, this will yield to understand the efficiency
and completeness of the association method based on IR colours. One more aim is to search for redshift estimates of the potential UGS counterparts.

During our campaign we also observed several active galaxies of uncertain type as defined according to the \fer\ catalogs
(Ackerman et al. 2011a; Nolan et al. 2012; Abdo et al. 2014)
to verify if they are blazars. 
In addition we observed several sources that already belong to the \bzcat\ because either there were no optical spectra available in the literature, or their estimated redshifts were still uncertain when the
catalog was released.

The total number of targets presented is 25. The results of this part of the spectroscopic campaign can be reported as follows:

\begin{itemize}
\item In the BCU subsample, all of the sources have a blazar nature. One of them, namely WISE J014935.28+860115.4	is dominated by absorption of the host galaxy, and were able to detect
absorption lines in the optical spectrum leading to a redshift measurement of $z$ =  0.15. We have also been  able to set a lower limit for WISE J065344.26+281547.5 of 0.45 thanks to the detection
of a Mg II intervening system.
\item We obtained the spectra of 3 sources classified as BL Lacs in the literature but with no spectra published at the time of the observations. We found that 2 of them are indeed BL Lacs but the optical
spectrum of WISE J101336.51+344003.6	shows this source is a QSO at $z$ = 0.208. 
\item For the 5 BZBs listed in the \bzcat\ with uncertain redshift estimation we were not able to obtain any z value with our observations. 
\item We also analysed 2 BL Lac candidates listed in the \bzcat. According to our results 5BZB J0814-1012 is confirmed as a BZB. The spectra of the source 5BZB J0724+2621 observed by White et al. (2000)
showed a featureless continuum corresponding to the classification as a BL Lac, but in our observations we detected a broad emission line. This strongly indicates that the source
 was previously observed during a state 
dominated by non-thermal radiation that did not allow to detect emission lines. 
During our observation we observed this previously classified BL Lac showing a broad 
emission line that led to a QSO classification.

It is a `changing look' blazar (see e.g. Giommi et al. 2012).
and we classify it as a QSO at a redshift $z$ = 1.17.
It has been suggested that this transition could happen due to a  change in the bulk Lorentz factor of the jet \citep{bianchin09}.
Other interpretation is that these blazars are instead FSRQs, whose emission lines are swamped by the relativistically boosted jet flux \citep{ghisellini12}.
\end{itemize}

We thank the anonymous referee for useful comments that led to improvements in the paper.
We are grateful to Dr. D. Hammer and Dr. W. Boschin for their help to schedule, prepare and perform the KPNO and the TNG observations, respectively.
We thank E. Cavazzuti for her help checking tables.
This investigation is supported by the NASA grants NNX12AO97G and NNX13AP20G.
H. A. Smith acknowledges partial support from NASA/JPL grant RSA 1369566.
The work by G. Tosti is supported by the ASI/INAF contract I/005/12/0.
V.C. and J.T. are supported by the CONACyT research grant 151494 (Mexico). 
Part of this work is based on archival data, software or on-line services provided by the ASI Science Data Center.
This research has made use of data obtained from the high-energy Astrophysics Science Archive
Research Center (HEASARC) provided by NASA's Goddard Space Flight Center; 
the SIMBAD database operated at CDS,
Strasbourg, France; the NASA/IPAC Extragalactic Database
(NED) operated by the Jet Propulsion Laboratory, California
Institute of Technology, under contract with the National Aeronautics and Space Administration.
Part of this work is based on the NVSS (NRAO VLA Sky Survey):
The National Radio Astronomy Observatory is operated by Associated Universities,
Inc., under contract with the National Science Foundation and on the VLA low-frequency Sky Survey (VLSS).
The Molonglo Observatory site manager, Duncan Campbell-Wilson, and the staff, Jeff Webb, Michael White and John Barry, 
are responsible for the smooth operation of Molonglo Observatory Synthesis Telescope (MOST) and the day-to-day observing programme of SUMSS. 
The SUMSS survey is dedicated to Michael Large whose expertise and vision made the project possible. 
The MOST is operated by the School of Physics with the support of the Australian Research Council and the Science Foundation for Physics within the University of Sydney.
This publication makes use of data products from the Wide-field Infrared Survey Explorer, 
which is a joint project of the University of California, Los Angeles, and 
the Jet Propulsion Laboratory/California Institute of Technology, 
funded by the National Aeronautics and Space Administration.
This publication makes use of data products from the Two Micron All Sky Survey, which is a joint project of the University of 
Massachusetts and the Infrared Processing and Analysis Center/California Institute of Technology, funded by the National Aeronautics 
and Space Administration and the National Science Foundation.
This research has made use of the USNOFS Image and Catalogue Archive
operated by the United States Naval Observatory, Flagstaff Station
(http://www.nofs.navy.mil/data/fchpix/).
Funding for the SDSS and SDSS-II has been provided by the Alfred P. Sloan Foundation, 
the Participating Institutions, the National Science Foundation, the U.S. Department of Energy, 
the National Aeronautics and Space Administration, the Japanese Monbukagakusho, 
the Max Planck Society, and the Higher Education Funding Council for England. 
The SDSS Web Site is http://www.sdss.org/.
The SDSS is managed by the Astrophysical Research Consortium for the Participating Institutions. 
The Participating Institutions are the American Museum of Natural History, 
Astrophysical Institute Potsdam, University of Basel, University of Cambridge, 
Case Western Reserve University, University of Chicago, Drexel University, 
Fermi lab, the Institute for Advanced Study, the Japan Participation Group, 
Johns Hopkins University, the Joint Institute for Nuclear Astrophysics, 
the Kavli Institute for Particle Astrophysics and Cosmology, the Korean Scientist Group, 
the Chinese Academy of Sciences (LAMOST), Los Alamos National Laboratory, 
the Max-Planck-Institute for Astronomy (MPIA), the Max-Planck-Institute for Astrophysics (MPA), 
New Mexico State University, Ohio State University, University of Pittsburgh, 
University of Portsmouth, Princeton University, the United States Naval Observatory, 
and the University of Washington.
The WENSS project was a collaboration between the Netherlands Foundation 
for Research in Astronomy and the Leiden Observatory. 
We acknowledge the WENSS team consisted of Ger de Bruyn, Yuan Tang, 
Roeland Rengelink, George Miley, Huub Rottgering, Malcolm Bremer, 
Martin Bremer, Wim Brouw, Ernst Raimond and David Fullagar 
for the extensive work aimed at producing the WENSS catalog.
TOPCAT\footnote{\underline{http://www.star.bris.ac.uk/$\sim$mbt/topcat/}} 
\citep{taylor05} for the preparation and manipulation of the tabular data and the images.
The Aladin Java applet\footnote{\underline{http://aladin.u-strasbg.fr/aladin.gml}}
was used to create the finding charts reported in this paper \citep{bonnarell00}. 
It can be started from the CDS (Strasbourg - France), from the CFA (Harvard - USA), from the ADAC (Tokyo - Japan), 
from the IUCAA (Pune - India), from the UKADC (Cambridge - UK), or from the CADC (Victoria - Canada).


\begin{figure*}
\begin{center}
\includegraphics[height=7.9cm,width=8.4cm,angle=0]{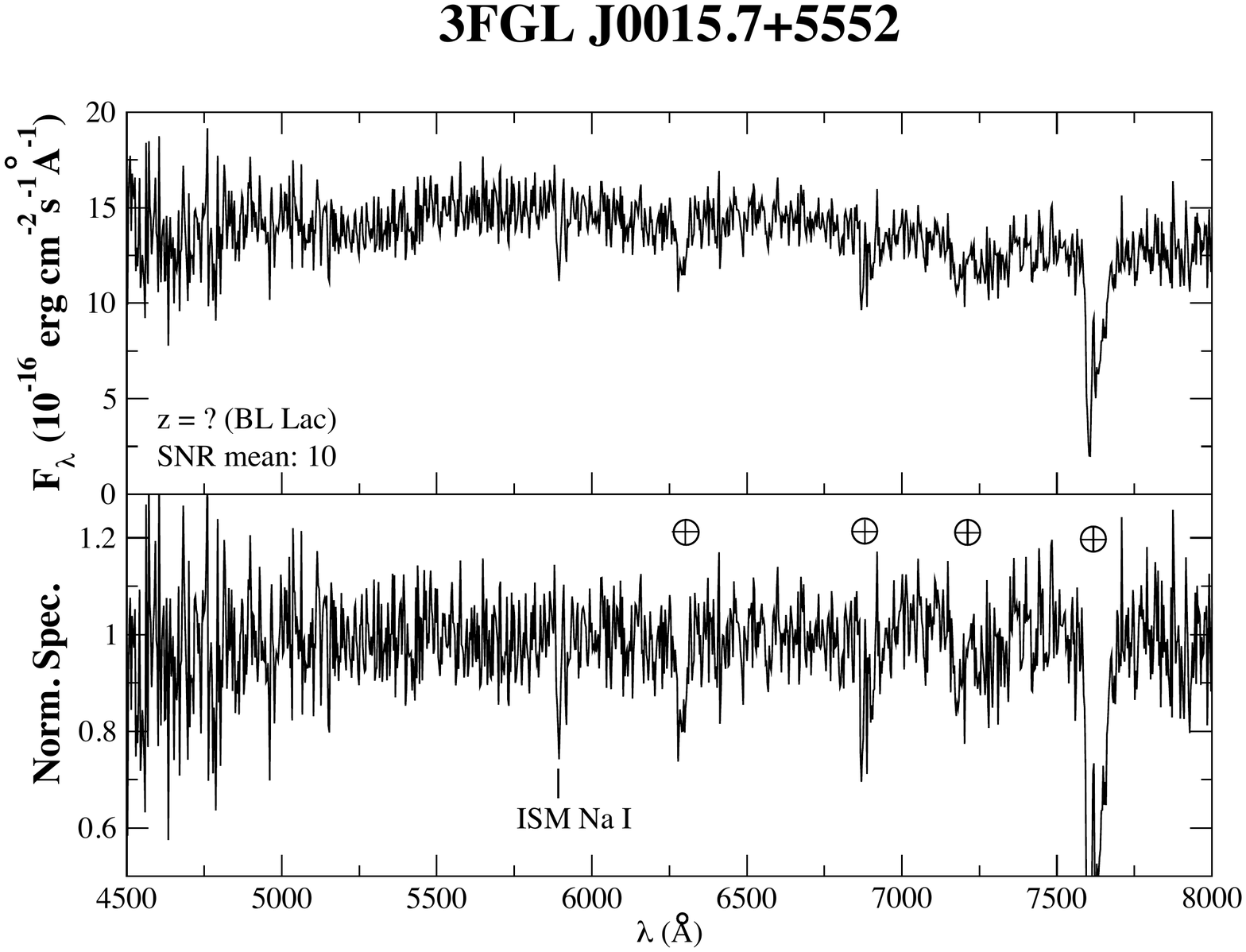} 
\includegraphics[height=6.5cm,width=8.0cm,angle=0]{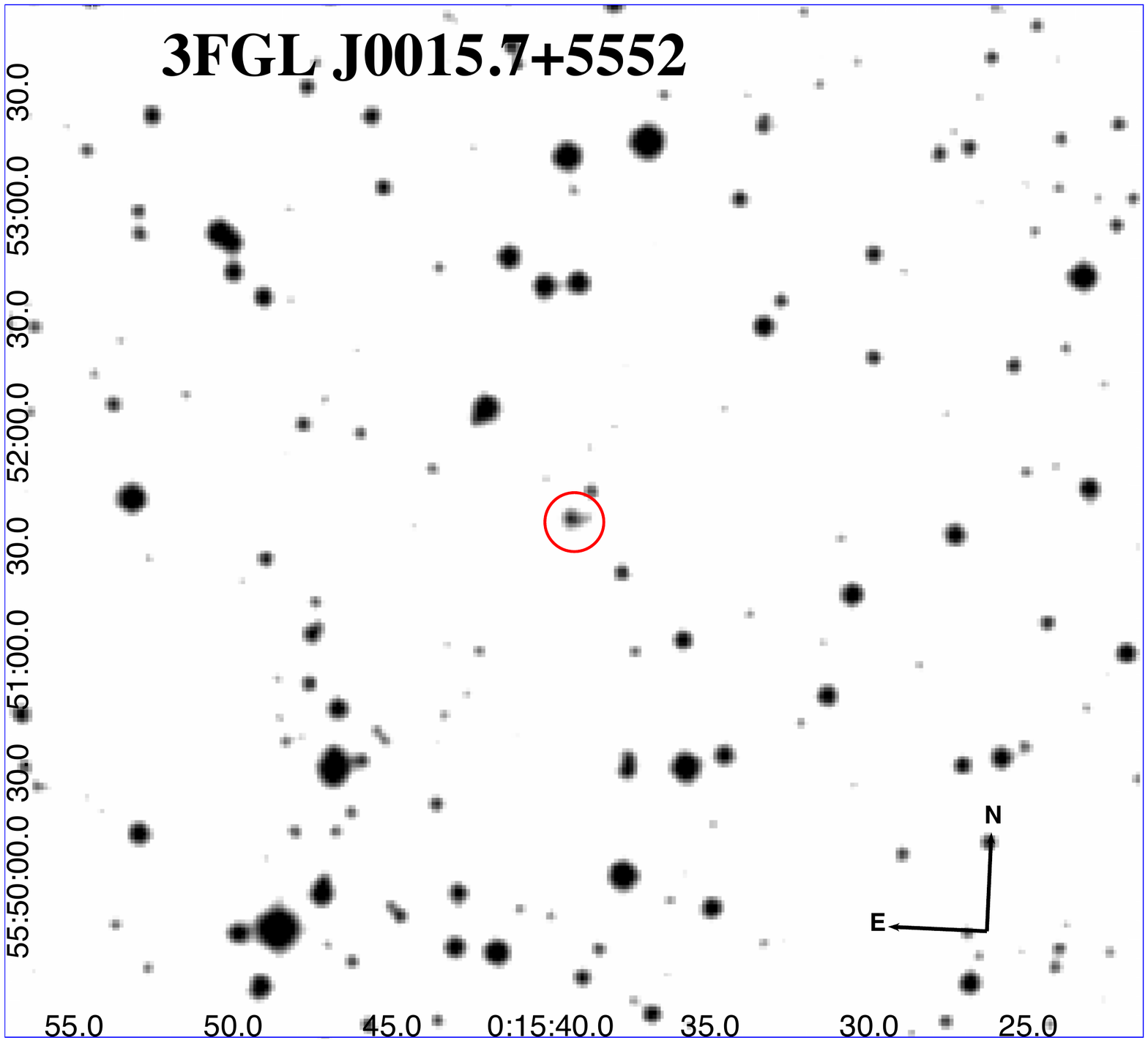} 
\end{center}
\caption{\emph{Left:} Upper panel) The optical spectra of WISE J001540.13+555144.7, potential counterpart of 
3FGL J0015.7+5552. It is classified as a BL Lac on the basis of its featureless continuum.
The average signal-to-noise ratio (SNR) is also indicated in the figure.
Lower panel) The normalized spectrum is shown here. Telluric lines are indicated with a symbol.
\emph{Right} The $5\arcmin\,x\,5\arcmin\,$ finding chart from the Digital Sky Survey (red filter). 
The potential counterpart of 3FGL J0015.7+5552
pointed during our observations is indicated by the red circle.}
\label{fig:J0015}
\end{figure*}

\begin{figure*}
\begin{center}
\includegraphics[height=7.9cm,width=8.4cm,angle=0]{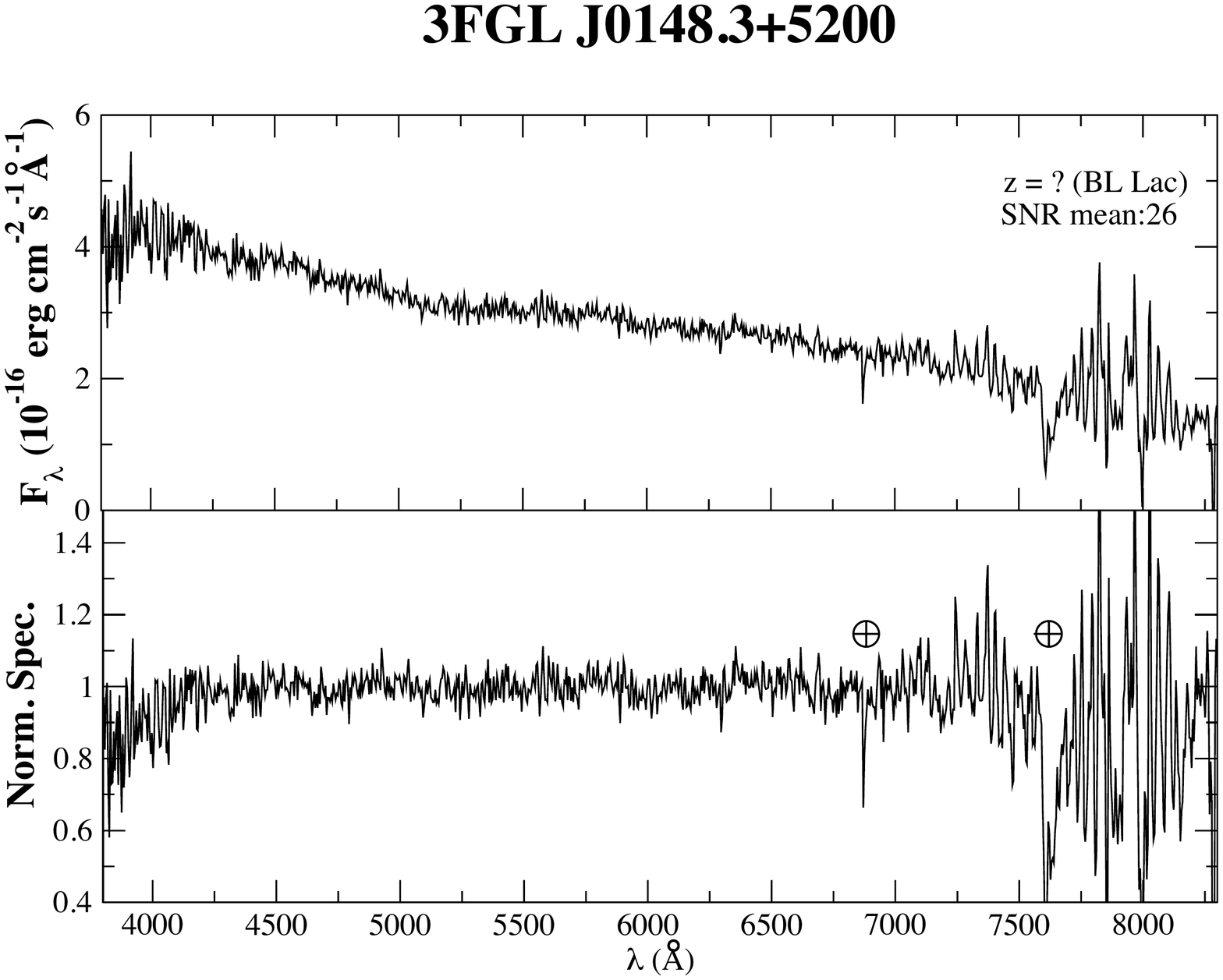} 
\includegraphics[height=6.5cm,width=8.0cm,angle=0]{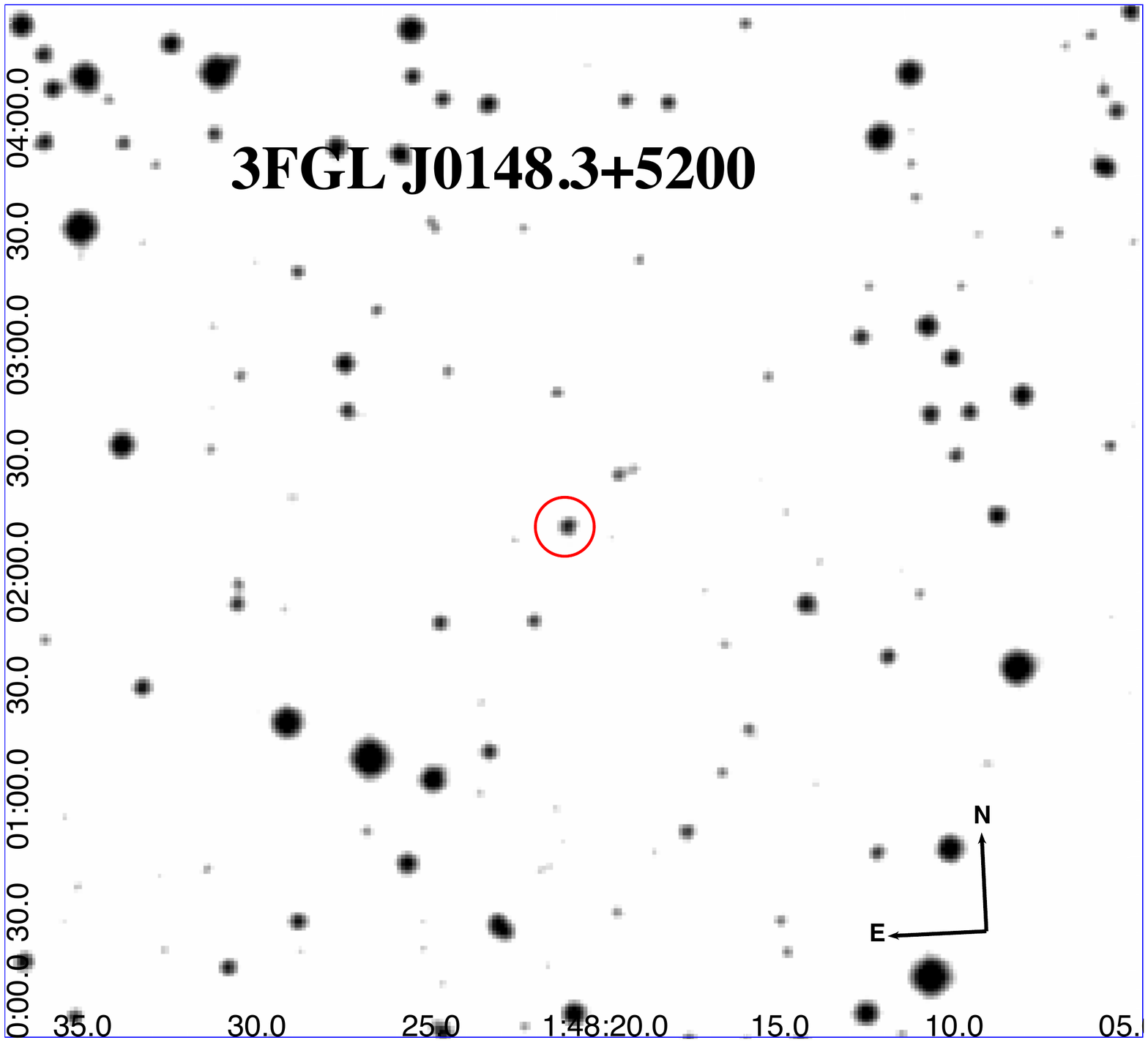} 
\end{center}
2
\caption{\emph{Leftt:} Upper panel) The optical spectra of WISE J014820.33+520204.9, potential counterpart of 
3FGL J0148.3+5200. It is classified as a BL Lac on the basis of its featureless continuum.
The average signal-to-noise ratio (SNR) is also indicated in the figure.
Lower panel) The normalized spectrum is shown here.
\emph{Right:} The $5\arcmin\,x\,5\arcmin\,$ finding chart from the Digitized Sky Survey (red filter) as in Fig. ~\ref{fig:J0015}.}
\label{fig:J0148}
\end{figure*}

\begin{figure*}
\begin{center}
\includegraphics[height=7.9cm,width=8.4cm,angle=0]{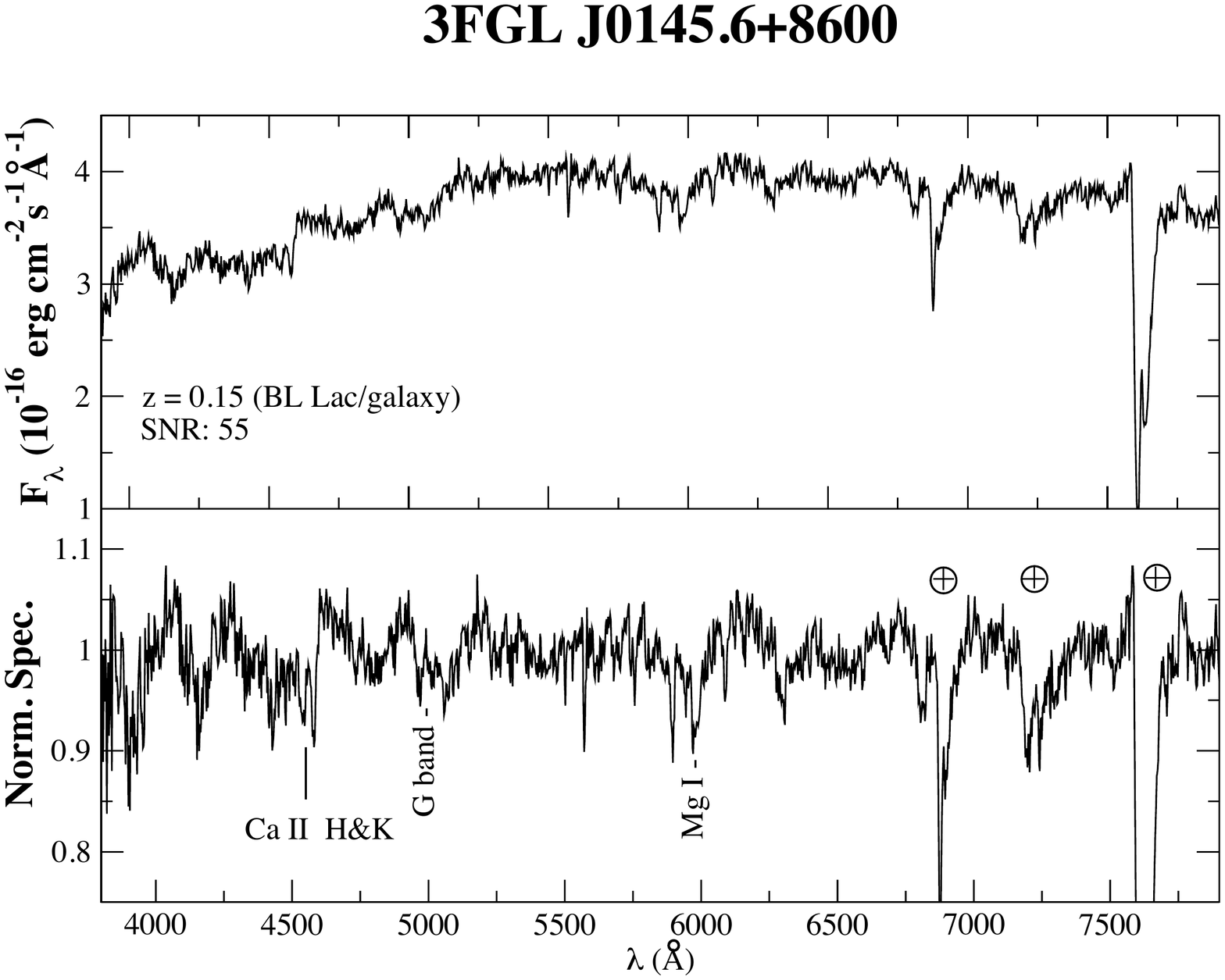} 
\includegraphics[height=7.5cm,width=8.0cm,angle=0]{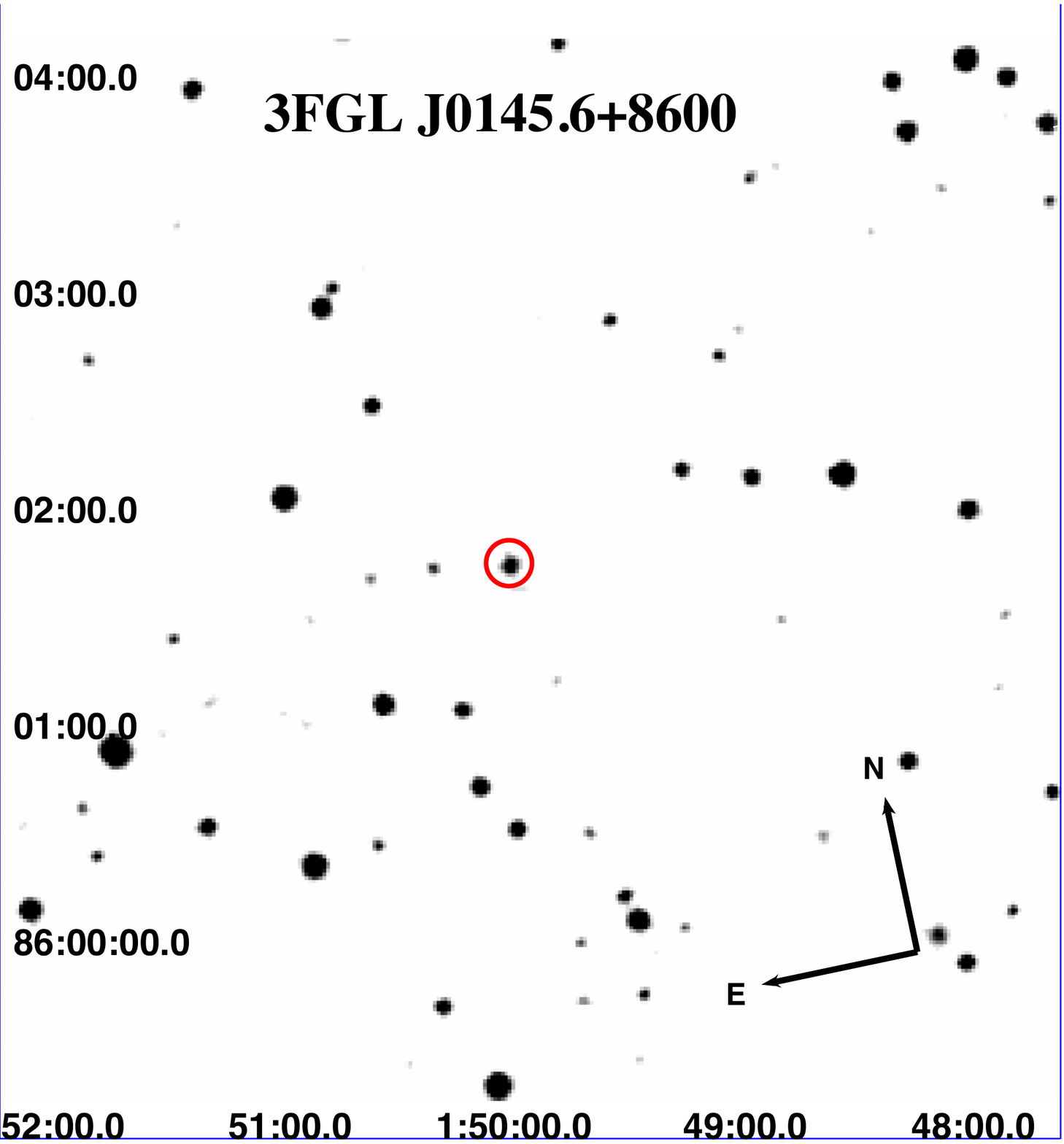} 
\end{center}

\caption{\emph{Left:} Upper panel) The optical spectra of WISE J014935.28+860115.4, potential counterpart of 
3FGL J0145.6+8600. The spectrum is dominated by the emission of the host elliptical galaxy and shows 
G band, doublet Ca H+K ($\lambda\lambda_{obs} = 4518 - 4578 \AA$) and Mg I ($\lambda_{obs} = 6028 \AA$). 
These features enable us to measure a redshift of $z$=0.15.
SNR is also indicated in the figure.
Lower panel) The normalized spectrum is shown here.
\emph{Right:} The 5$\arcmin\,x\,5\arcmin\,$ finding chart.}
\label{fig:J0145}
\end{figure*}

\begin{figure*}
\begin{center}
\includegraphics[height=7.9cm,width=8.4cm,angle=0]{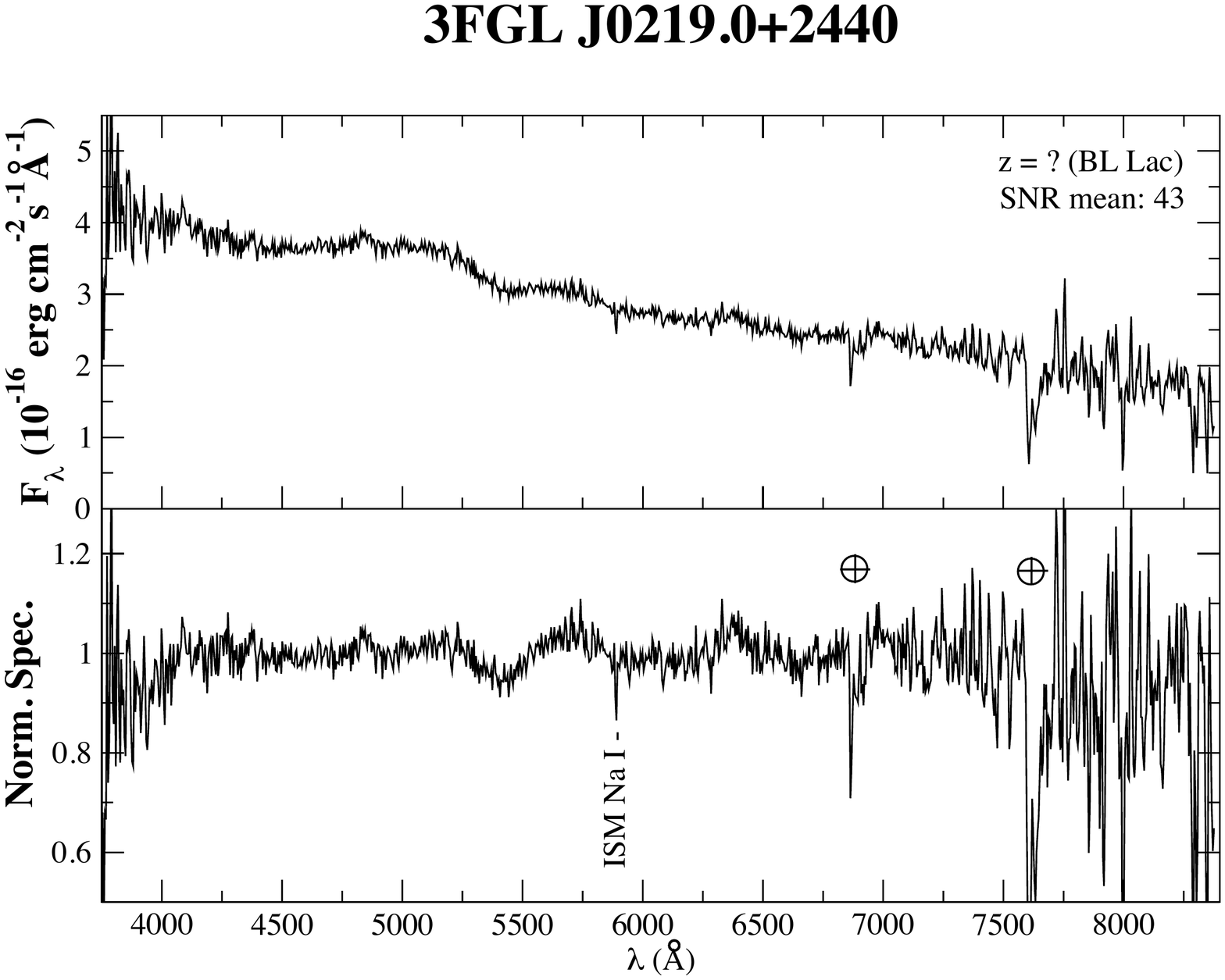} 
\includegraphics[height=6.5cm,width=8.0cm,angle=0]{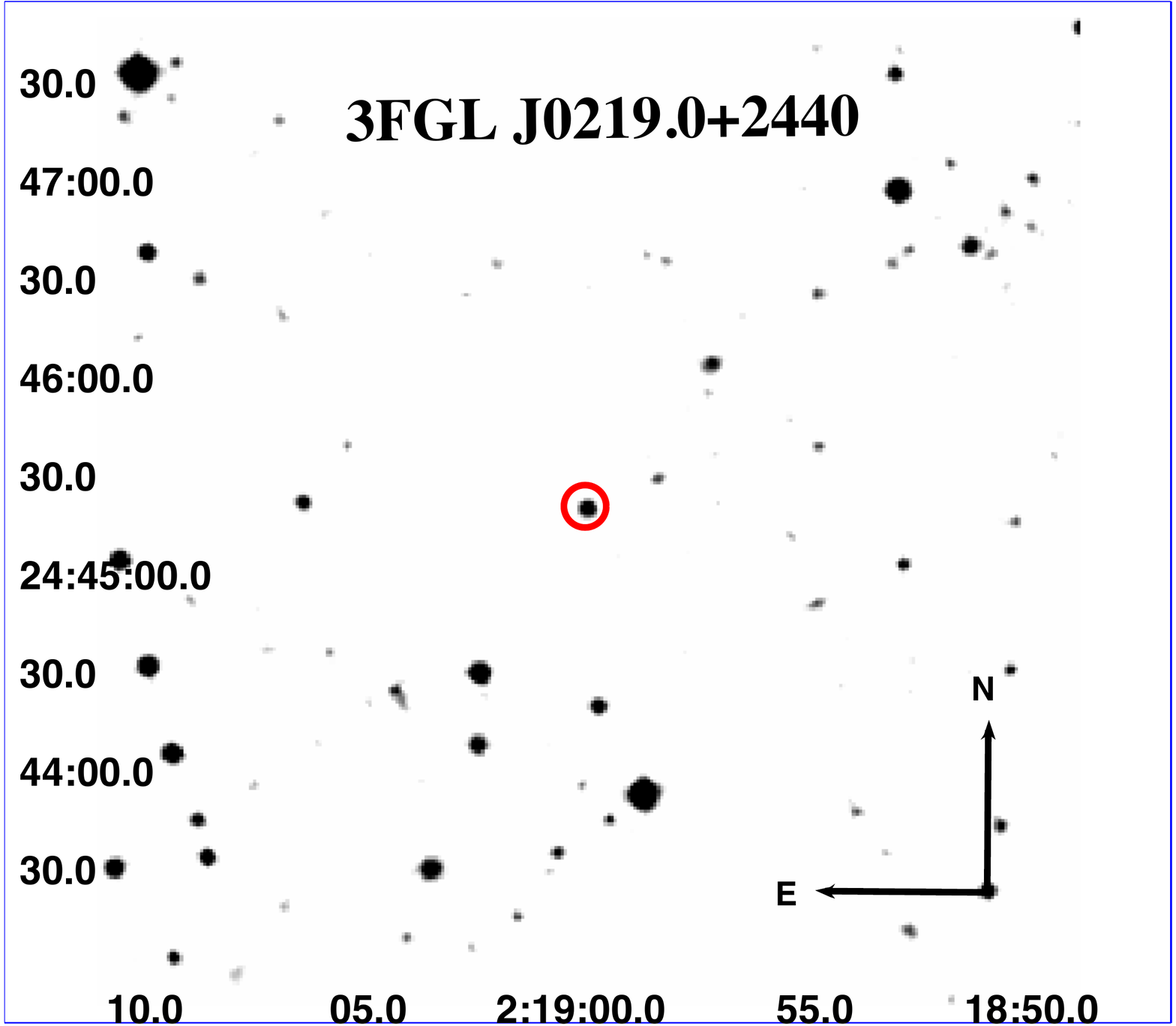} 
\end{center}

\caption{\emph{Left:} Upper panel) The optical spectra of WISE J021911.90+244052.6, potential counterpart of 
3FGL J0219.0+2440. It is classified as a BL Lac on the basis of its featureless continuum.
SNR is also indicated in the figure.
Lower panel) The normalized spectrum is shown here.
\emph{Right:} The $5\arcmin\,x\,5\arcmin\,$ finding chart.}
\label{fig:J0219}
\end{figure*}

\begin{figure*}
\begin{center}
\includegraphics[height=7.9cm,width=8.4cm,angle=0]{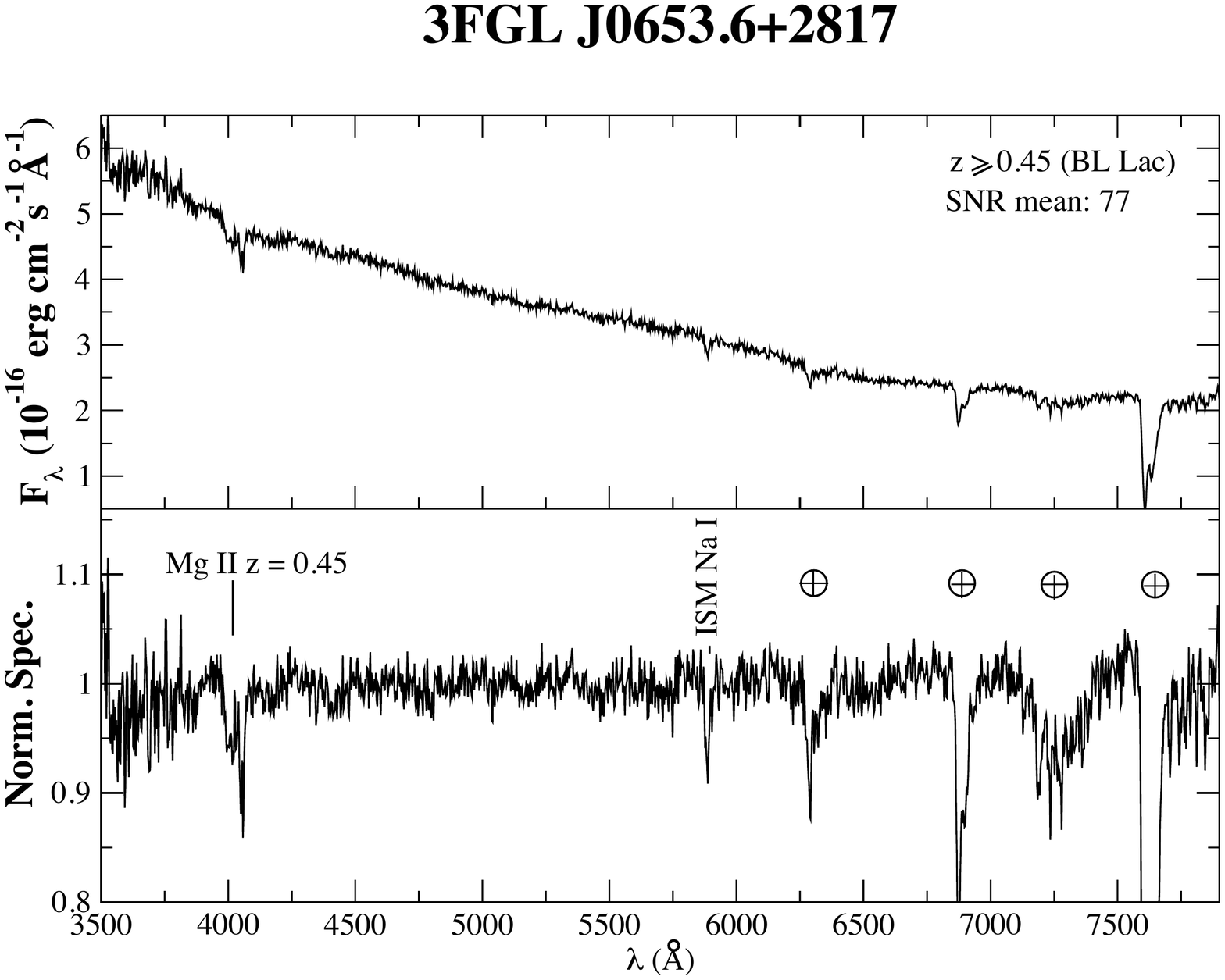} 
\includegraphics[height=6.5cm,width=8.0cm,angle=0]{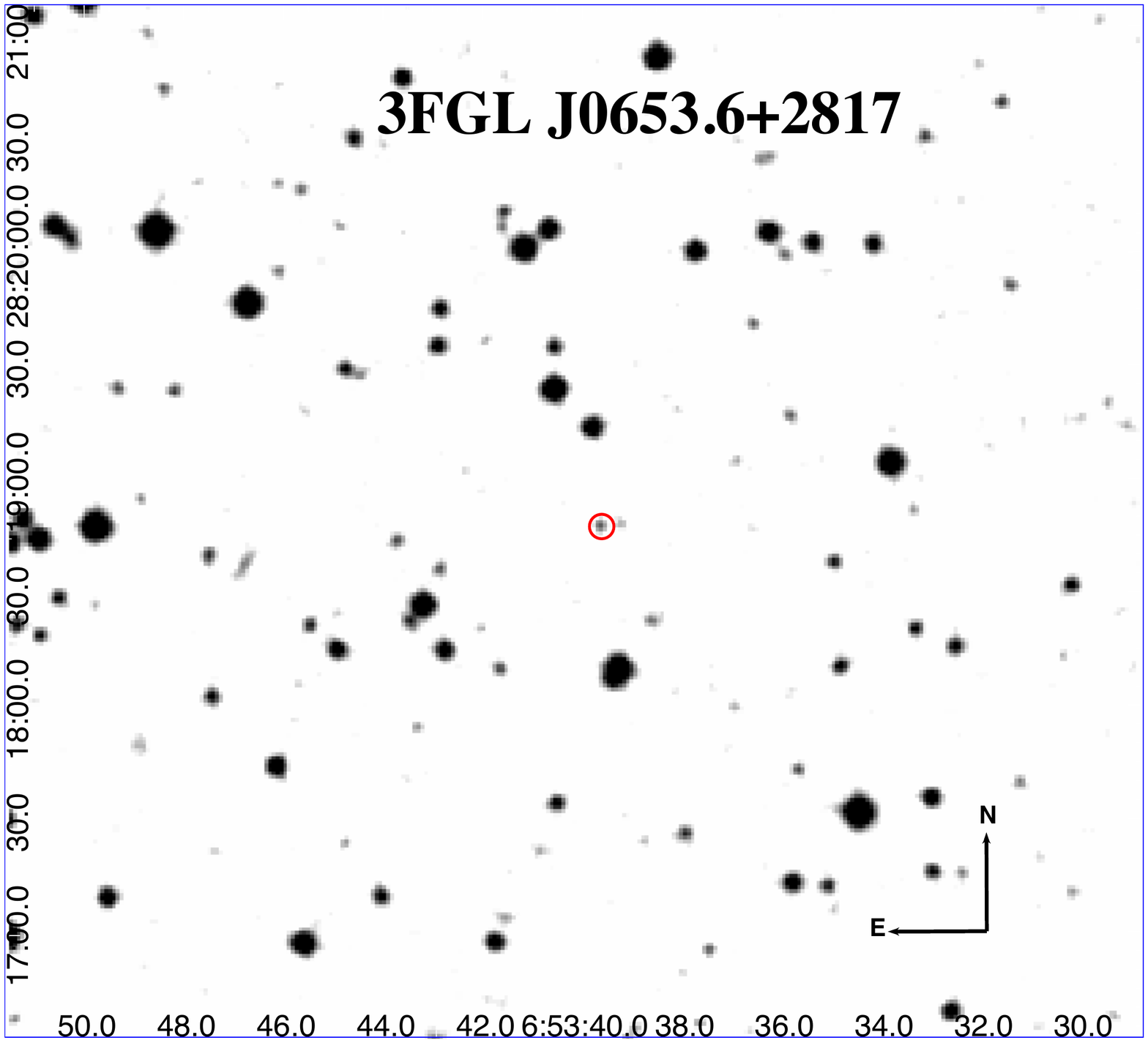} 
\end{center}

\caption{ \emph{Left:} Upper panel) The optical spectra of WISE J065344.26+281547.5, potential counterpart of 
3FGL J0653.6+2817. Our observation shows presence of an intervening doublet system
 of Mg II ($\lambda\lambda_{obs} = 4018 - 4055 \AA$) with a lower redshift limit of $z >$0.45.
SNR also indicated in the figure.
Lower panel) The normalized spectrum is shown here.
\emph{Right:} The $5\arcmin\,x\,5\arcmin\,$ finding chart.}
\label{fig:J0653}
\end{figure*}

\begin{figure*}
\begin{center}
\includegraphics[height=7.9cm,width=8.4cm,angle=0]{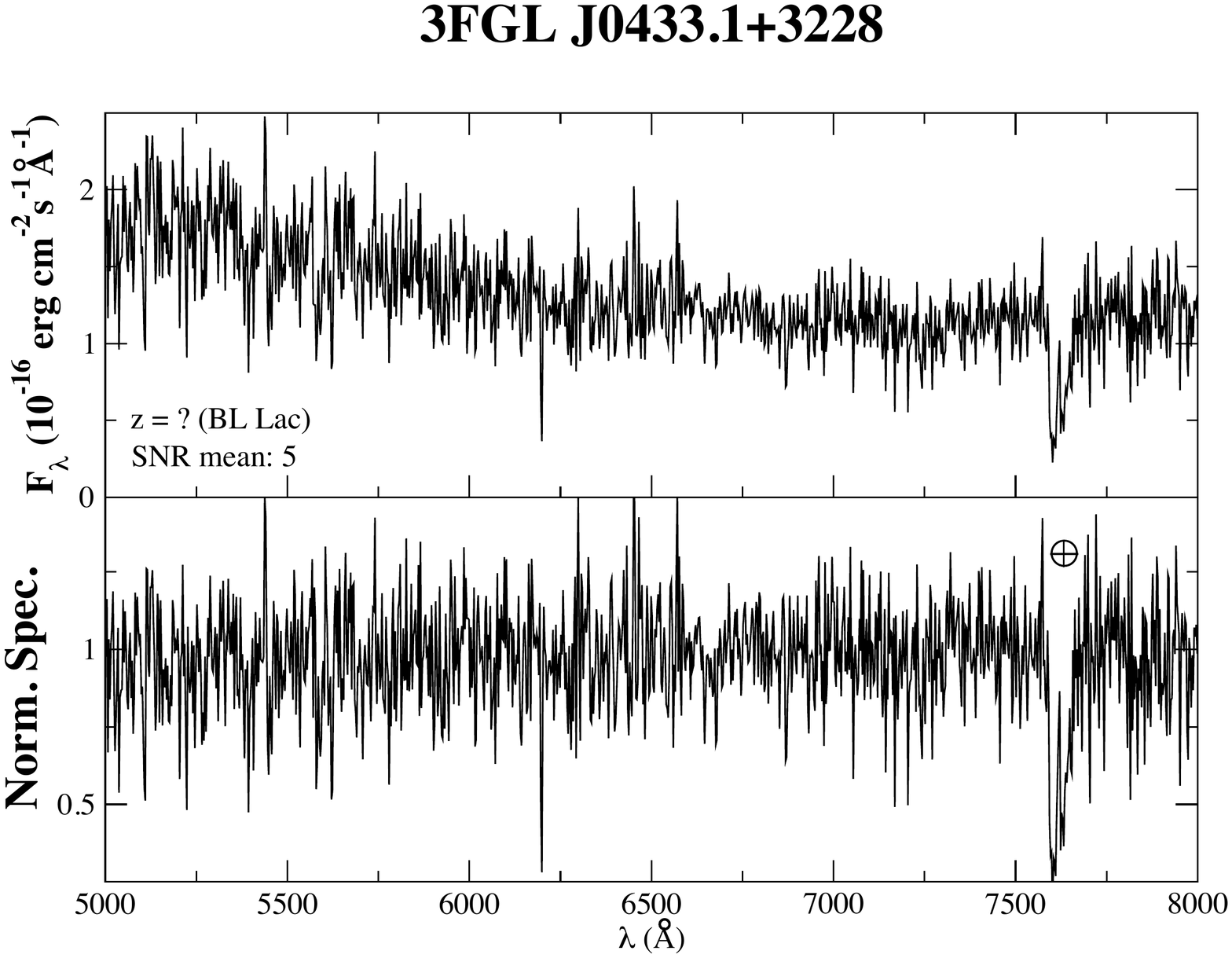} 
\includegraphics[height=6.5cm,width=8.0cm,angle=0]{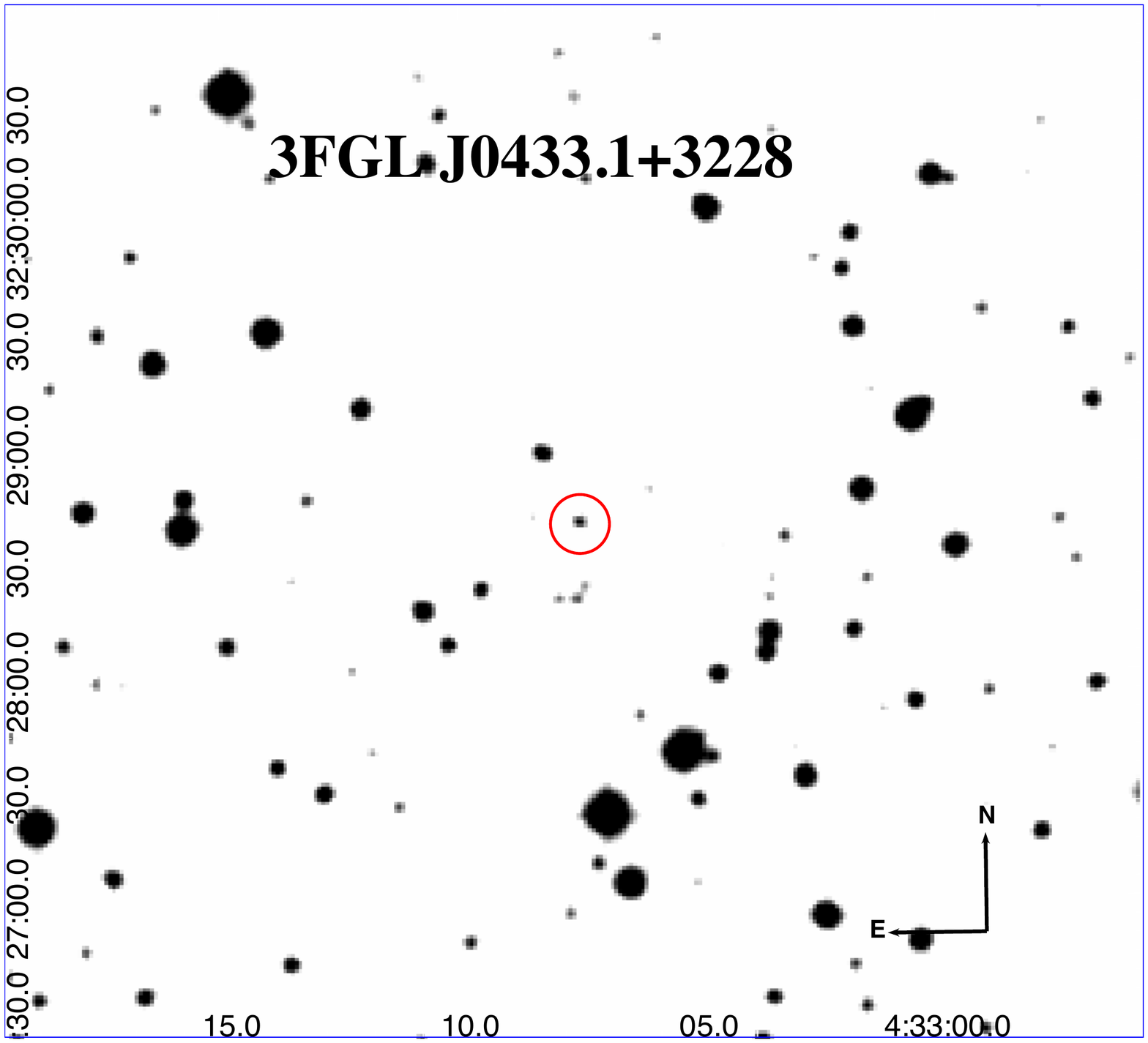} 
\end{center}

\caption{ \emph{Left:} Upper panel) The optical spectra of WISE J043307.54+322840.7, potential counterpart of 
3FGL J0433.1+3228. It is classified as a BL Lac on the basis of its featureless continuum.
SNR also indicated.
Lower panel) The normalized spectrum is shown here.
\emph{Right:}  The $5\arcmin\,x\,5\arcmin\,$ finding chart.} 
\label{fig:J0433}
\end{figure*}

\begin{figure*}
\begin{center}
\includegraphics[height=7.9cm,width=8.4cm,angle=0]{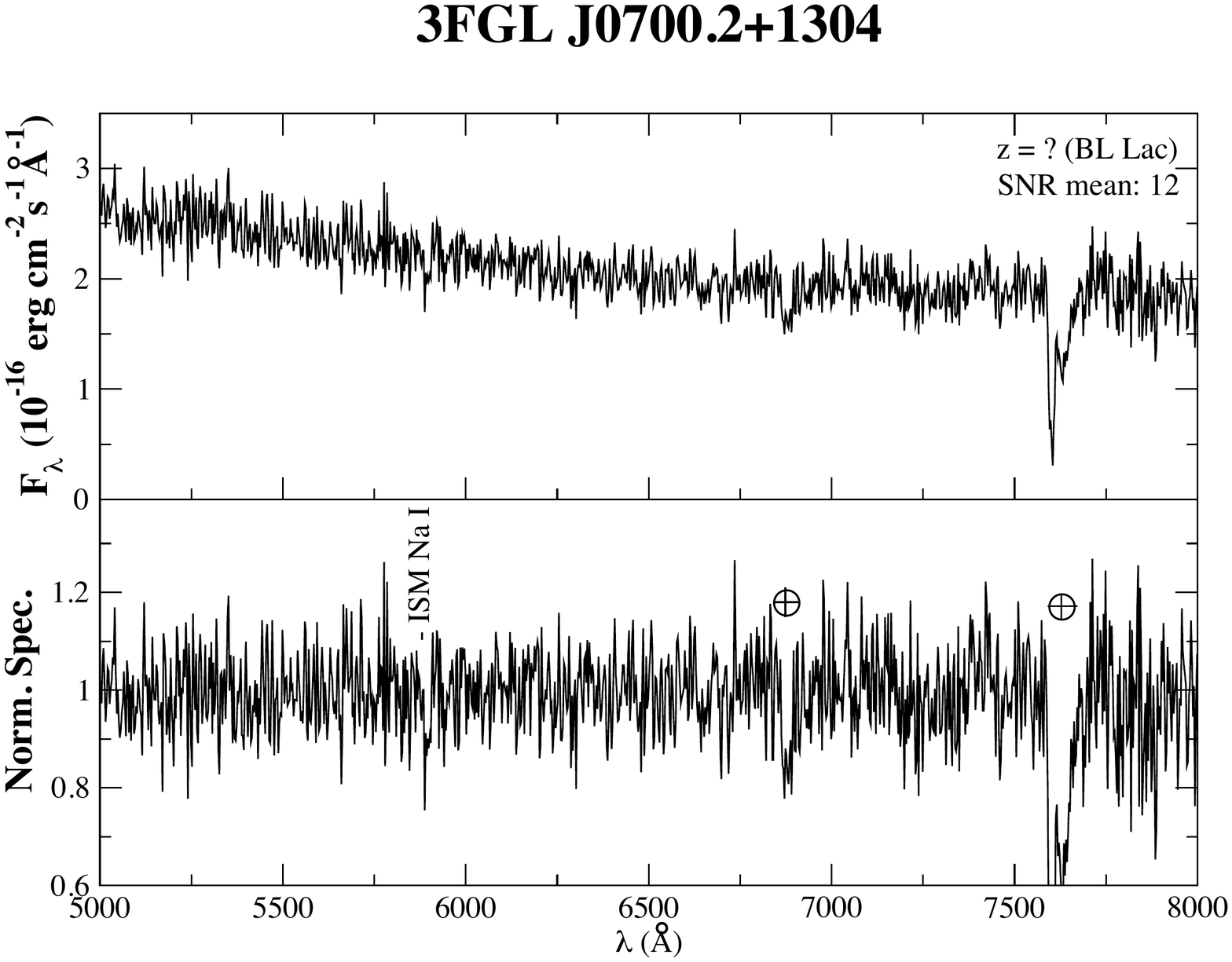} 
\includegraphics[height=6.5cm,width=8.0cm,angle=0]{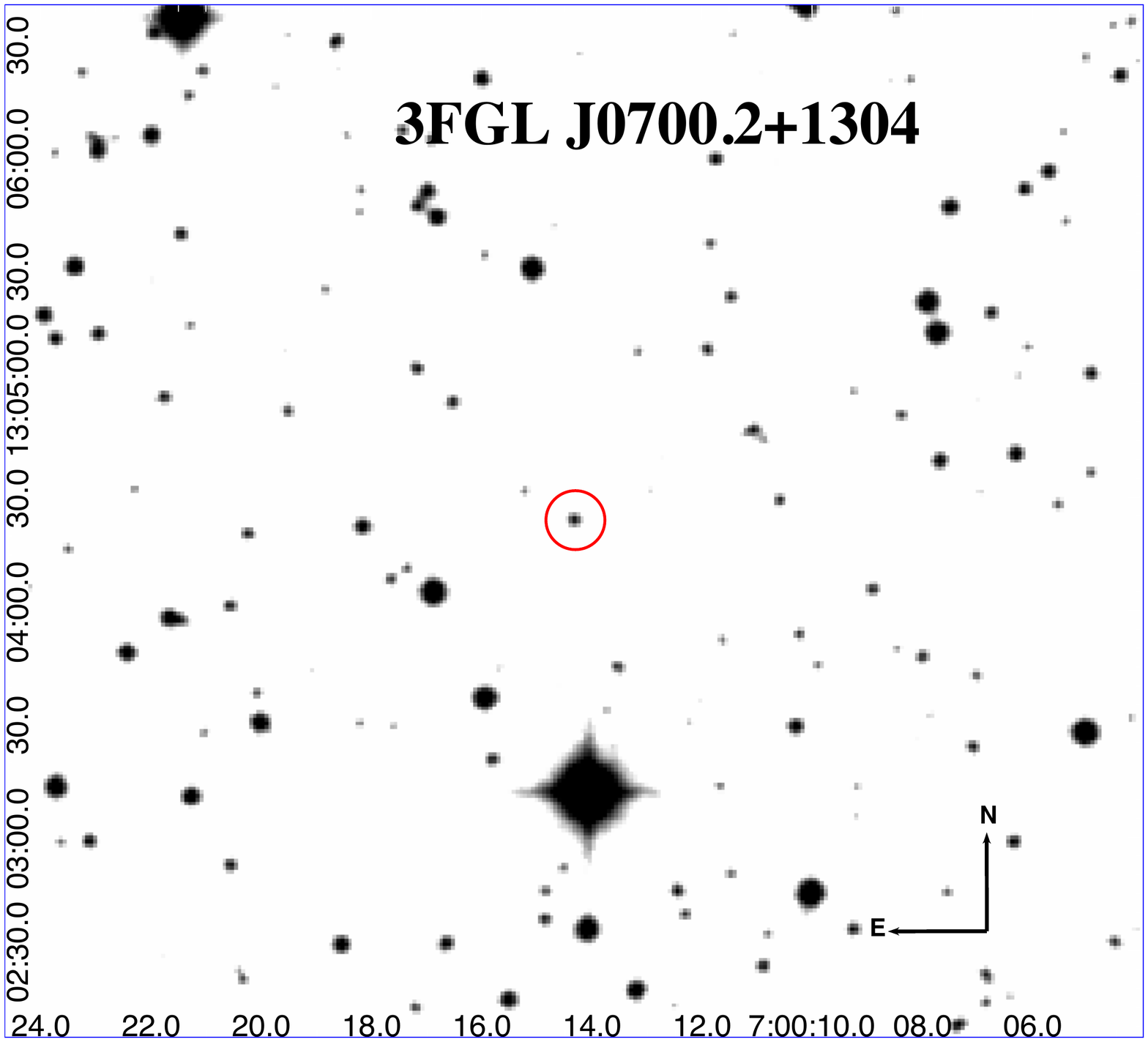} 
\end{center}
\caption{\emph{Left:} Upper panel) The optical spectra of WISE J070014.31+130424.4, potential counterpart of 
3FGL J0700.2+1304. It is classified as a BL Lac on the basis of its featureless continuum.
SNR is also indicated in the figure.
Lower panel) The normalized spectrum is shown here.
\emph{Right:} The $5\arcmin\,x\,5\arcmin\,$ finding chart.}
\label{fig:J0700}
\end{figure*}

\begin{figure*}
\begin{center}
\includegraphics[height=7.9cm,width=8.4cm,angle=0]{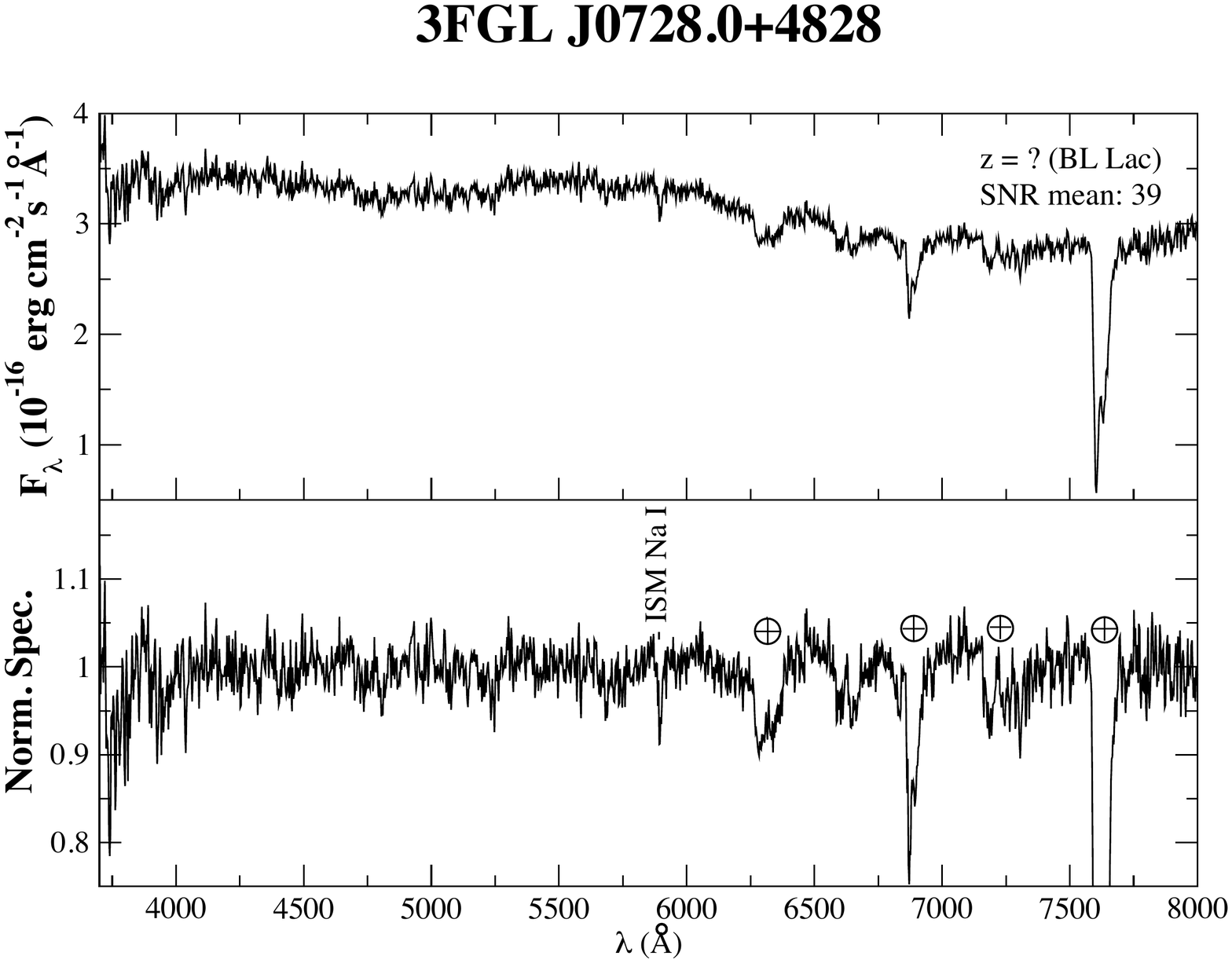} 
\includegraphics[height=6.5cm,width=8.0cm,angle=0]{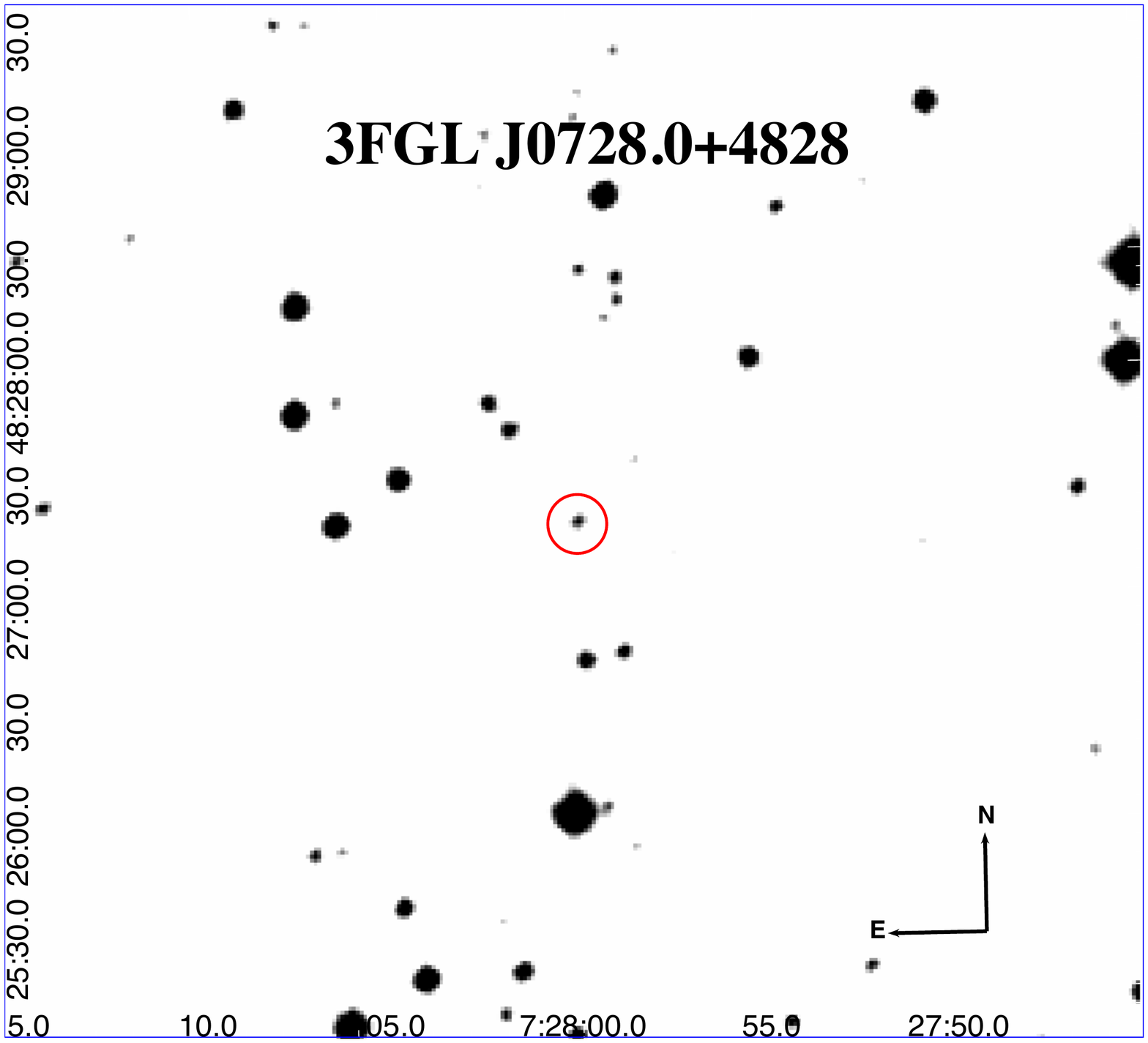} 
\end{center}

\caption{\emph{Left:} Upper panel) The optical spectra of WISE J072759.84+482720.3, potential counterpart of 
3FGL J0728.0+4828. It is classified as a BL Lac on the basis of its featureless continuum.
SNR also indicated in the figure.
Lower panel) The normalized spectrum is shown here.
\emph{Right:} The $5\arcmin\,x\,5\arcmin\,$ finding chart.}  
\label{fig:J0728}
\end{figure*}

\begin{figure*}
\begin{center}
\includegraphics[height=7.9cm,width=8.4cm,angle=0]{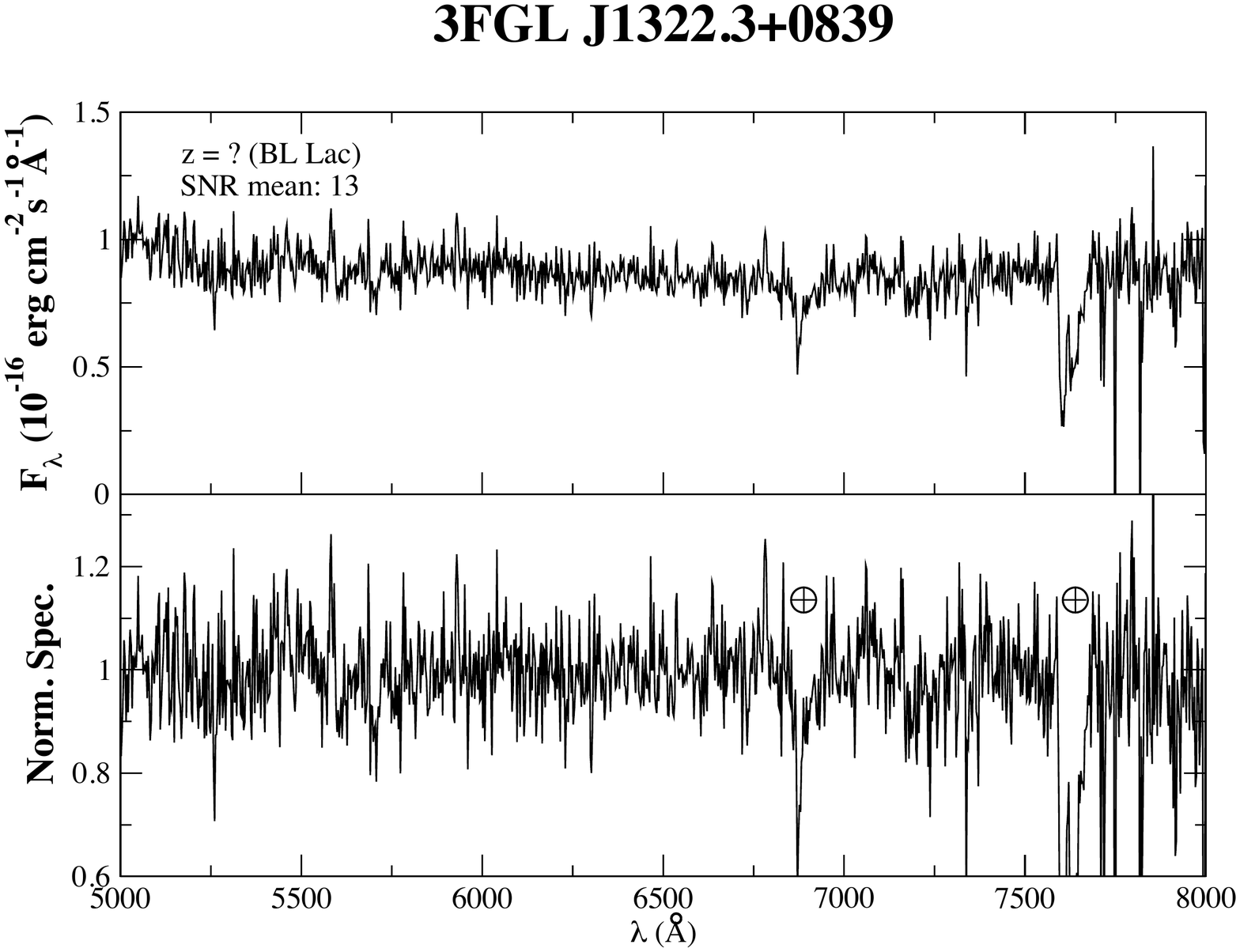} 
\includegraphics[height=6.5cm,width=8.0cm,angle=0]{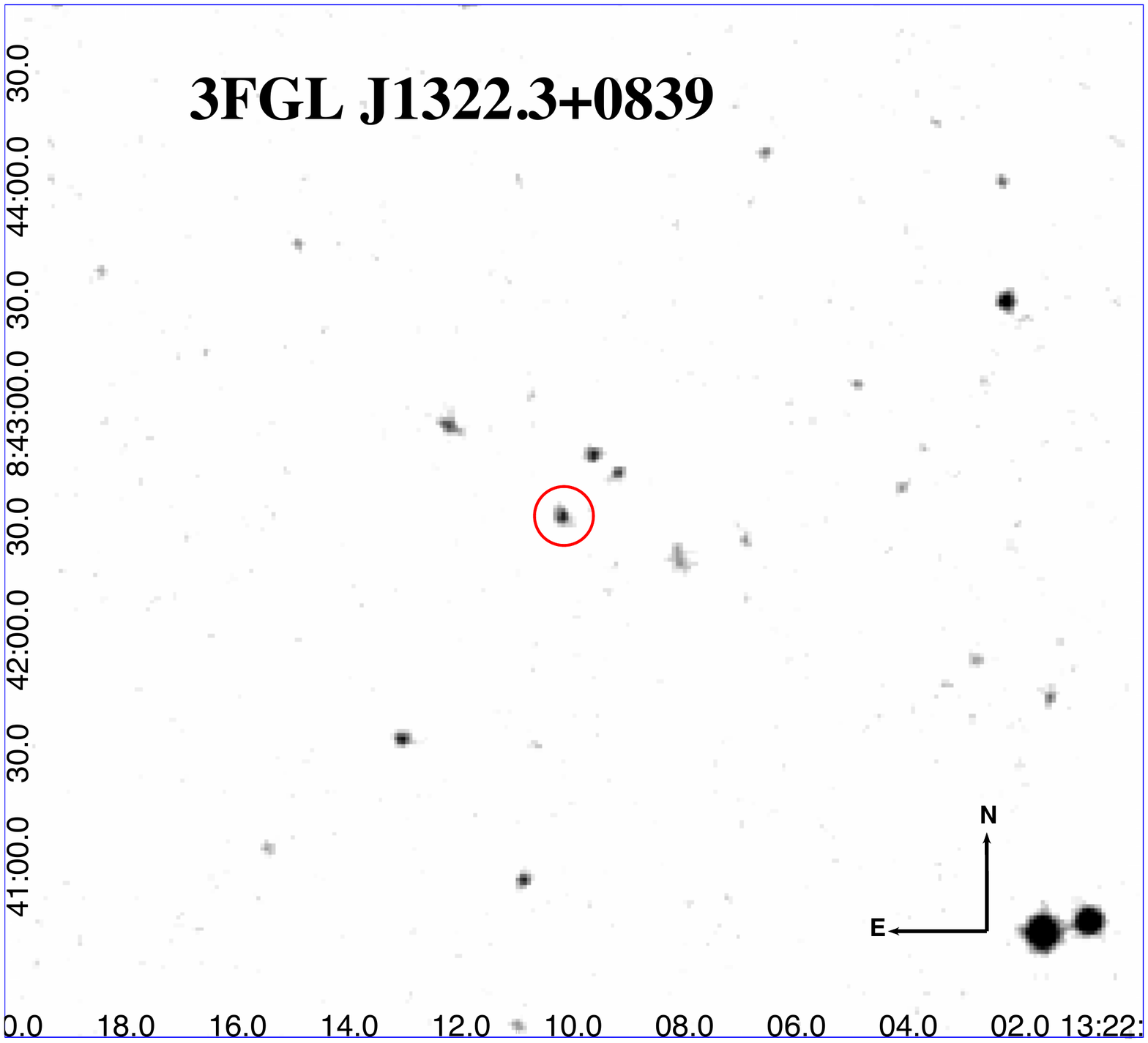} 
\end{center}

\caption{\emph{Left:} Upper panel) The optical spectra of WISE J132210.17+084232.9, potential counterpart of 
3FGL J1322.3+0839. It is classified as a BL Lac on the basis of its featureless continuum.
SNR is also indicated in the figure.
Lower panel) The normalized spectrum is shown here.
\emph{Right:} The $5\arcmin\,x\,5\arcmin\,$ finding chart.}
\label{fig:J1322}
\end{figure*}

\begin{figure*}
\begin{center}
\includegraphics[height=7.9cm,width=8.4cm,angle=0]{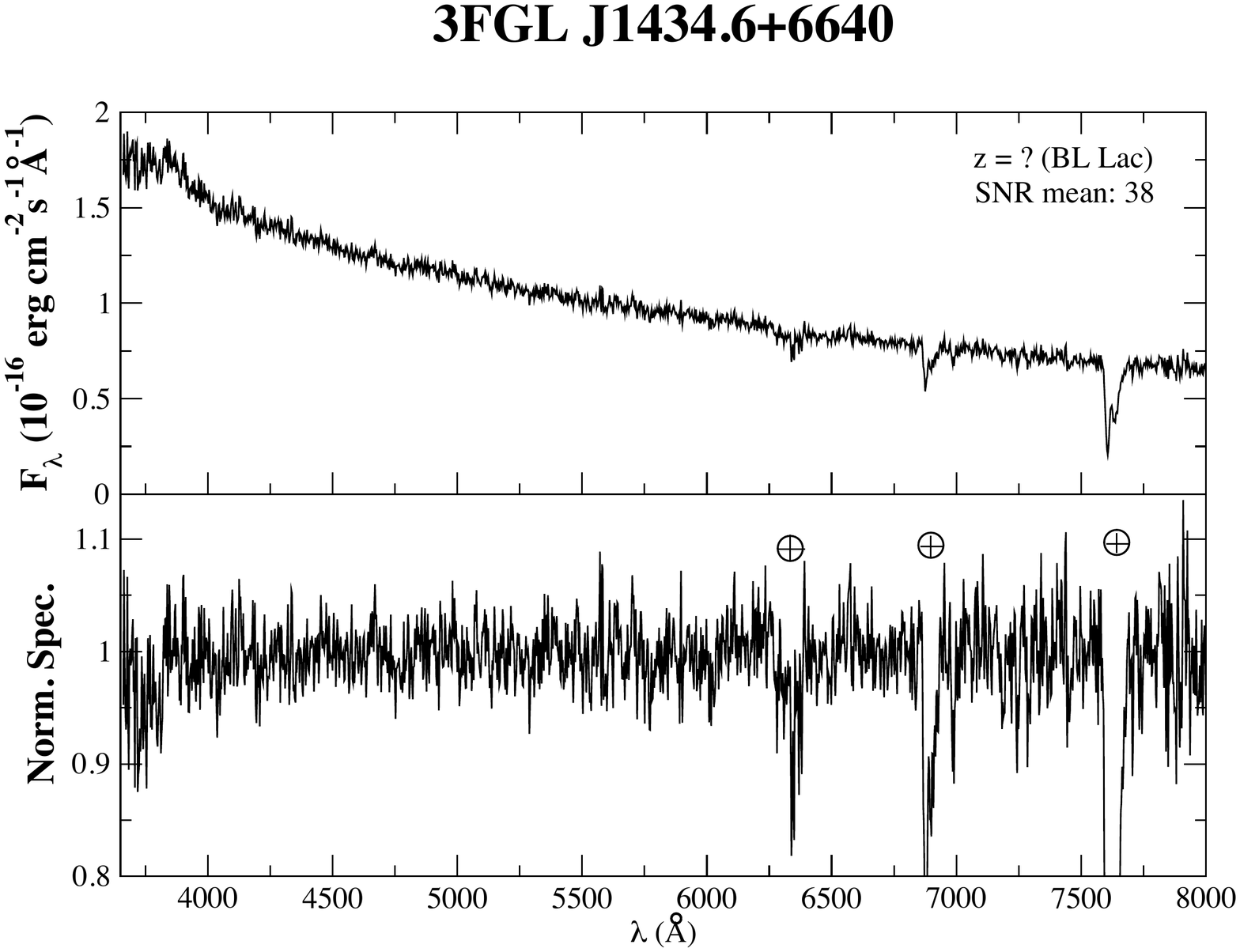} 
\includegraphics[height=6.5cm,width=8.0cm,angle=0]{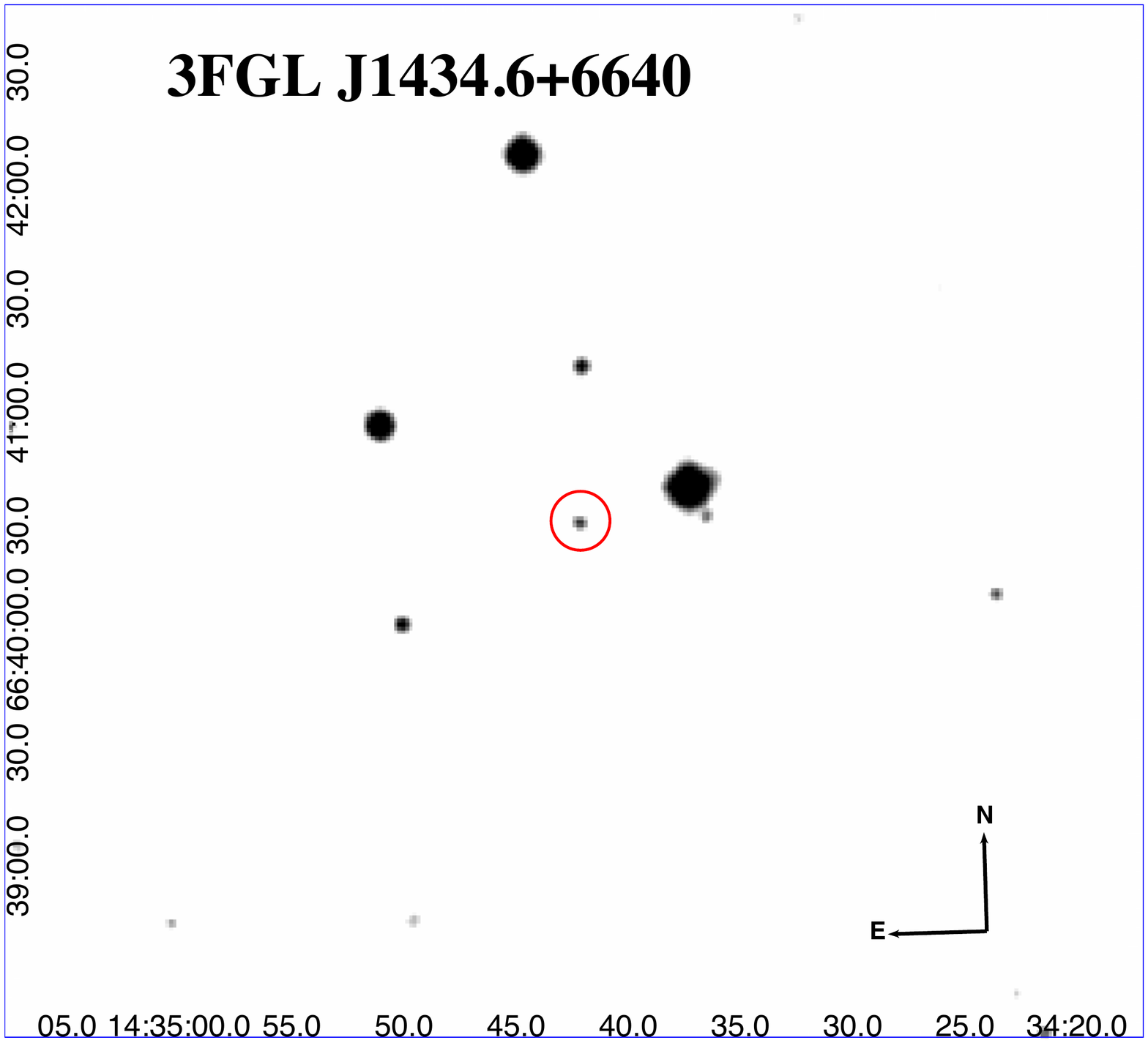} 
\end{center}

\caption{\emph{Left:} Upper panel) The optical spectra of WISE J143441.46+664026.5, potential counterpart of 
3FGL J1434.6+6640. It is classified as a BL Lac on the basis of its featureless continuum.
The average signal-to-noise ratio (SNR) is also indicated in the figure.
Lower panel) The normalized spectrum is shown here.
\emph{Right:}  The $5\arcmin\,x\,5\arcmin\,$ finding chart.}
\label{fig:J1434}
\end{figure*}

\begin{figure*}
\begin{center}s
\includegraphics[height=7.9cm,width=8.4cm,angle=0]{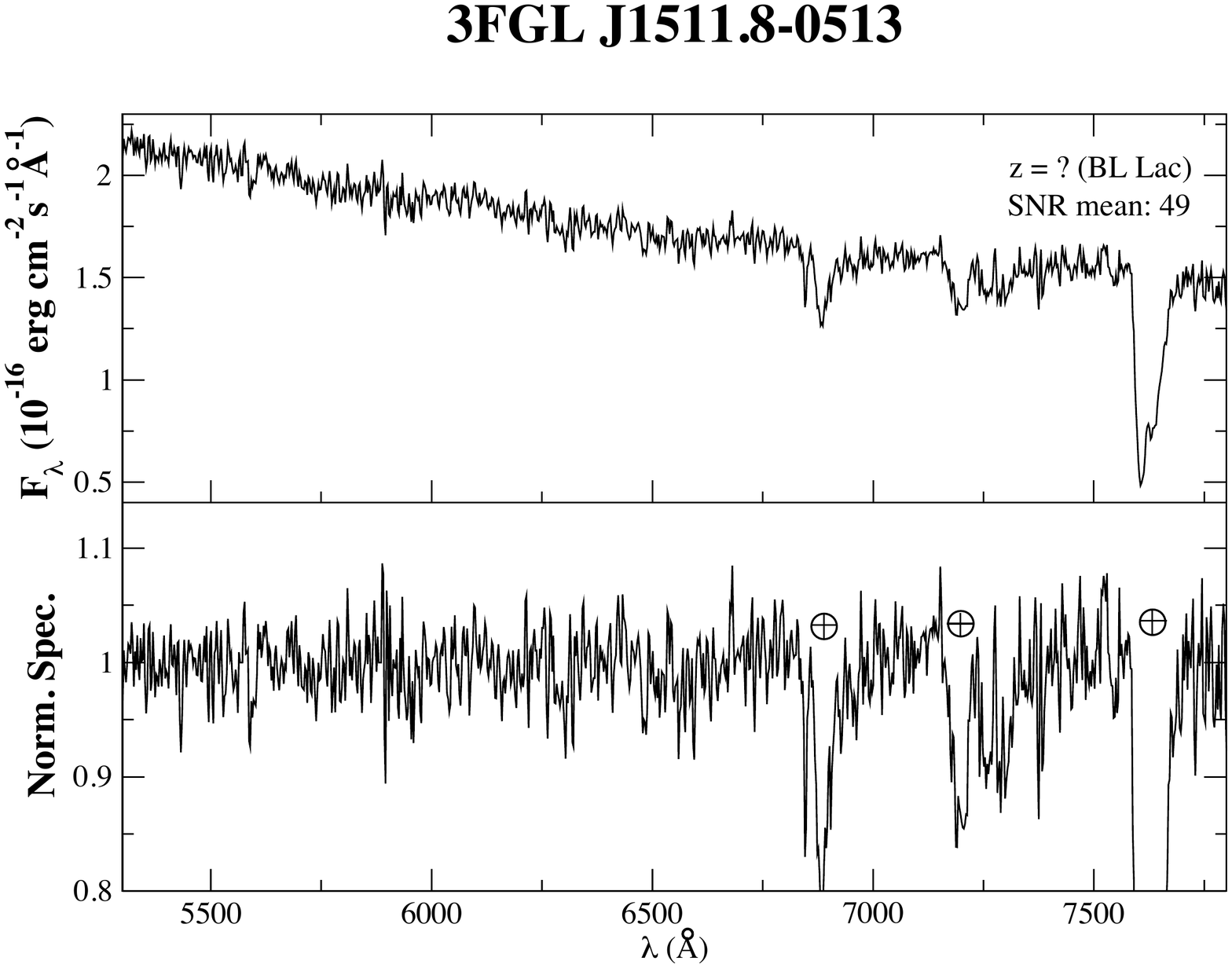} 
\includegraphics[height=6.5cm,width=7.5cm,angle=0]{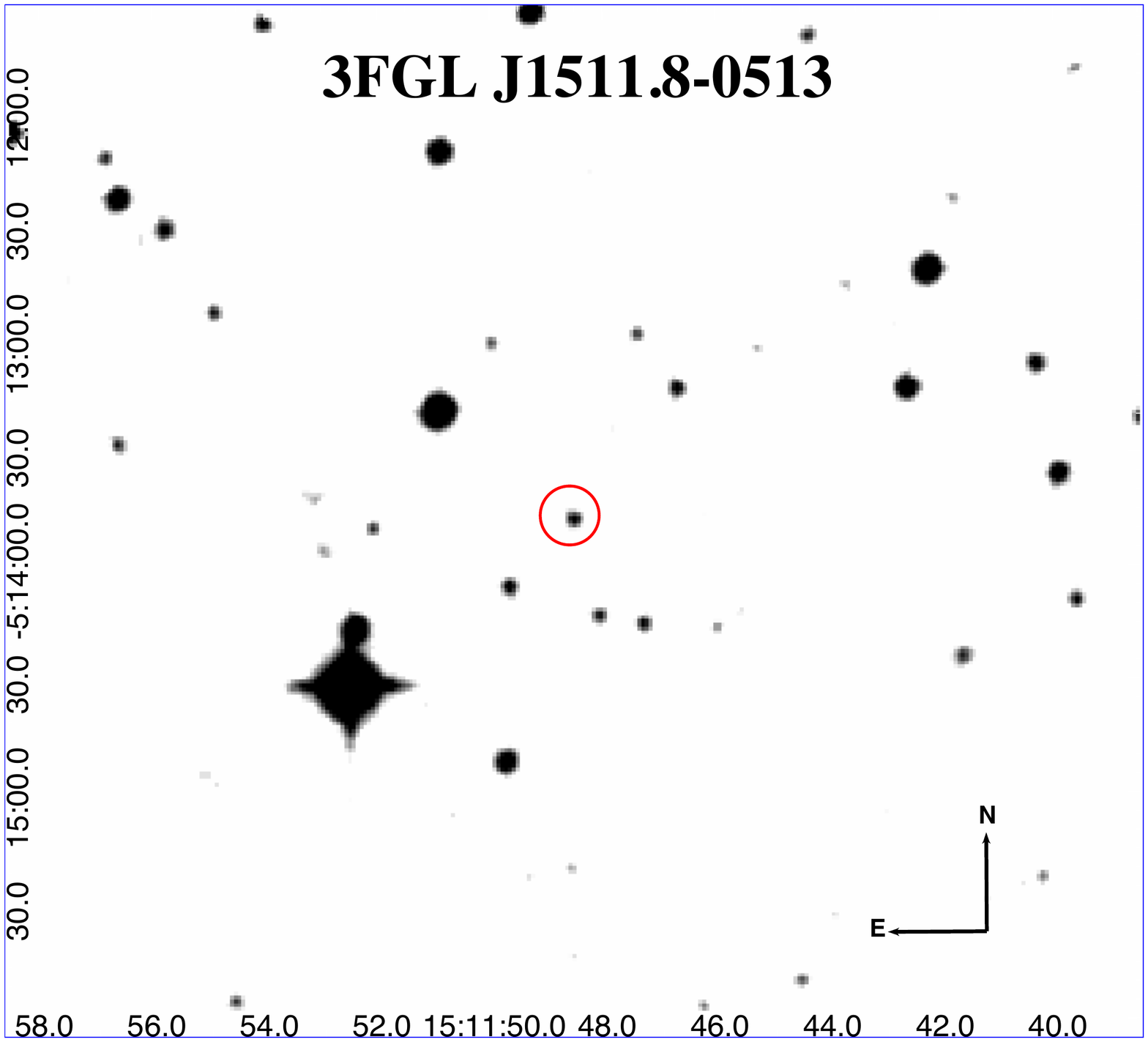} 
\end{center}

\caption{\emph{Left:} Upper panel) The optical spectra of WISE J151148.56-051346.9, potential counterpart of 
3FGL J1511.8-0513. It is classified as a BL Lac on the basis of its featureless continuum.
SNR is also indicated in the figure.
Lower panel) The normalized spectrum is shown here.
\emph{Right:}  The $5\arcmin\,x\,5\arcmin\,$ finding chart.}
\label{fig:J1511}
\end{figure*}

\begin{figure*}
\begin{center}
\includegraphics[height=7.9cm,width=8.4cm,angle=0]{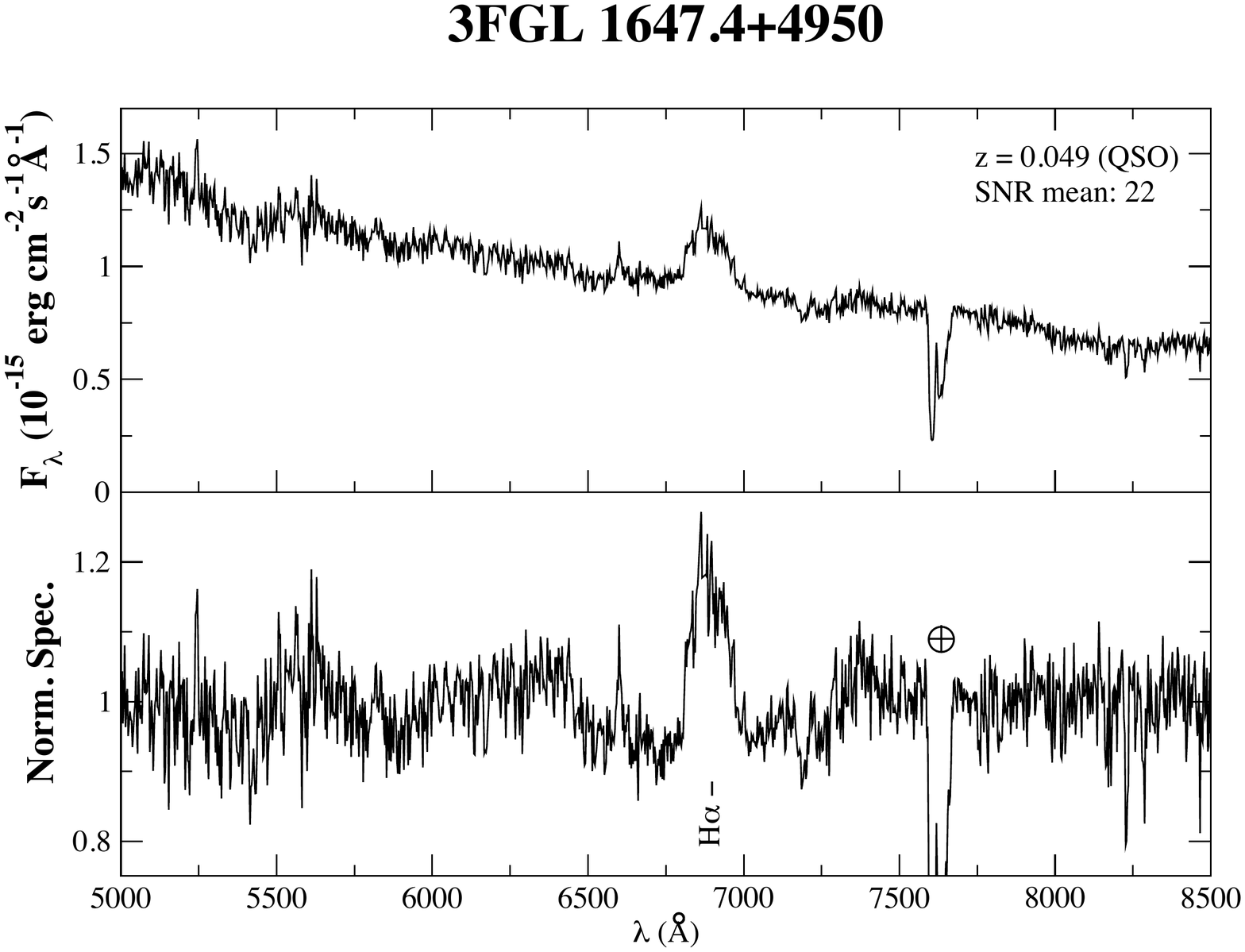} 
\includegraphics[height=6.5cm,width=8.0cm,angle=0]{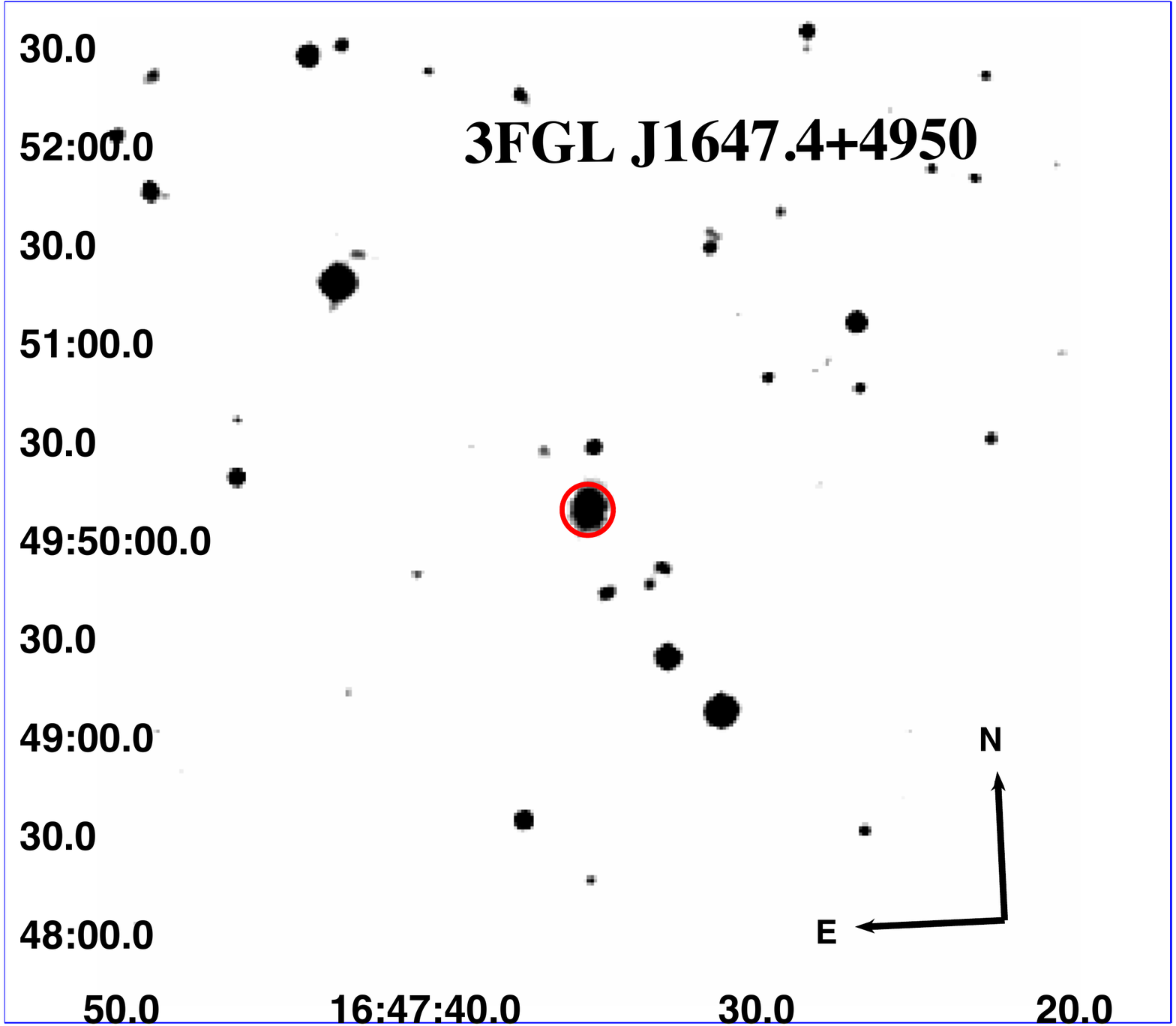} 
\end{center}

\caption{\emph{Left:} Upper panel) The optical spectra of WISE J164734.91+495000.5, potential counterpart of 
3FGL J1647.4+4950 Our observation shows an emission line of $H\alpha$
($\lambda_{obs} = 6887 \AA$). The source is a FSRQ at $z$=0.049.
SNR is also indicated in the figure.
Lower panel) The normalized spectrum is shown here.
\emph{Right:} The $5\arcmin\,x\,5\arcmin\,$ finding chart.}
\label{fig:J1647}
\end{figure*}

\begin{figure*}
\begin{center}
\includegraphics[height=7.9cm,width=8.4cm,angle=0]{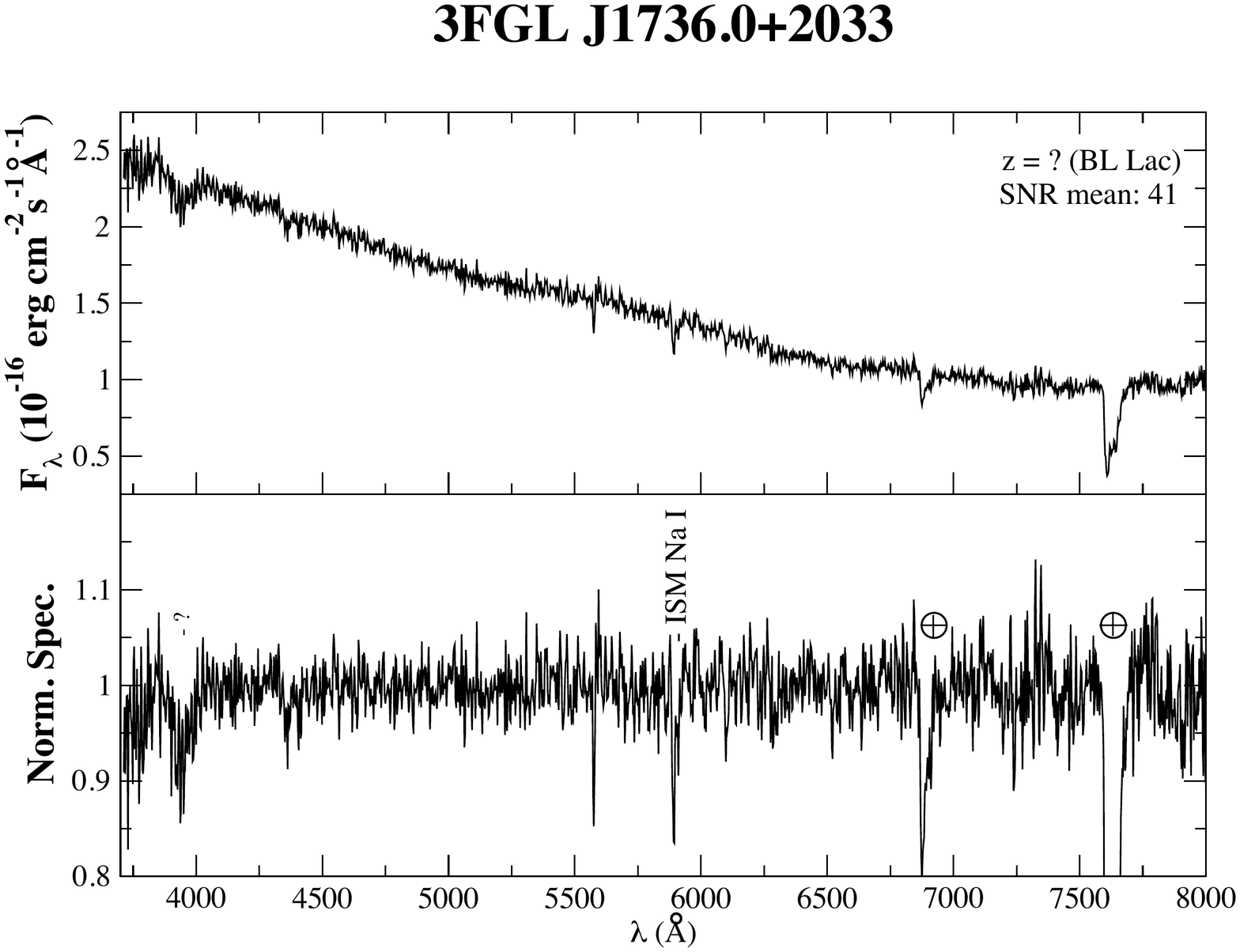} 
\includegraphics[height=6.5cm,width=8.0cm,angle=0]{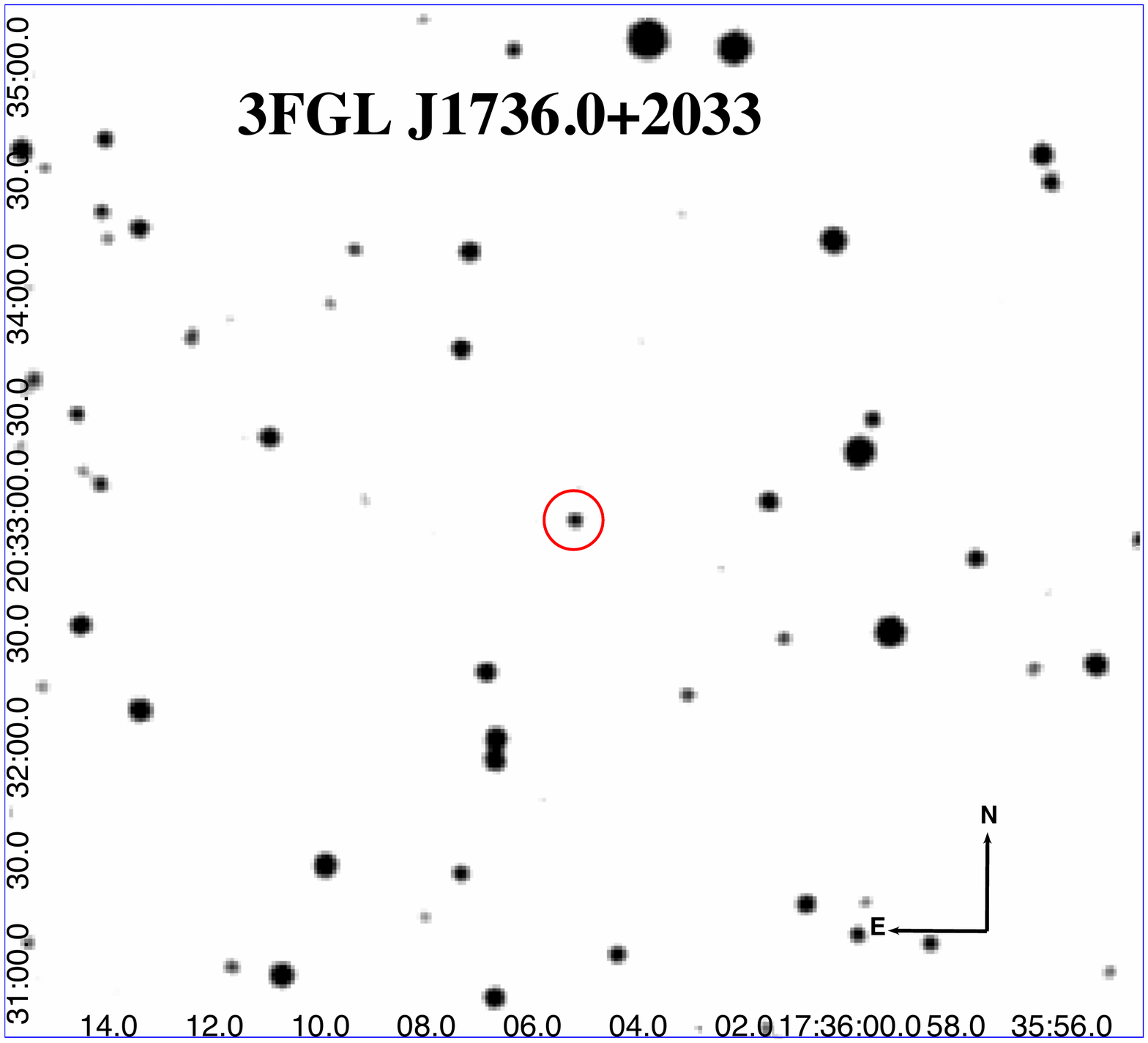} 
\end{center}

\caption{\emph{Left:} Upper panel) The optical spectra of WISE J173605.25+203301.1, potential counterpart of 
3FGL J1736.0+2033. It is classified as a BL Lac on the basis of its featureless continuum.
SNR is also indicated in the figure. Lower panel) The normalized spectrum is shown here.
\emph{Right:} The $5\arcmin\,x\,5\arcmin\,$ finding chart.}
\label{fig:J1736}
\end{figure*}

\begin{figure*}
\begin{center}
\includegraphics[height=7.9cm,width=8.4cm,angle=0]{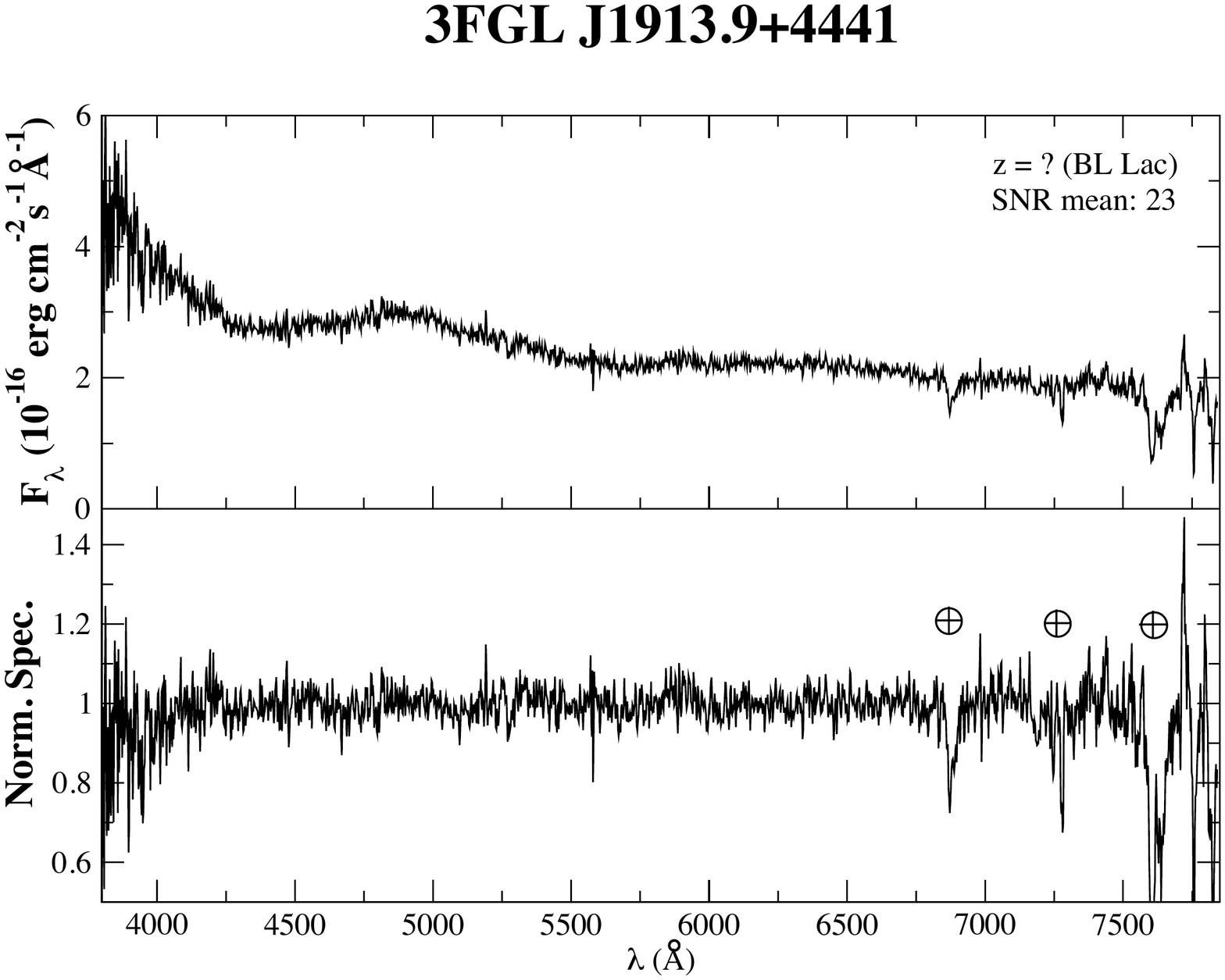} 
\includegraphics[height=6.5cm,width=8.0cm,angle=0]{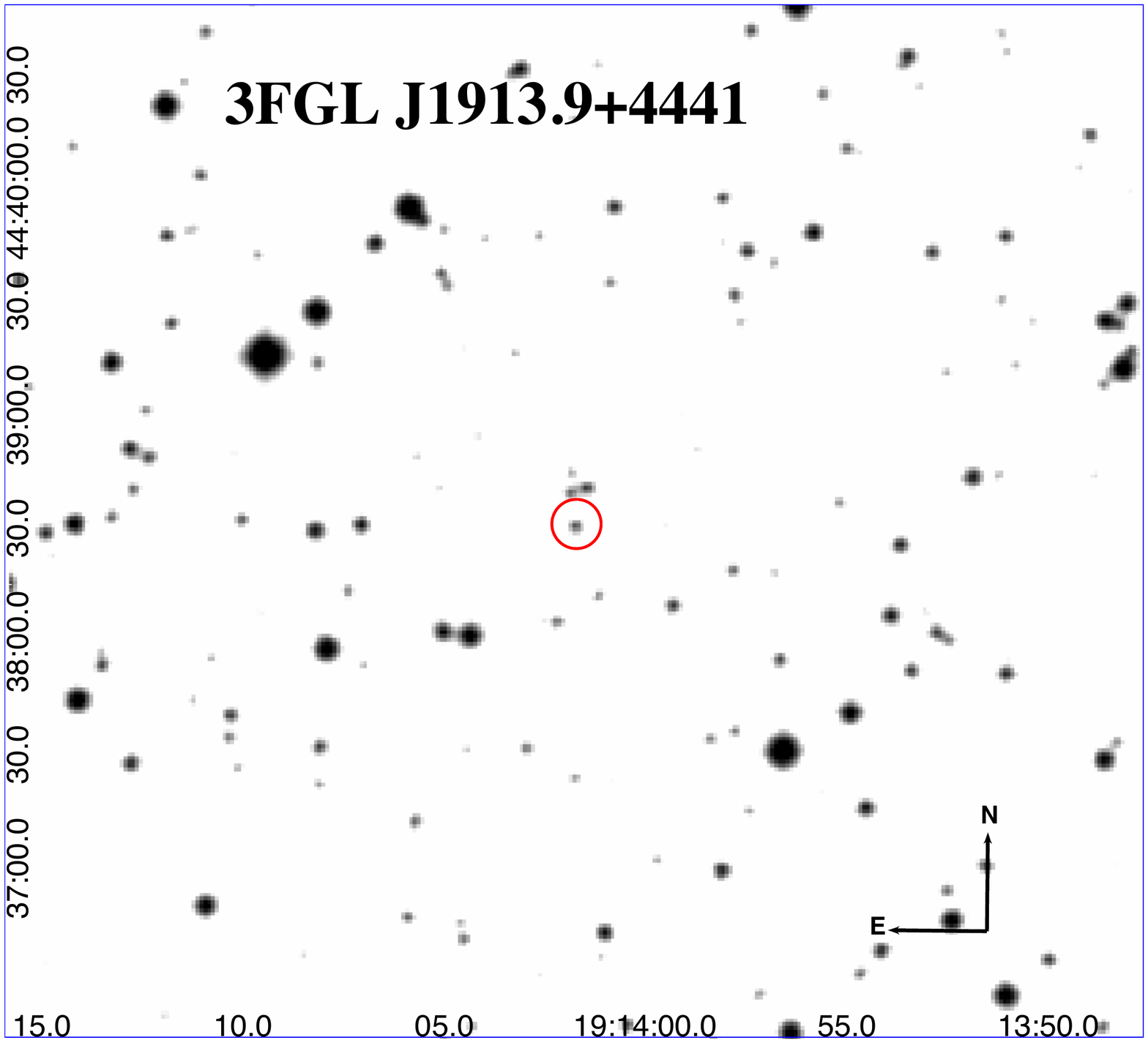} 
\end{center}

\caption{\emph{Left:} Upper panel) The optical spectra of WISE J191401.88+443832.2, potential counterpart of 
3FGL J1913.9+4441. It is classified as a BL Lac on the basis of its featureless continuum.
SNR is also indicated in the figure.
Lower panel) The normalized spectrum is shown here.
\emph{Right:} The $5\arcmin\,x\,5\arcmin\,$ finding chart.}
\label{fig:J1913}
\end{figure*}

\begin{figure*}
\begin{center}
\includegraphics[height=7.9cm,width=8.4cm,angle=0]{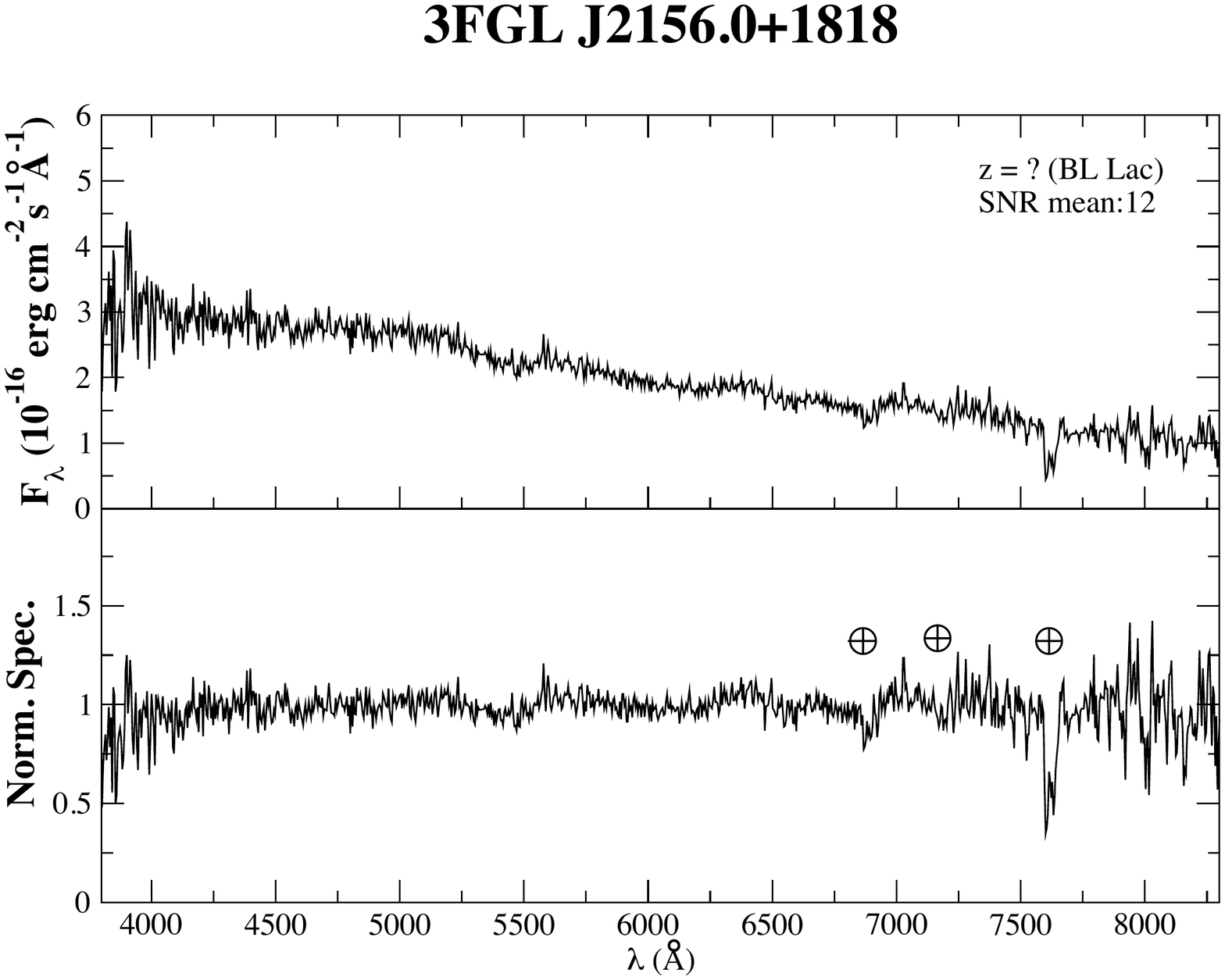} 
\includegraphics[height=6.5cm,width=8.0cm,angle=0]{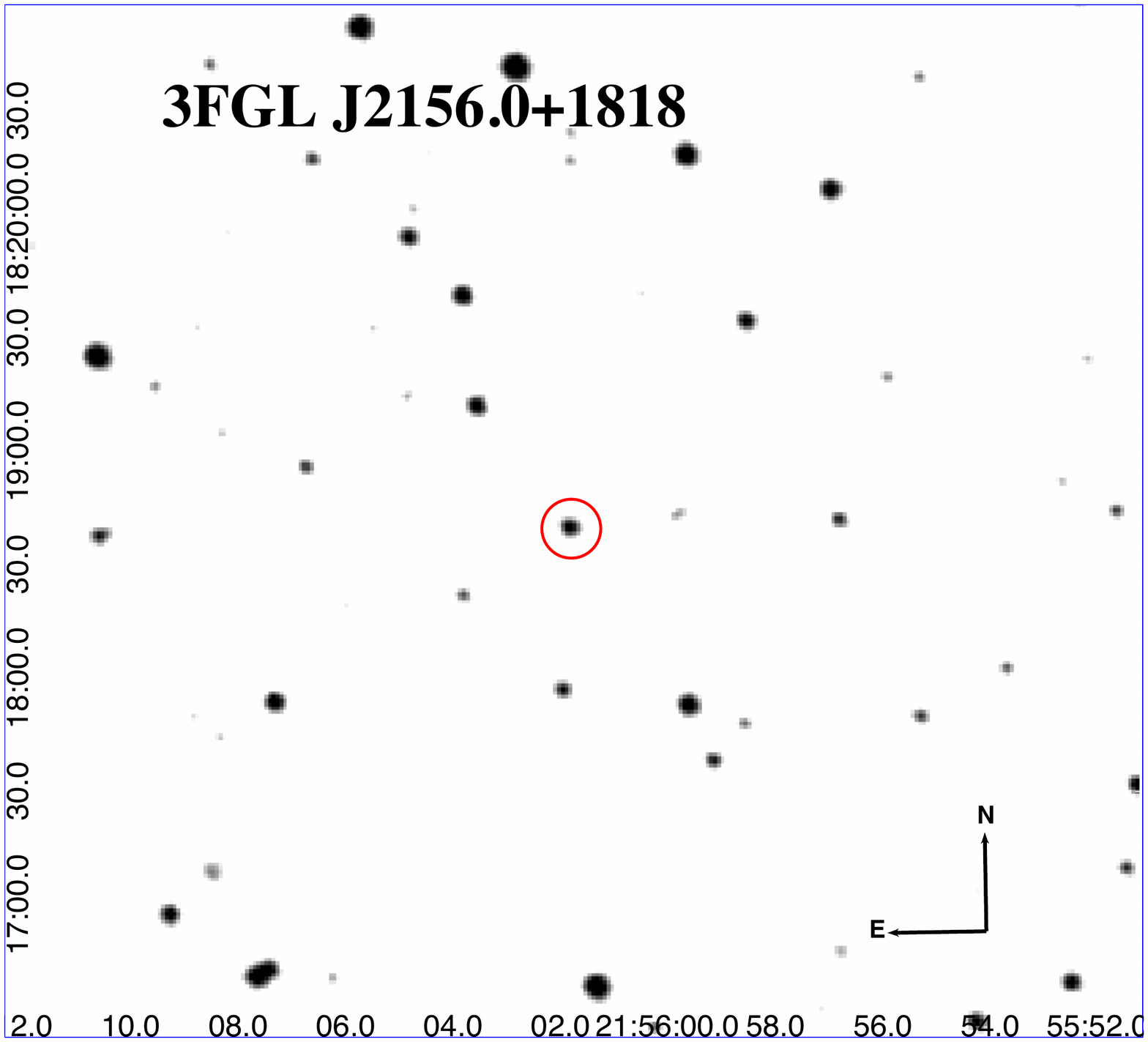} 
\end{center}

\caption{\emph{Left:} Upper panel) The optical spectra of WISE J215601.64+181837.1, potential counterpart of 
3FGL J2156.0+1818. It is classified as a BL Lac on the basis of its featureless continuum.
The average signal-to-noise ratio (SNR) is also indicated in the figure.
Lower panel) The normalized spectrum is shown here.
\emph{Right:} The $5\arcmin\,x\,5\arcmin\,$ finding chart.}
\label{fig:J2156}
\end{figure*}

\begin{figure*}
\begin{center}
\includegraphics[height=7.9cm,width=8.4cm,angle=0]{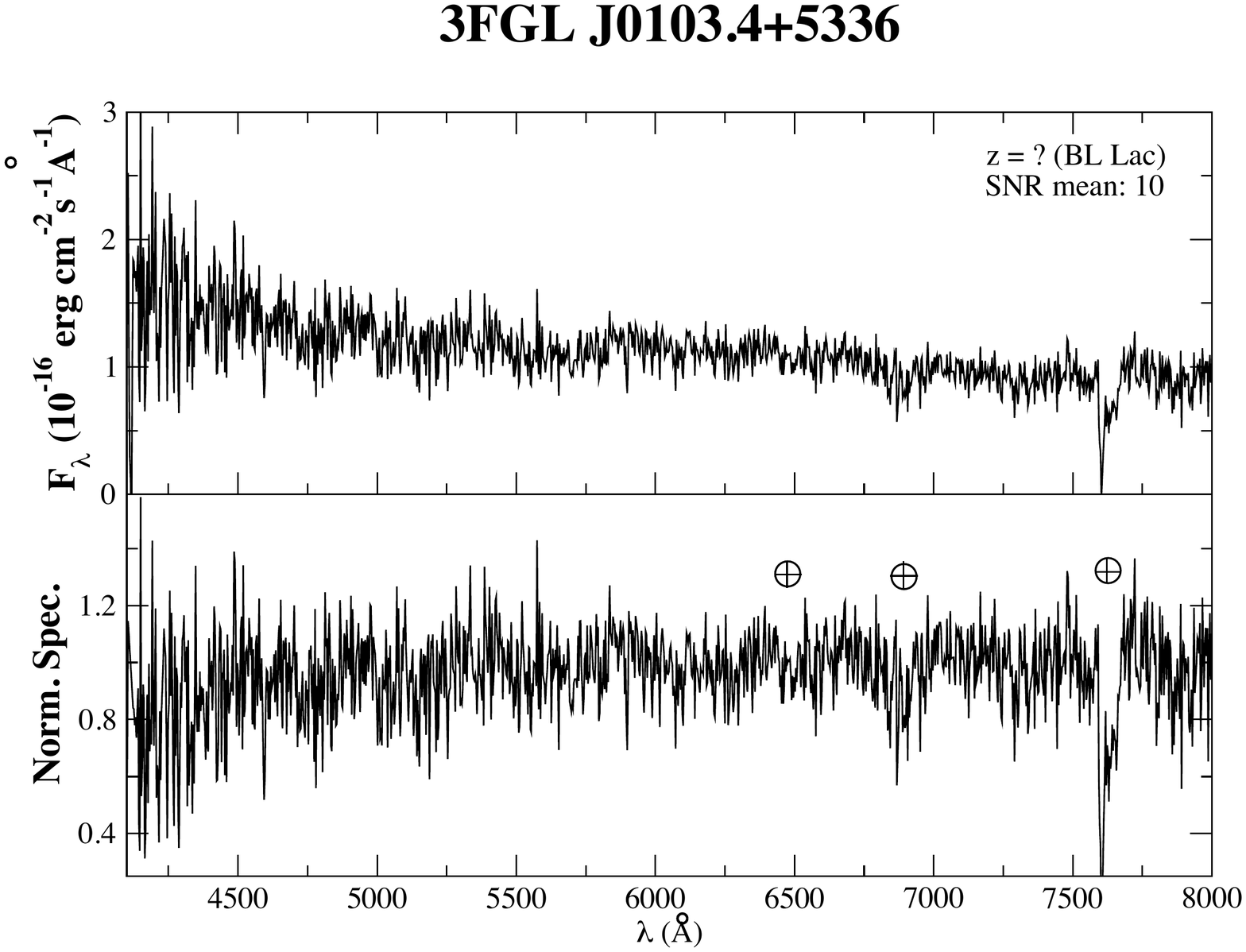} 
\includegraphics[height=6.5cm,width=8.0cm,angle=0]{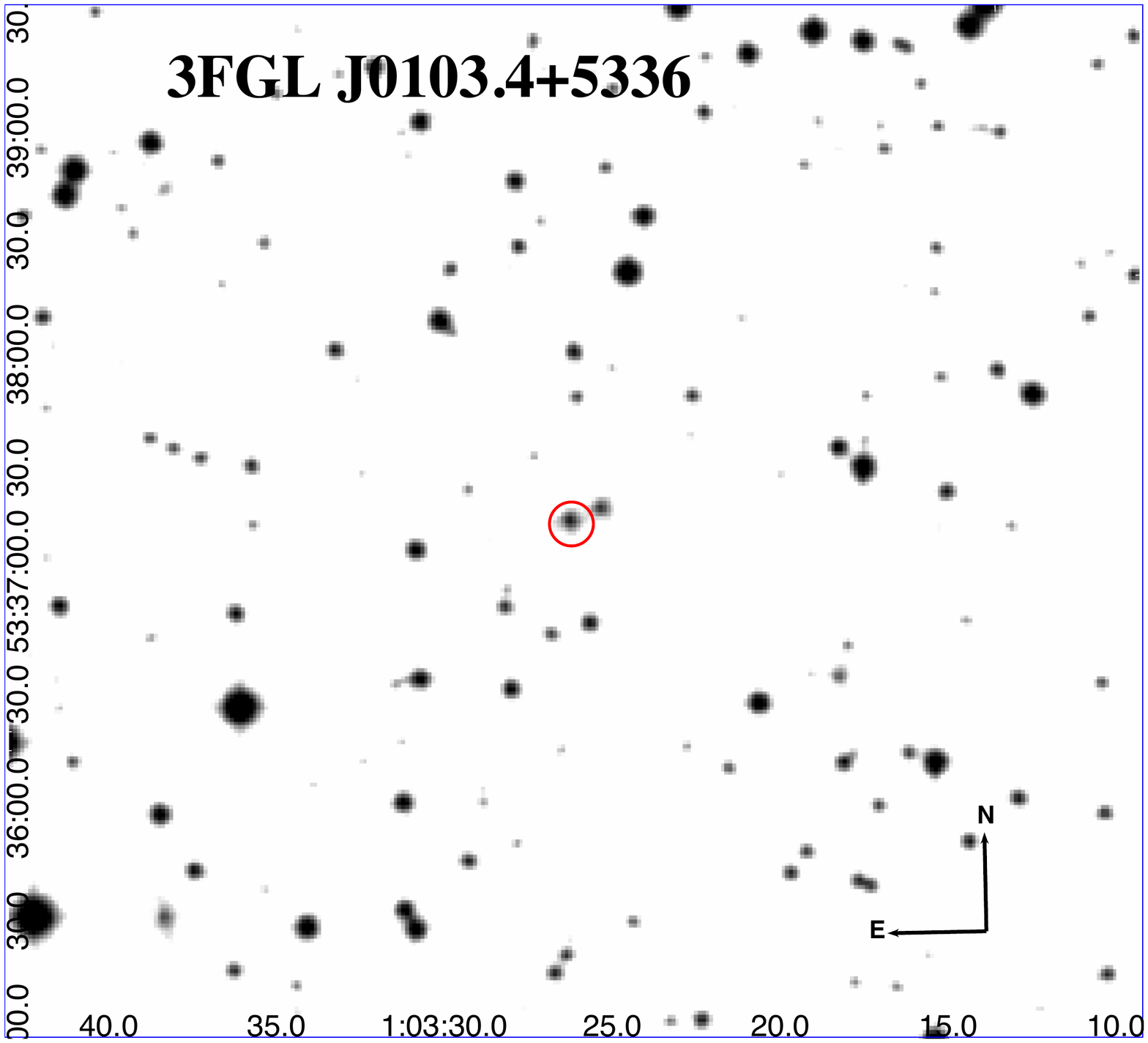} 
\end{center}

\caption{\emph{Left:} Upper panel) The optical spectra of WISE J010325.89+533713.4, potential counterpart of 
3FGL J0103.4+5336. It is classified as a BL Lac on the basis of its featureless continuum.
SNR is also indicated in the figure.
Lower panel) The normalized spectrum is shown here.
\emph{Right:} The $5\arcmin\,x\,5\arcmin\,$ finding chart.}
\label{fig:J0103}
\end{figure*}

\begin{figure*}
\begin{center}
\includegraphics[height=7.9cm,width=8.4cm,angle=0]{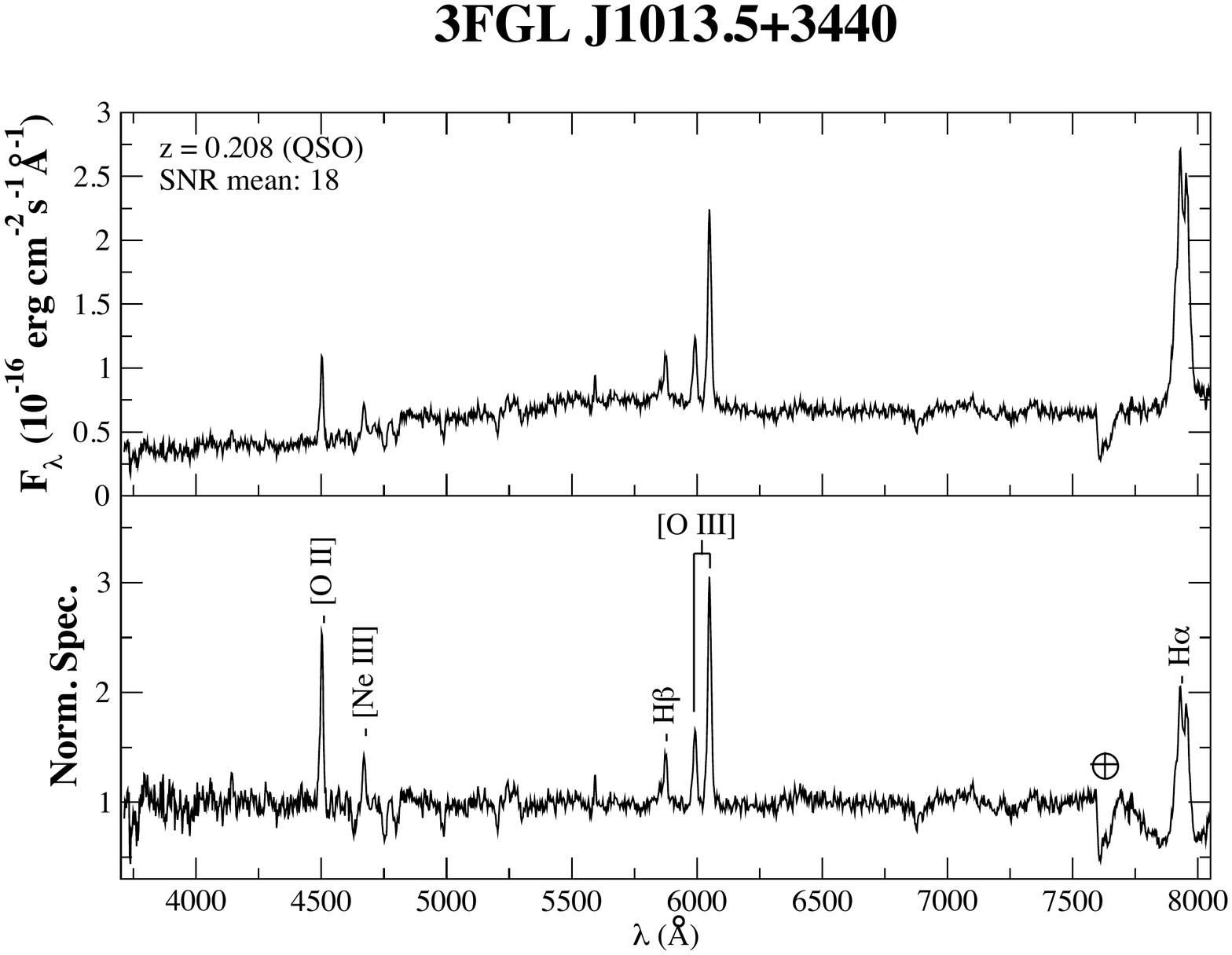} 
\includegraphics[height=6.5cm,width=8.0cm,angle=0]{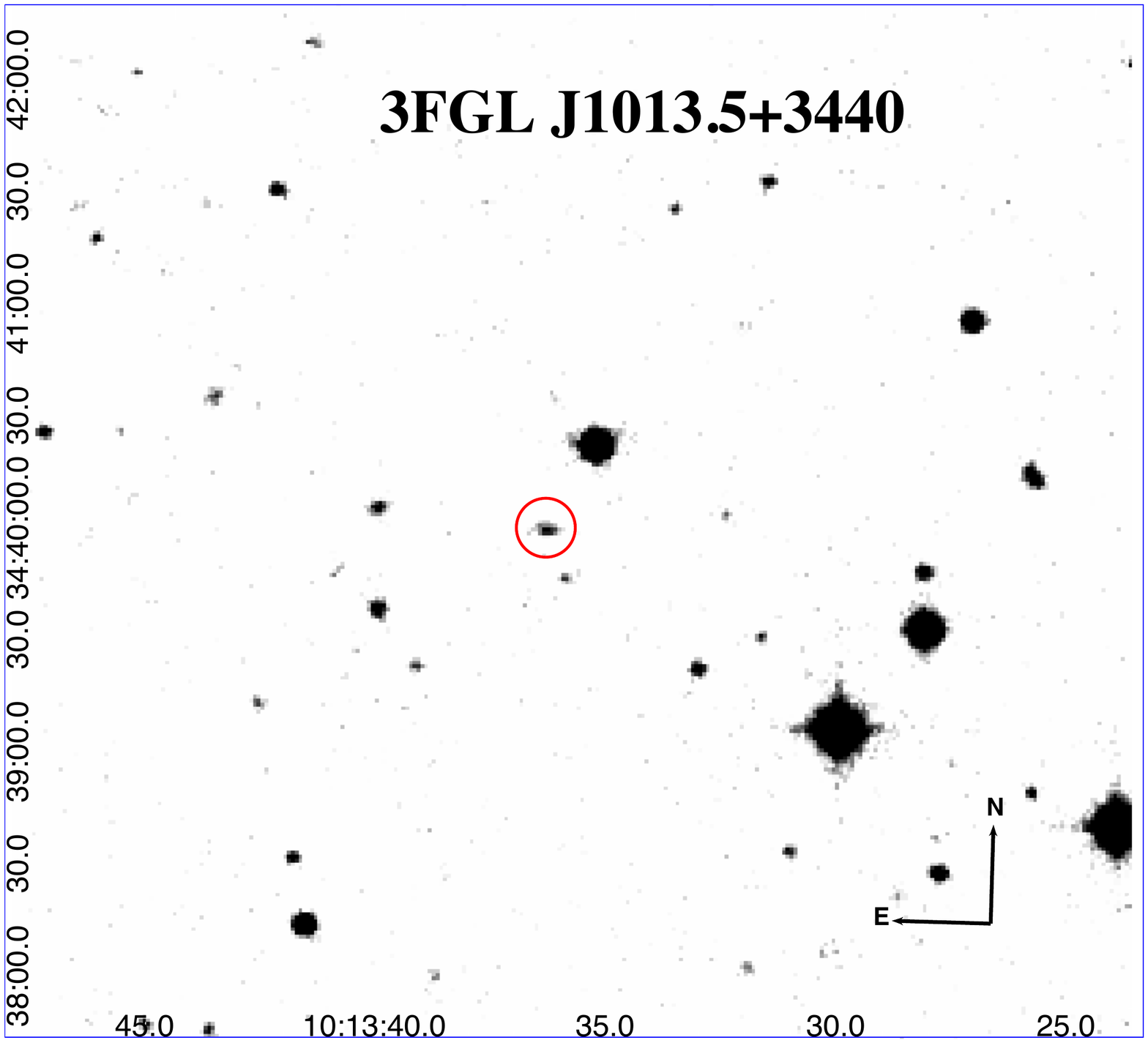} 
\end{center}

\caption{\emph{Left:} Upper panel) The optical spectra of WISE J101336.51+344003.6, potential counterpart of 
3FGL J1013.5+3440. It is classified as a QSO at $z$ = 0.208. Identification of the lines
[O II] ($\lambda_{obs} = 4502 \AA$), [Ne III] ($\lambda_{obs} = 4671 \AA$), 
$H\beta$ ($\lambda_{obs} = 5874 \AA$), the doublet [O III] ($\lambda\lambda_{obs} = 5992 - 6050 \AA$) 
and $H\alpha$ ($\lambda_{obs} = 7939 \AA$).
SNR is also indicated in the figure.
Lower panel) The normalized spectrum is shown here.
\emph{Right:}  The $5\arcmin\,x\,5\arcmin\,$ finding chart.}
\label{fig:J1013}
\end{figure*}

\begin{figure*}
\begin{center}
\includegraphics[height=7.9cm,width=8.4cm,angle=0]{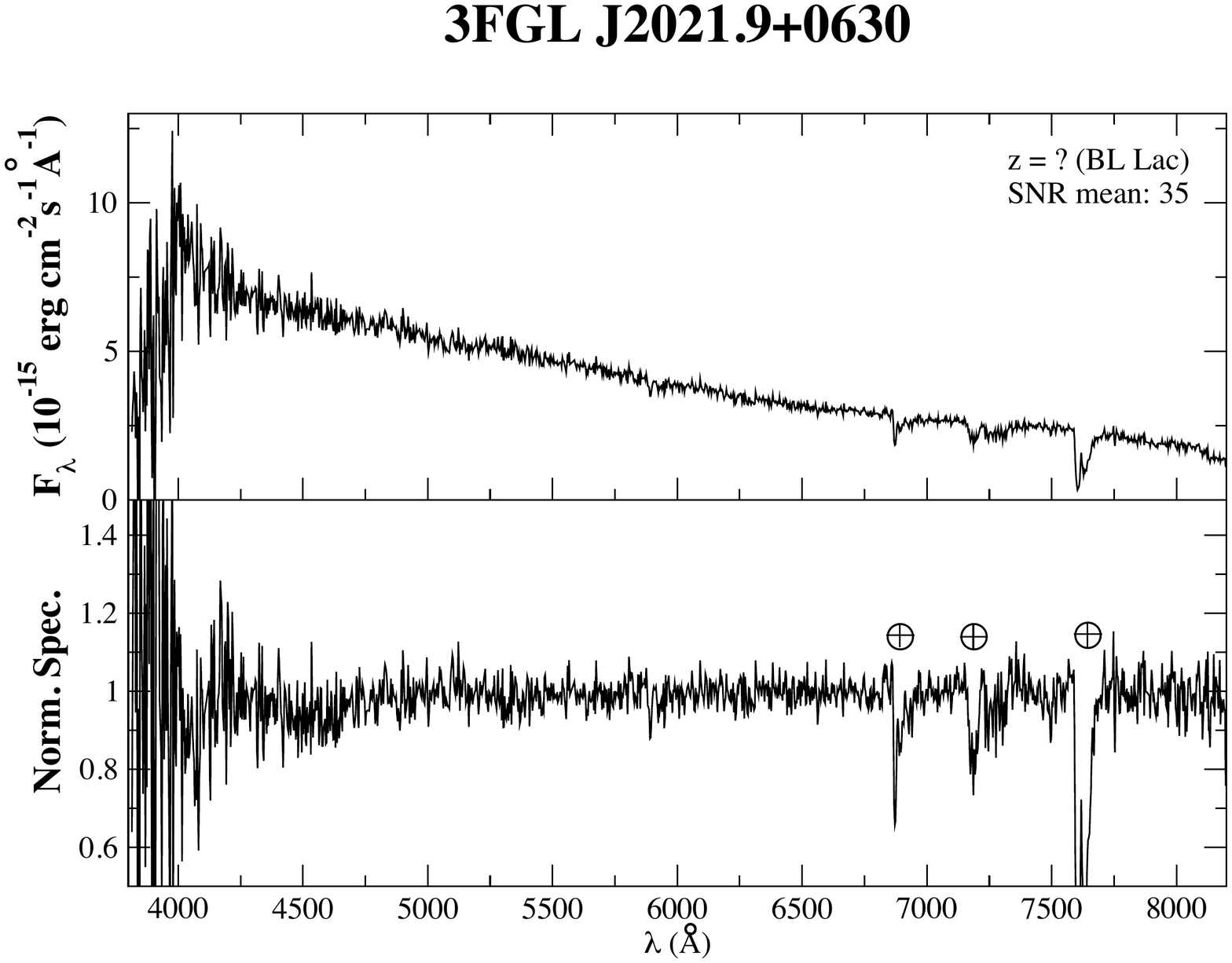} 
\includegraphics[height=6.5cm,width=8.0cm,angle=0]{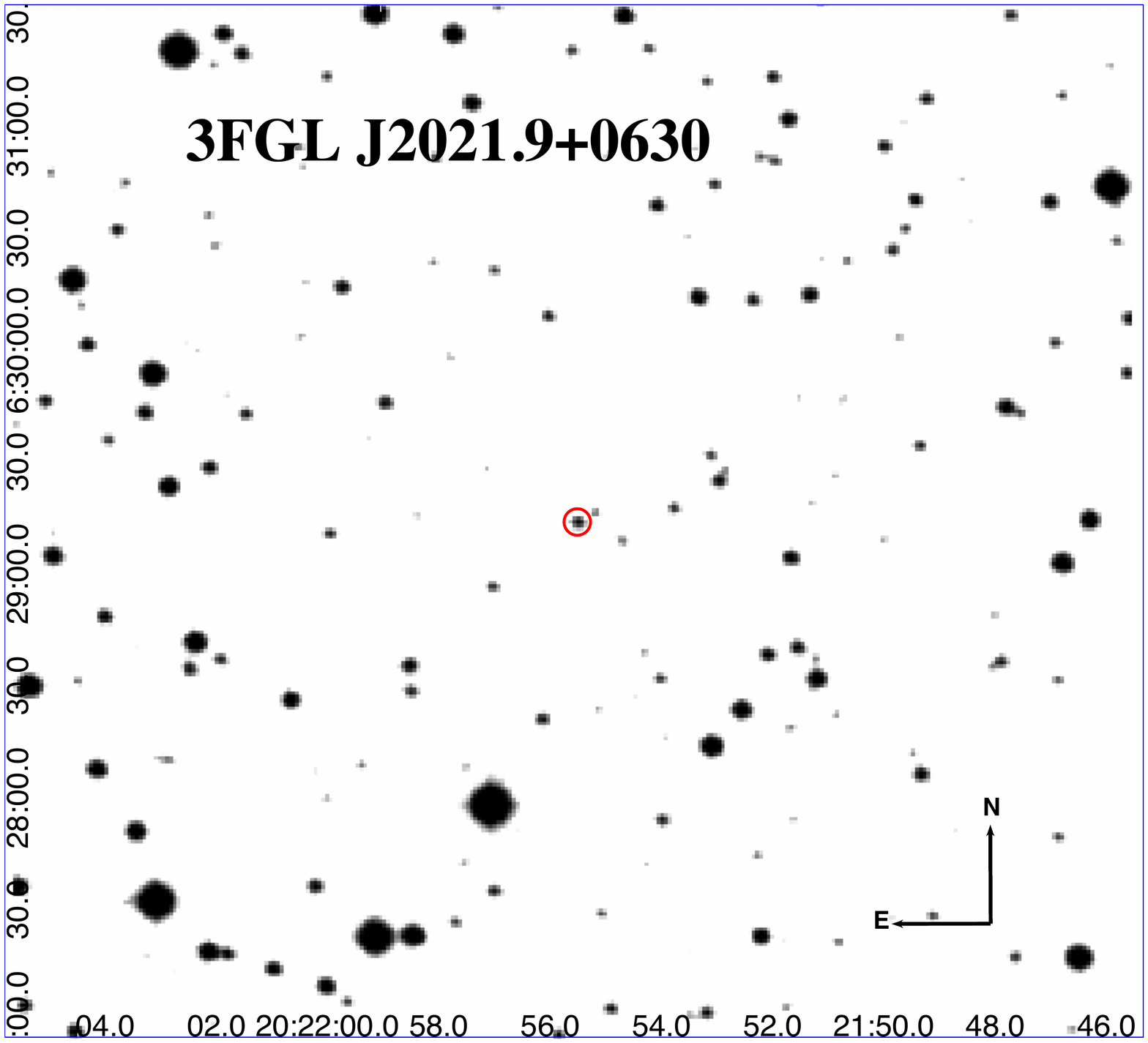} 
\end{center}

\caption{\emph{Left:} Upper panel) The optical spectra of WISE J202155.45+062913.6 potential counterpart of 
3FGL J2021.9+0630. It is classified as a BL Lac on the basis of its featureless continuum.
SNR is also indicated in the figure.
Lower panel) The normalized spectrum is shown here.
\emph{Right:}  The $5\arcmin\,x\,5\arcmin\,$ finding chart.}
\label{fig:J2021}
\end{figure*}

\begin{figure*}
\begin{center}
\includegraphics[height=7.9cm,width=8.4cm,angle=0]{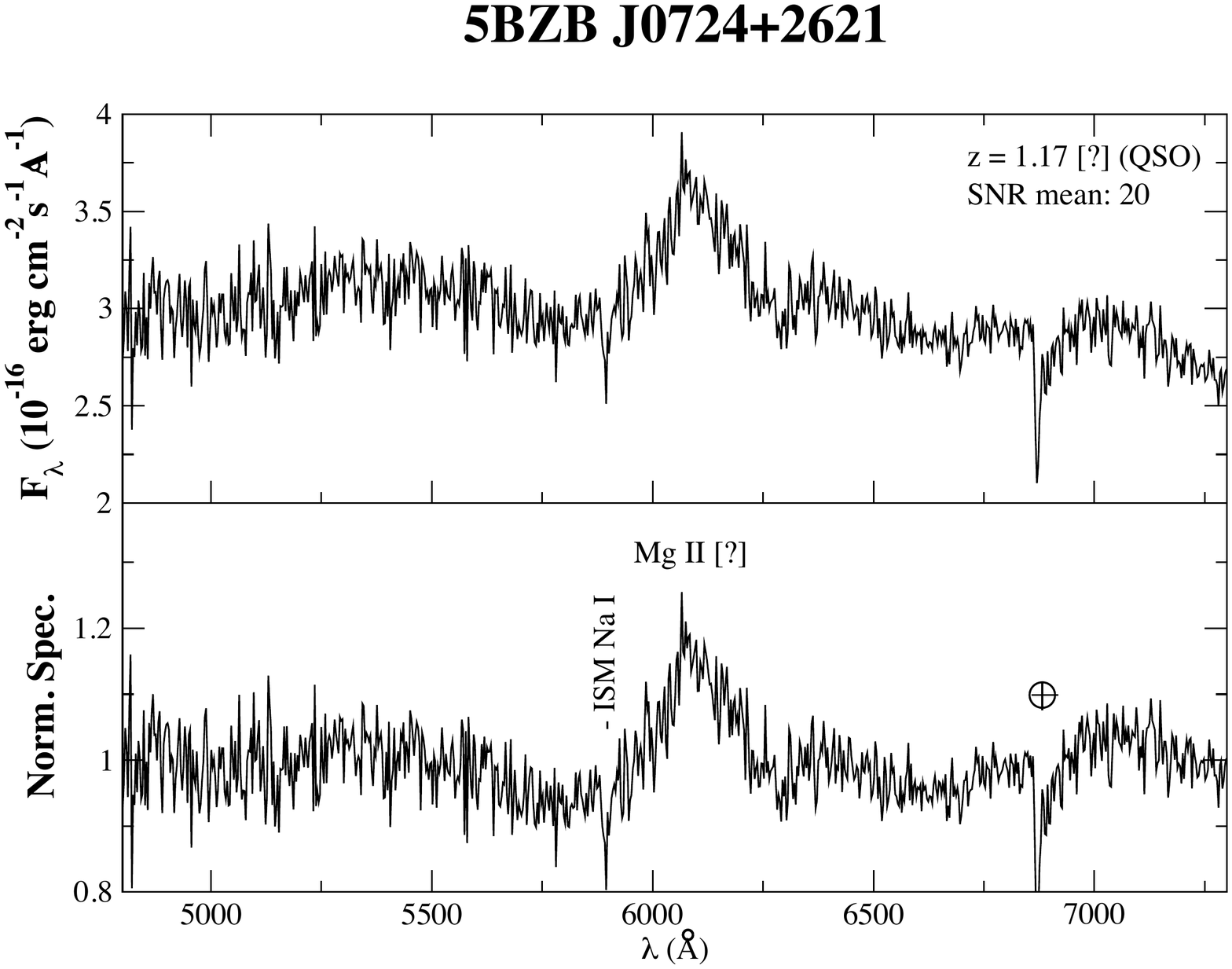} 
\includegraphics[height=6.5cm,width=8.0cm,angle=0]{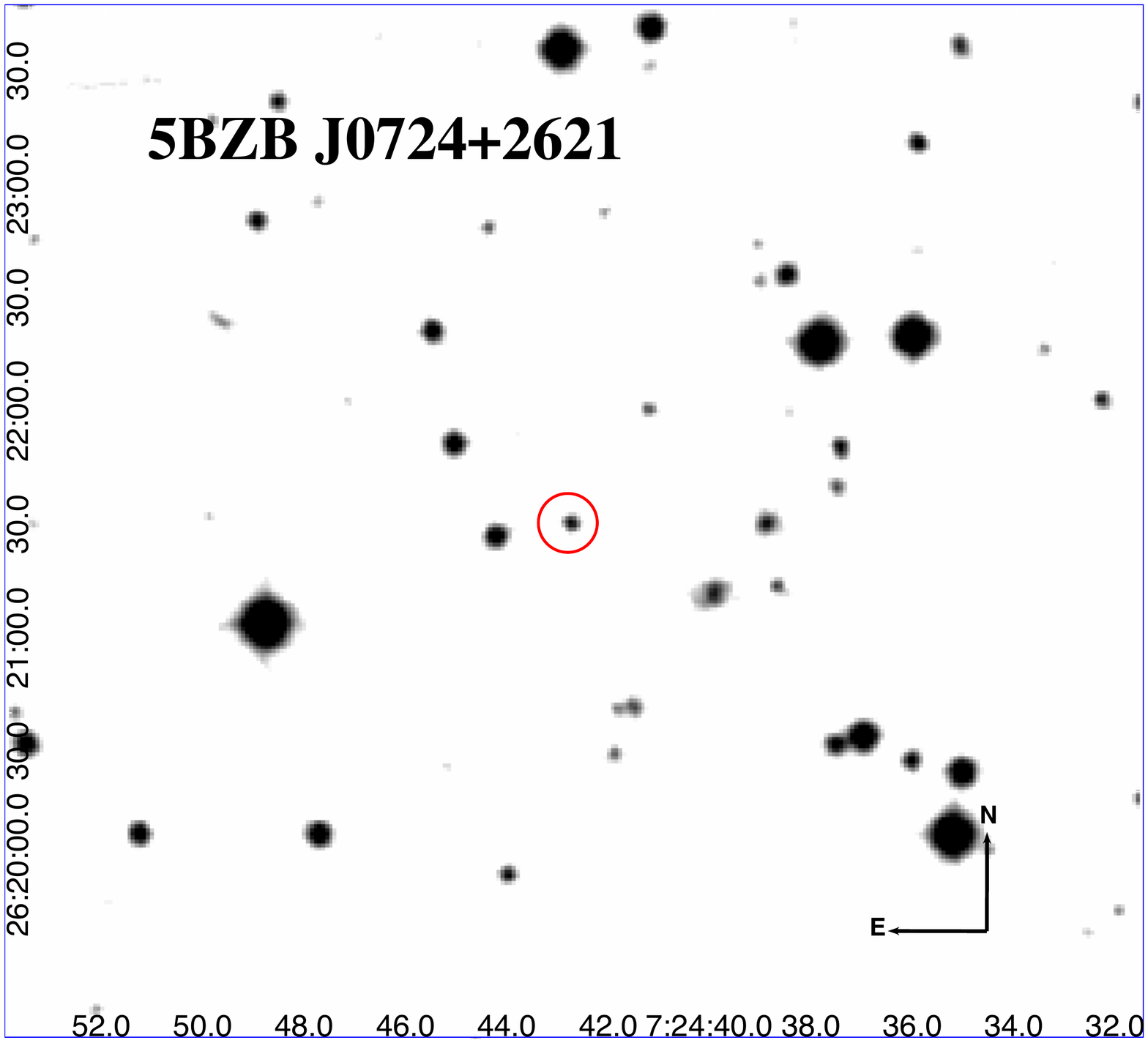} 
\end{center}

\caption{\emph{Left:} Upper panel) The optical spectra of 5BZBJ0724+262 listed in the \bzcat. It is classified as a QSO at $z $= 1.17. 
Broad emission line Mg ($\lambda_{obs} =  6099 \AA$). 
SNR is also indicated in the figure.
Lower panel) The normalized spectrum is shown here.
\emph{Right:} The $5\arcmin\,x\,5\arcmin\,$ finding chart.}
\label{fig:b0724}
\end{figure*}

\begin{figure*}
\begin{center}
\includegraphics[height=7.9cm,width=8.4cm,angle=0]{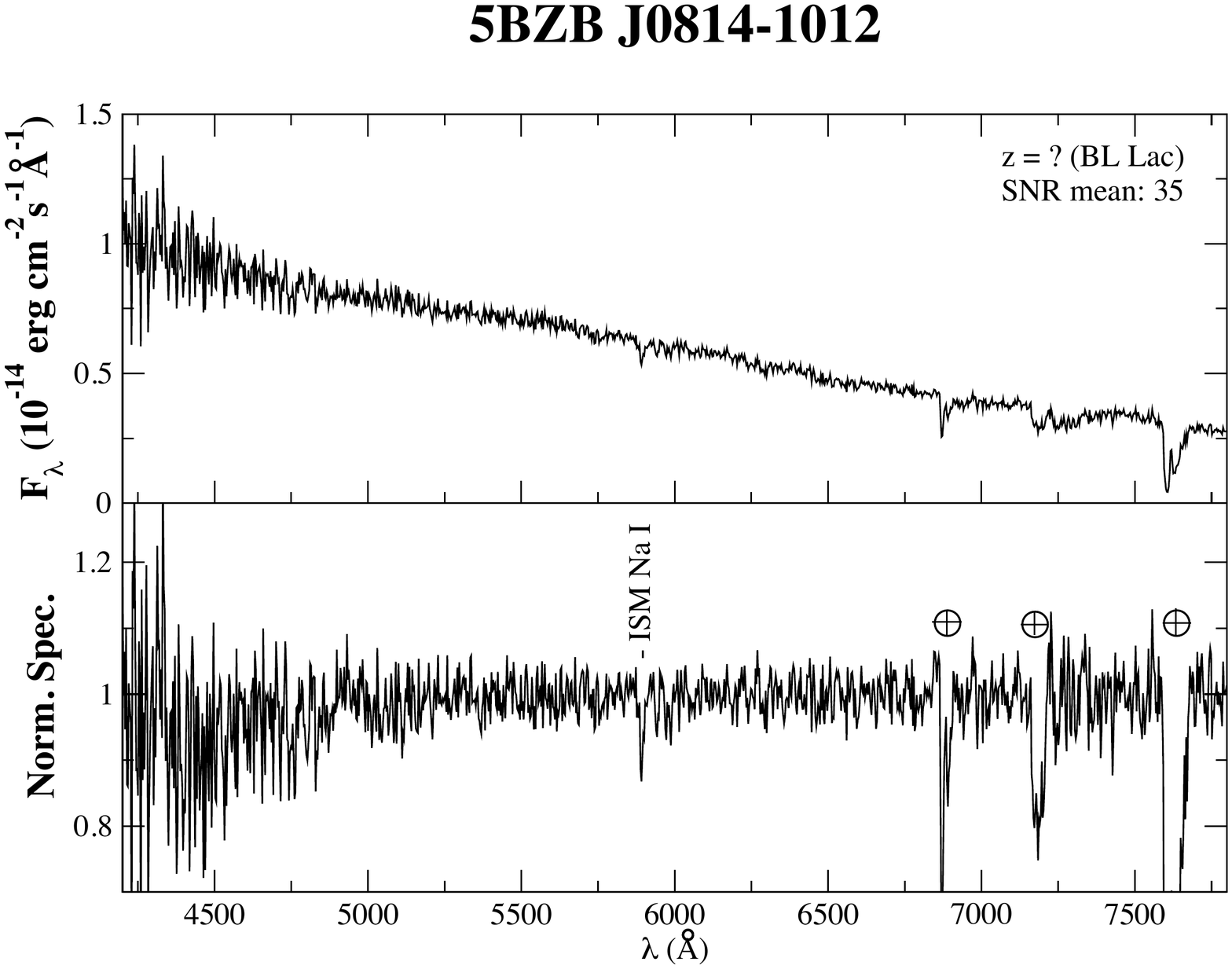} 
\includegraphics[height=6.5cm,width=8.0cm,angle=0]{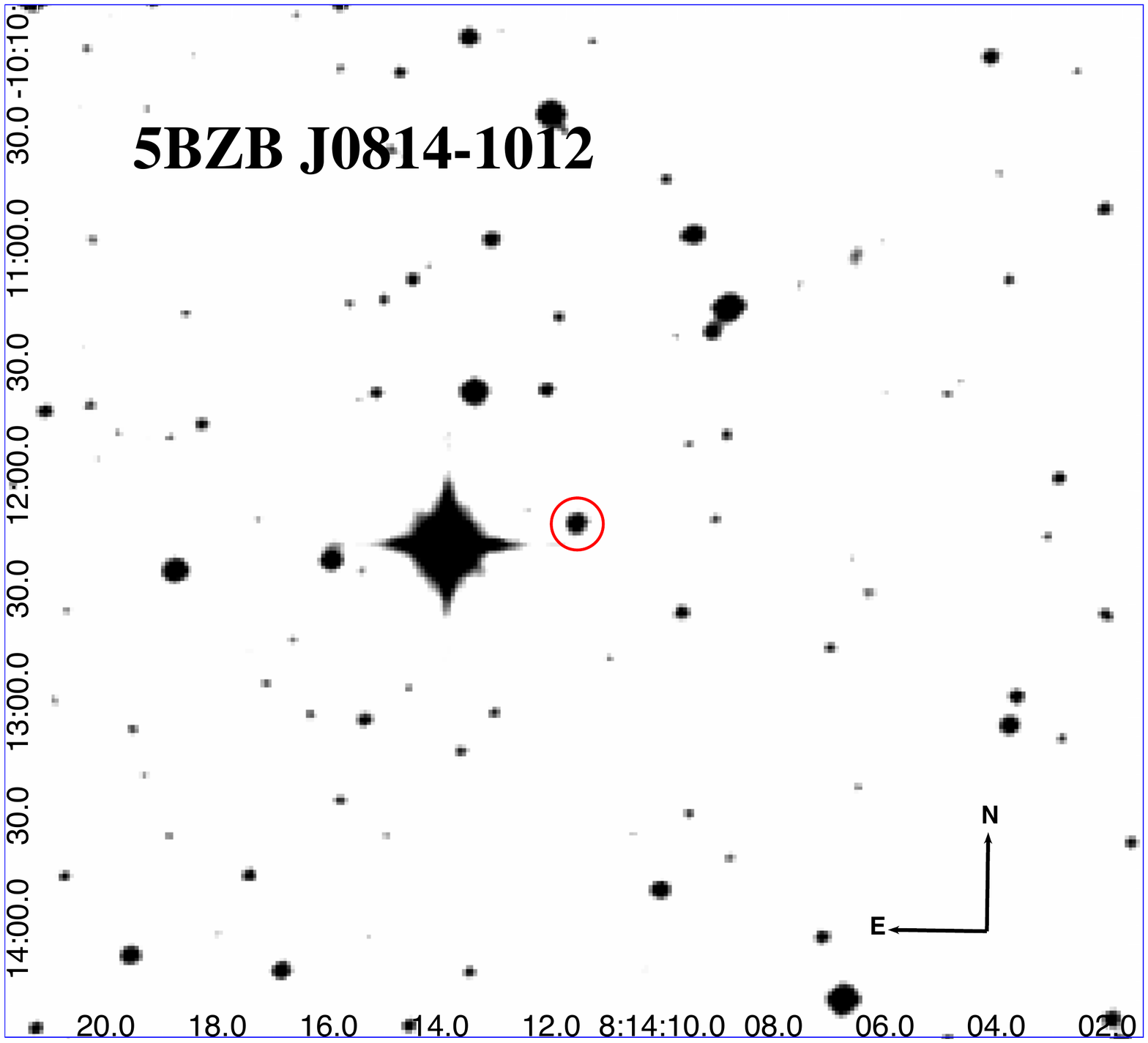} 
\end{center}

\caption{\emph{Left:}  Upper panel) The optical spectra of 5BZBJ0814-1012 listed in the \bzcat\, potential counterpart of 
3FGL J0814.1-1012. It is classified as a BZB on the basis of its featureless continuum.
SNR is also indicated in the figure.
Lower panel) The normalized spectrum is shown here.
\emph{Right:}  The $5\arcmin\,x\,5\arcmin\,$ finding chart.}
\label{fig:b0814}
\end{figure*}

\begin{figure*}
\begin{center}
\includegraphics[height=7.9cm,width=8.4cm,angle=0]{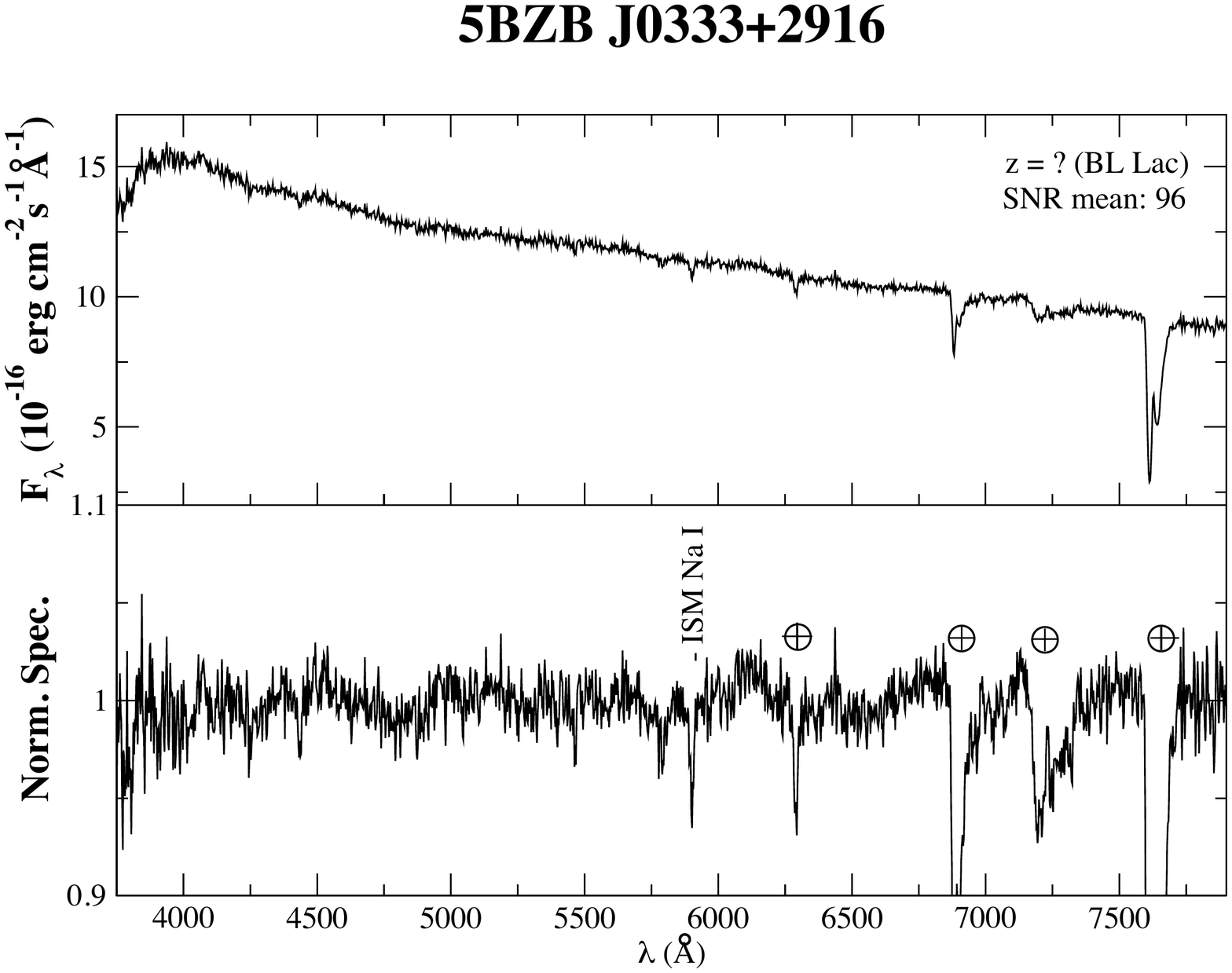} 
\includegraphics[height=6.5cm,width=8.0cm,angle=0]{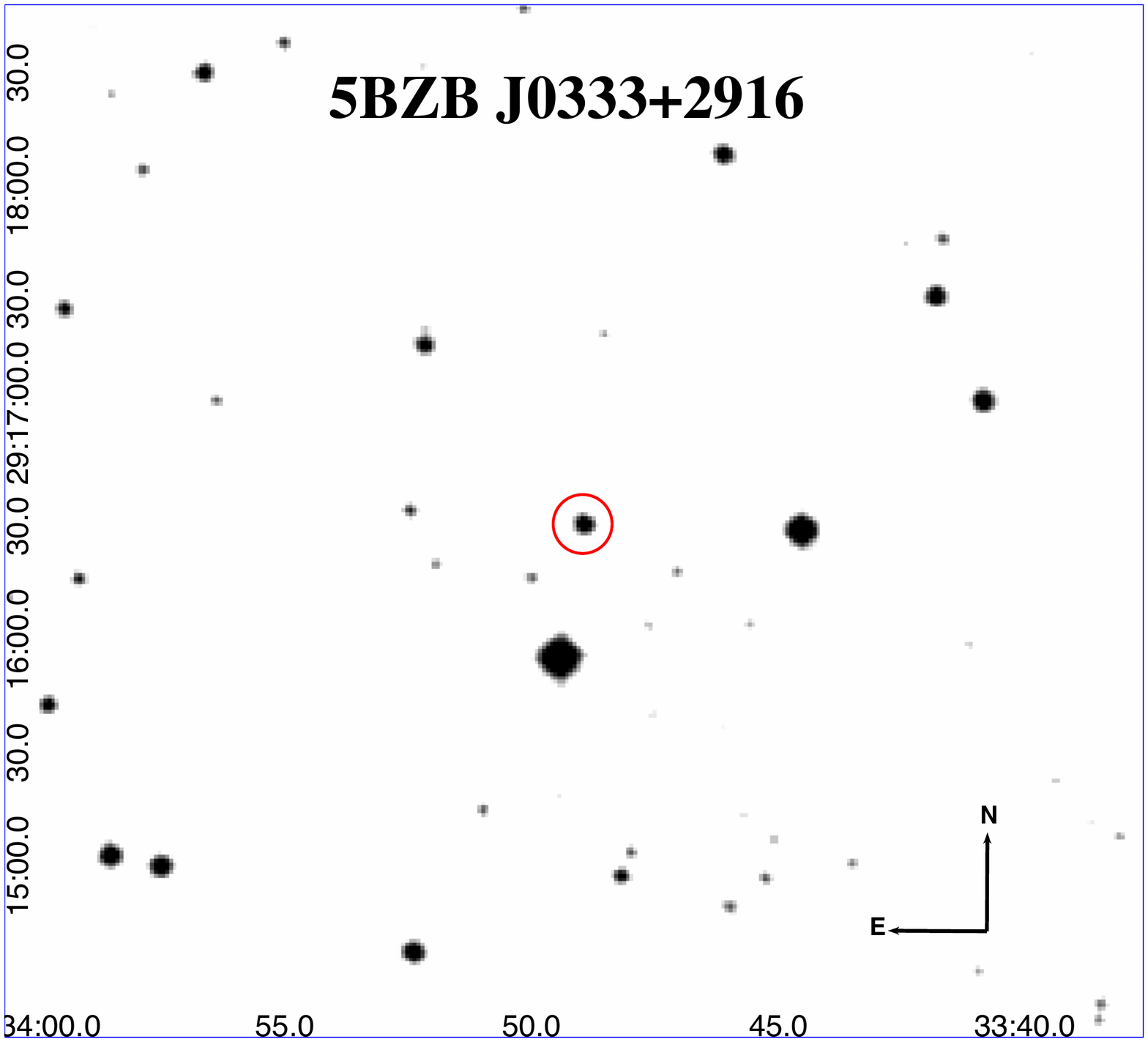} 
\end{center}

\caption{\emph{Left:}   Upper panel) The optical spectra of 5BZB J0333+2916 listed in the \bzcat\, potential counterpart of 
3FGL J0333.6+2916. It is classified as a BZB on the basis of its featureless continuum.
SNR is also indicated in the figure.
Lower panel) The normalized spectrum is shown here.
\emph{Right:}  The $5\arcmin\,x\,5\arcmin\,$ finding chart.}
\label{fig:J0333}
\end{figure*}

\begin{figure*}
\begin{center}
\includegraphics[height=7.9cm,width=8.4cm,angle=0]{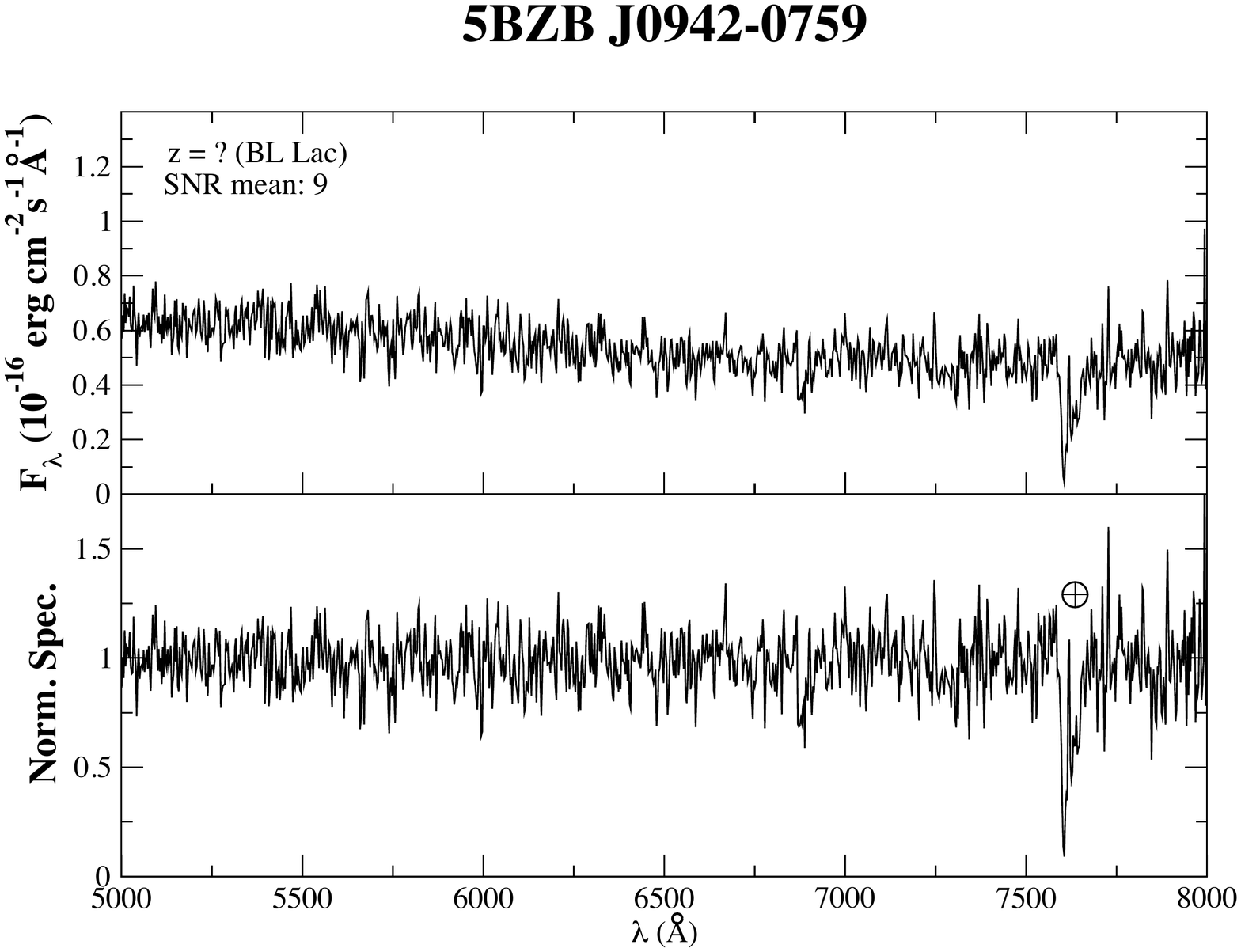} 
\includegraphics[height=6.5cm,width=8.0cm,angle=0]{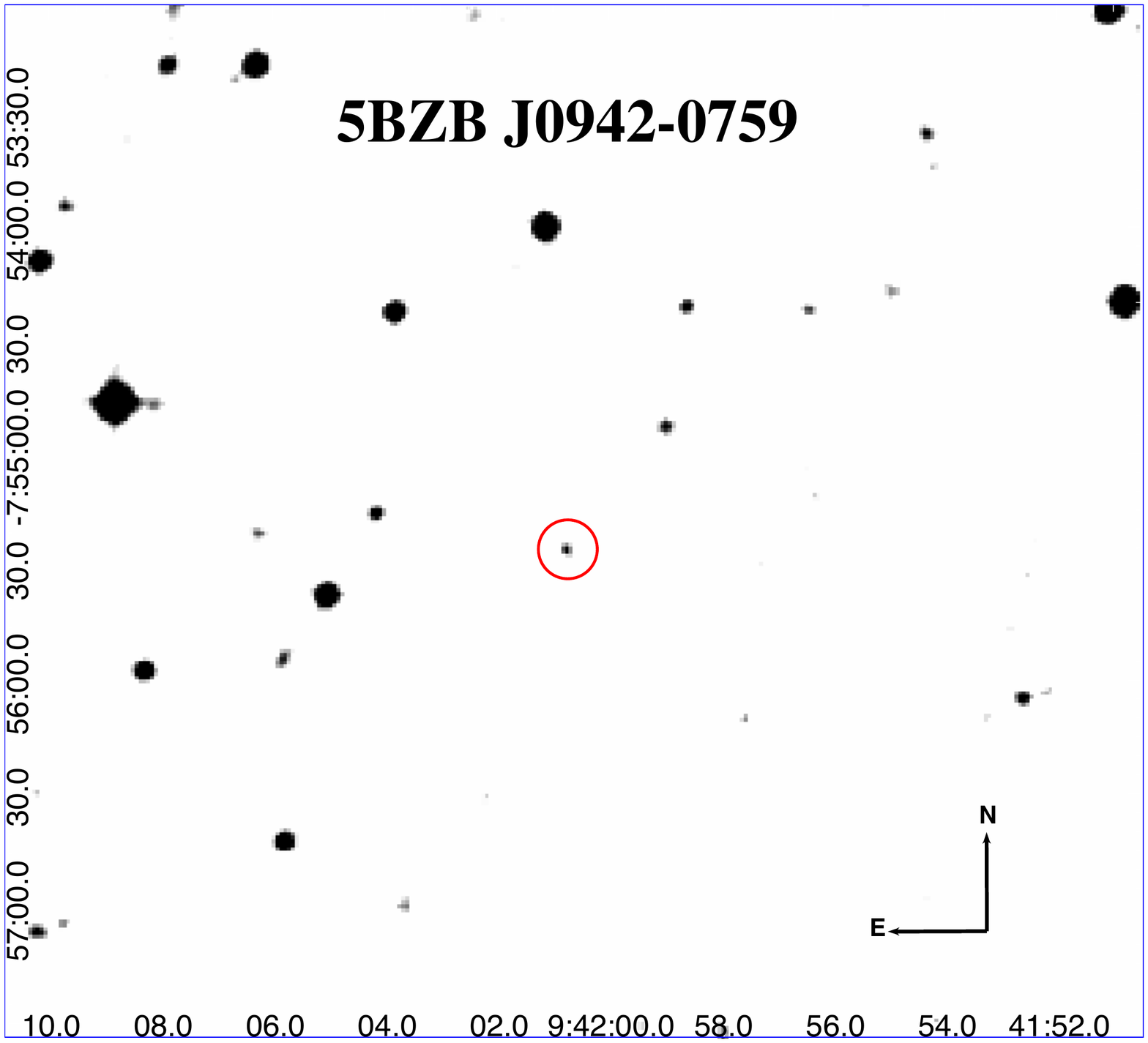} 
\end{center}

\caption{\emph{Left:} Upper panel) The optical spectra of 5BZB J0942-0759 listed in the \bzcat\, potential counterpart of 
3FGL J0942.1-0756. It is classified as a BZB on the basis of its featureless continuum.
SNR is also indicated in the figure.
Lower panel) The normalized spectrum is shown here.
\emph{Right:}   The $5\arcmin\,x\,5\arcmin\,$ finding chart.}
\label{fig:J0942}
\end{figure*}

\begin{figure*}
\begin{center}
\includegraphics[height=7.9cm,width=8.4cm,angle=0]{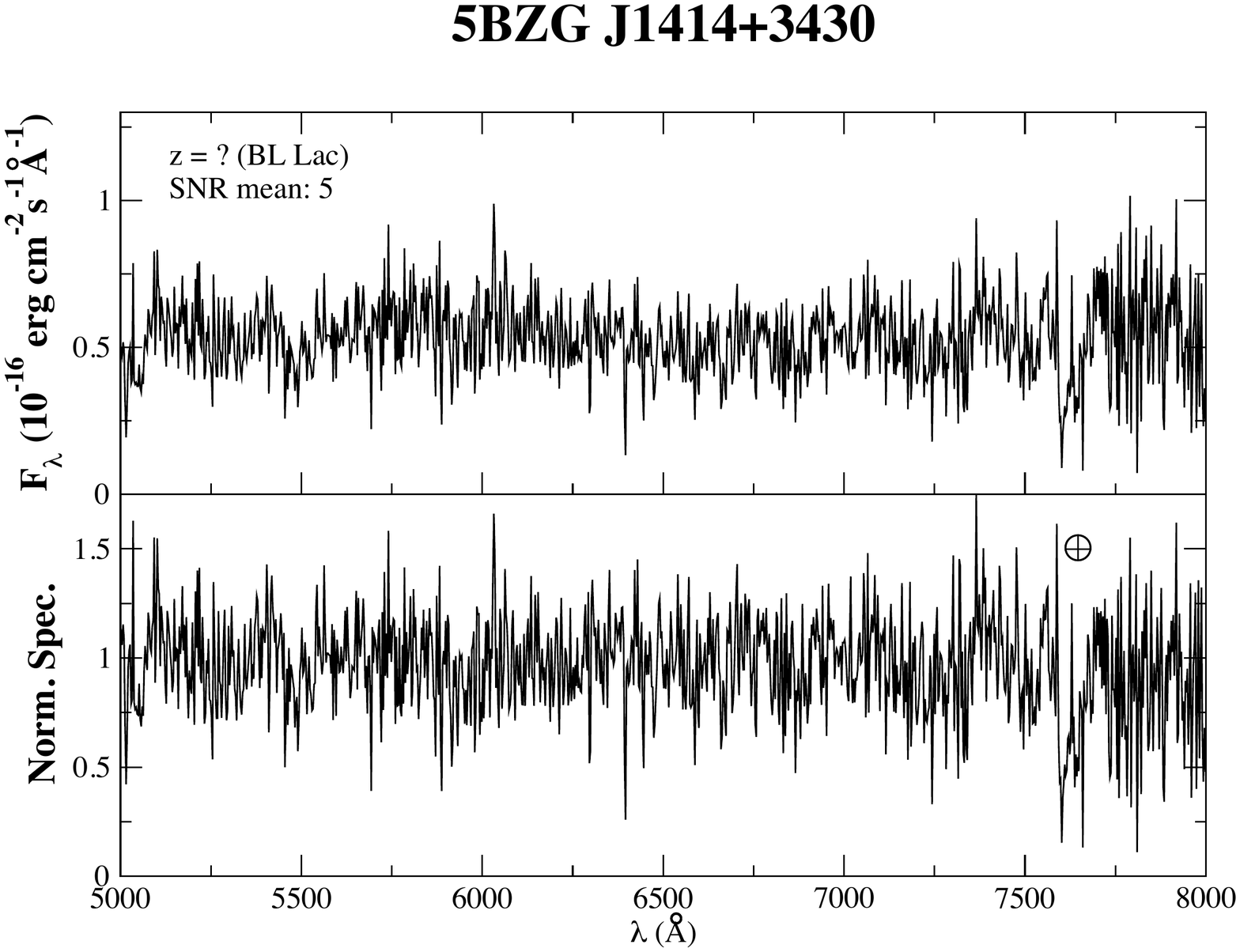} 
\includegraphics[height=6.5cm,width=8.0cm,angle=0]{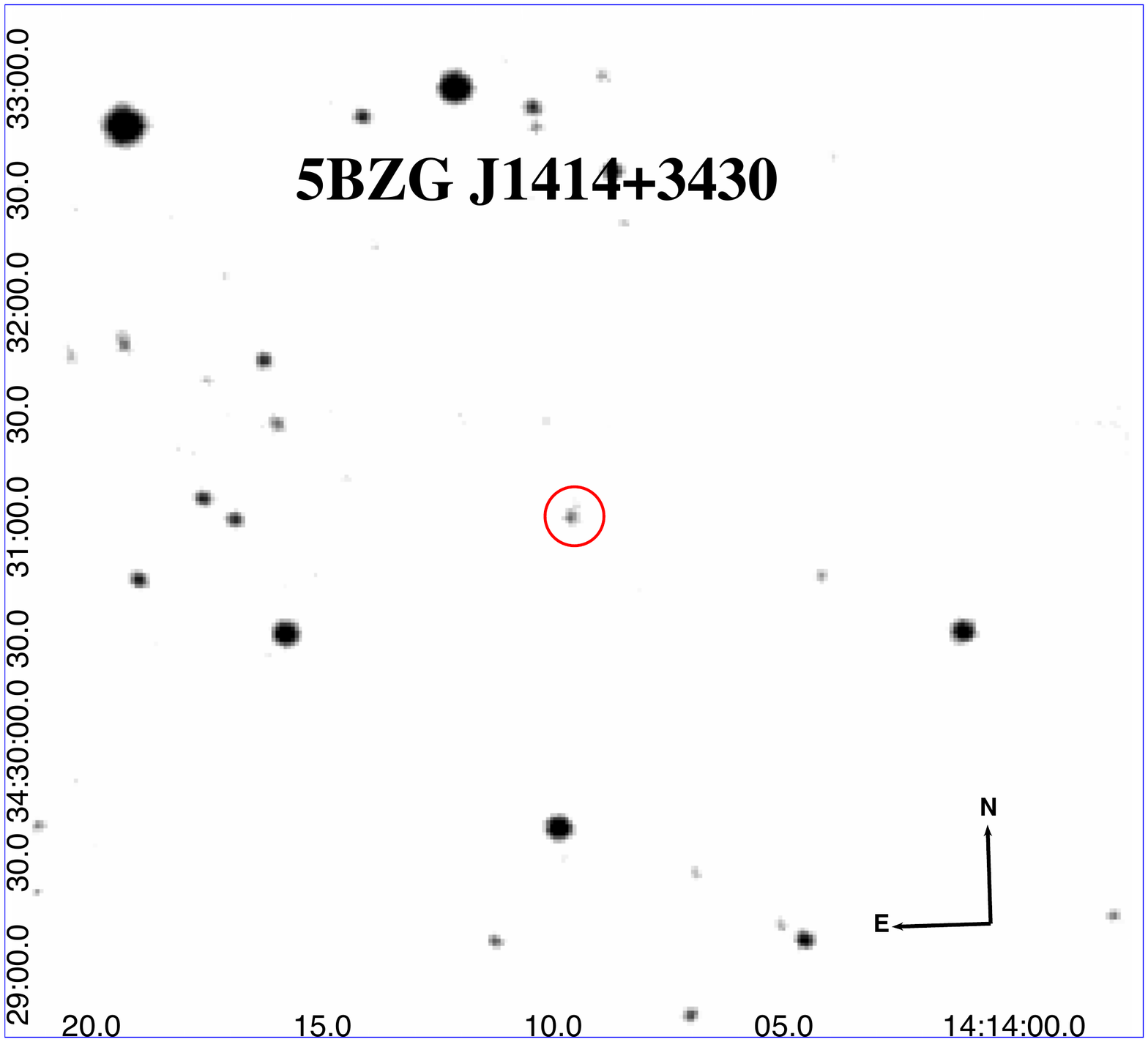} 
\end{center}

\caption{\emph{Left:}  Upper panel) The optical spectra of 5BZG J1414+3430 listed in the \bzcat\. 
It is classified as a BZB on the basis of its featureless continuum.
SNR is also indicated in the figure.
Lower panel) The normalized spectrum is shown here.
\emph{Right:}  The $5\arcmin\,x\,5\arcmin\,$ finding chart.}
\label{fig:b1414}
\end{figure*}

\begin{figure*}
\begin{center}
\includegraphics[height=7.9cm,width=8.4cm,angle=0]{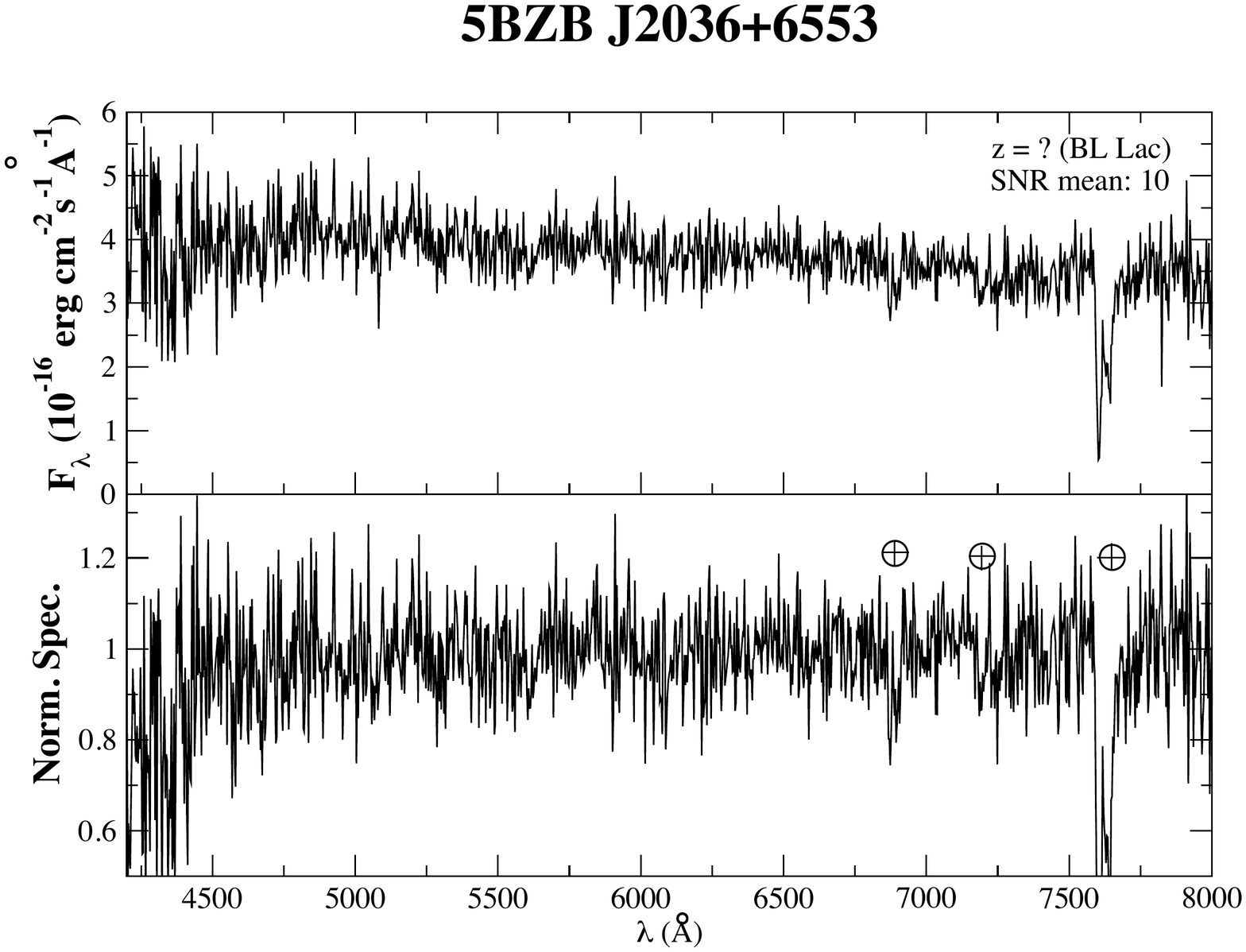} 
\includegraphics[height=6.5cm,width=8.0cm,angle=0]{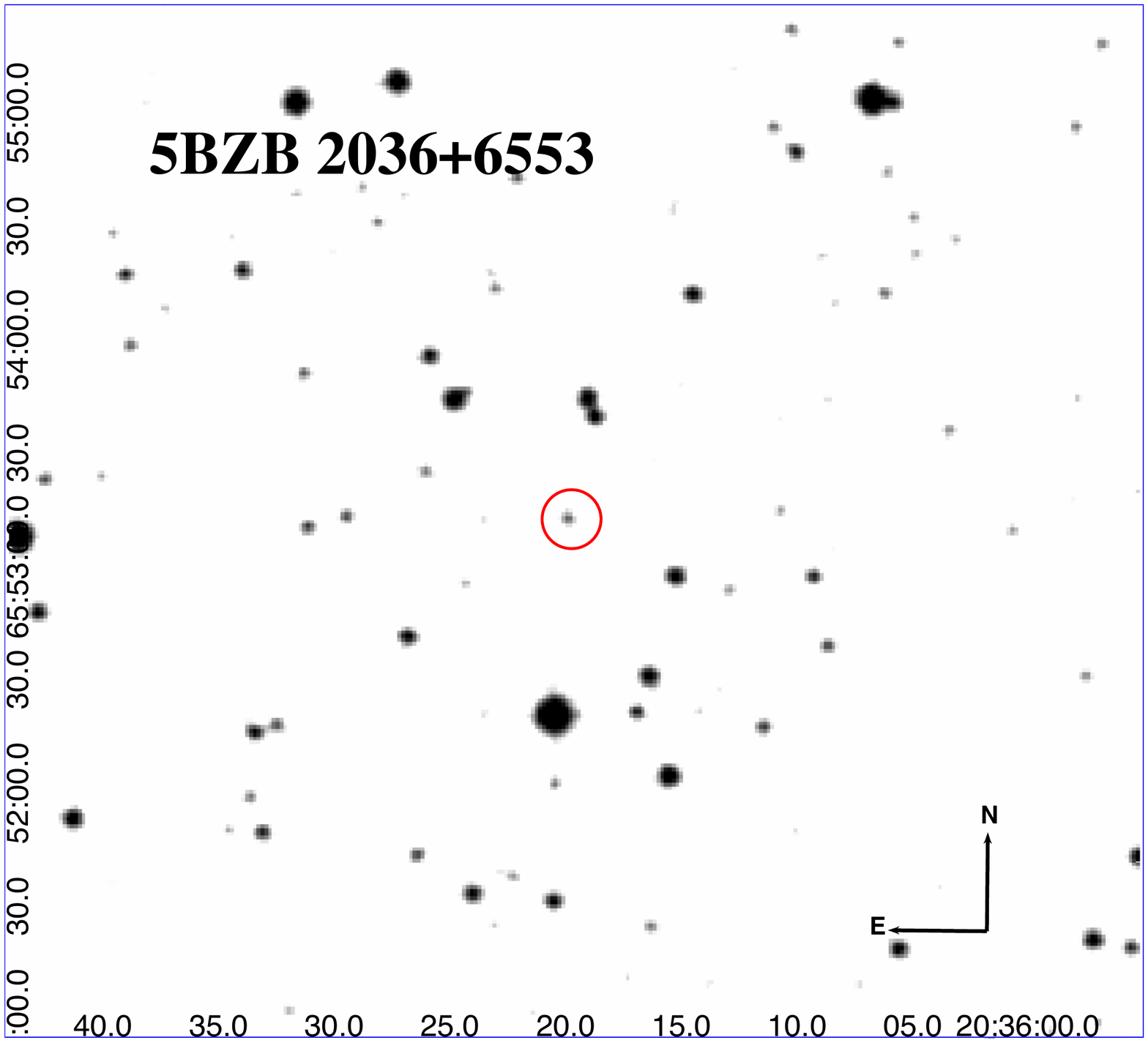} 
\end{center}

\caption{\emph{Left:}  Upper panel) The optical spectra of 5BZB J2036+6553	 listed in the \bzcat\
potential counterpart of 3FGL J2036.4+6551.
It is classified as a BZB on the basis of its featureless continuum.
SNR is also indicated in the figure.
Lower panel) The normalized spectrum is shown here.
\emph{Right:}  The $5\arcmin\,x\,5\arcmin\,$ finding chart.}
\label{fig:J2036}
\end{figure*}

\begin{figure*}
\begin{center}
\includegraphics[height=7.9cm,width=8.4cm,angle=0]{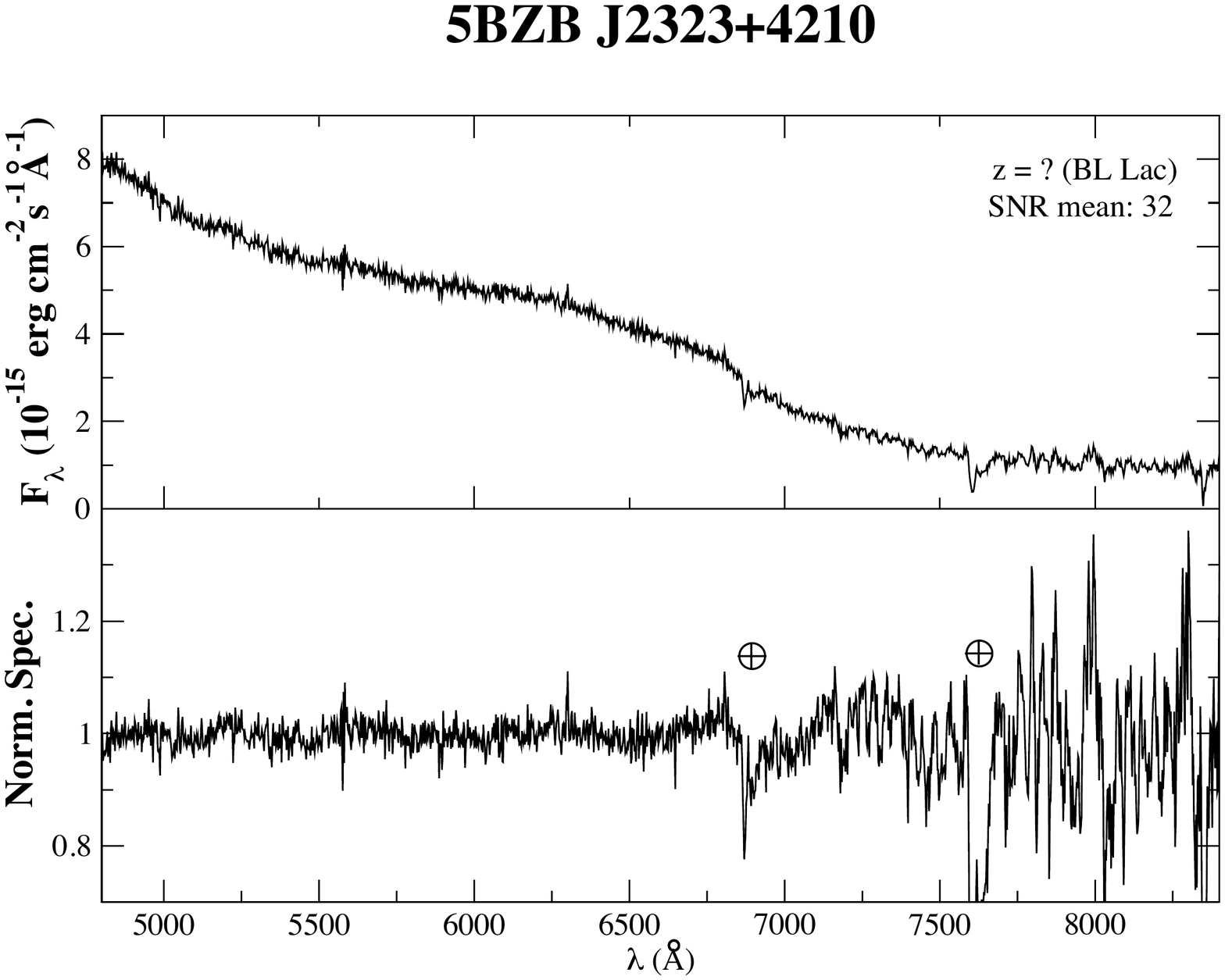} 
\includegraphics[height=6.5cm,width=8.0cm,angle=0]{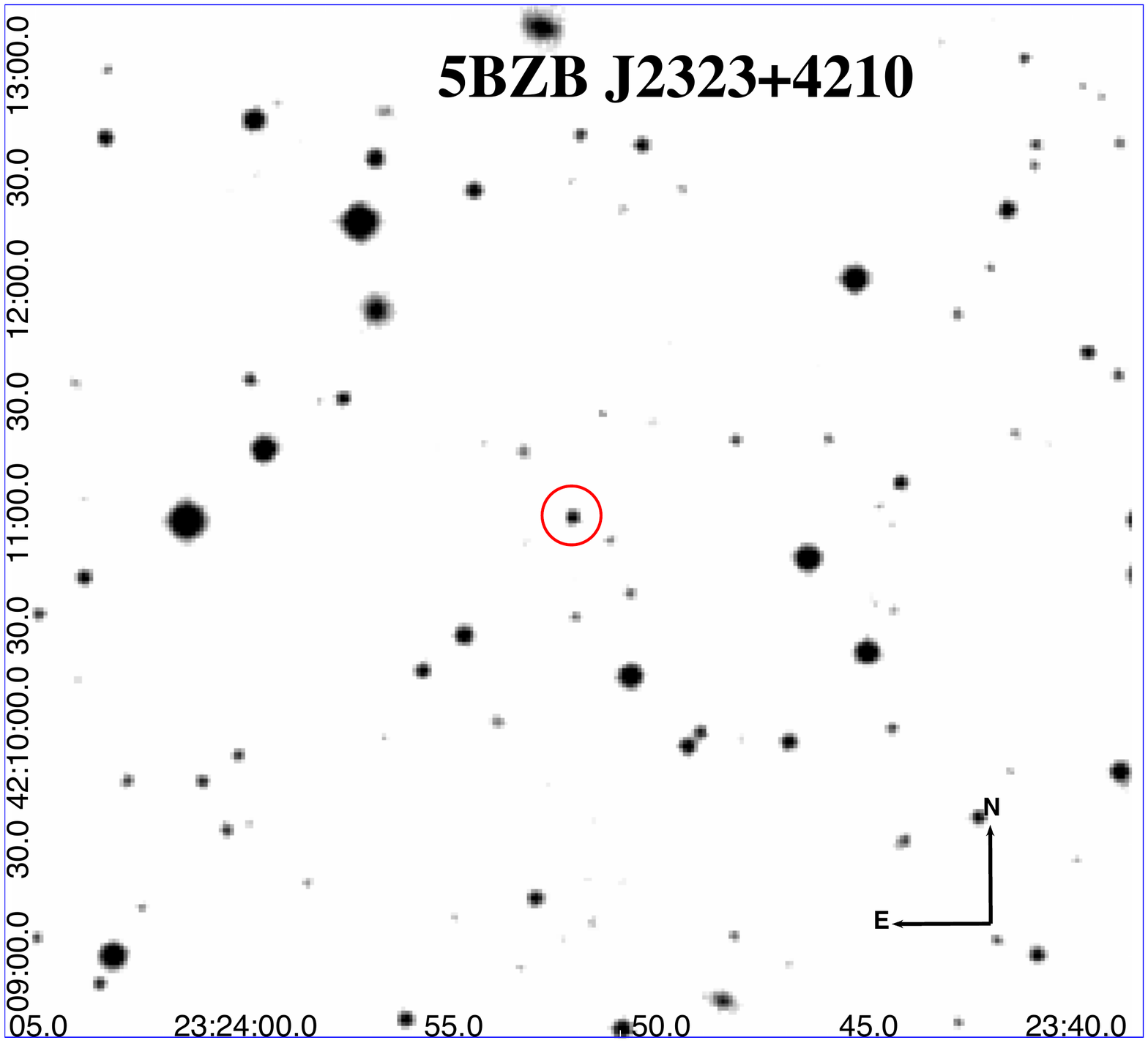} 
\end{center}

\caption{\emph{Left:} Upper panel) The optical spectra of 5BZB J2323+4210 listed in the \bzcat\
potential counterpart of 3FGL J2323.9+4211.
It is classified as a BZB on the basis of its featureless continuum. SNR is also indicated in the figure.
Lower panel) The normalized spectrum is shown here.
\emph{Right:}   The $5\arcmin\,x\,5\arcmin\,$,finding chart.}
\label{fig:b2323}
\end{figure*}


\end{document}